\documentclass[conference]{IEEEtran}

\usepackage[space,nocompress]{cite}
\usepackage{balance,amsfonts,amssymb,booktabs,graphicx,hyperref,cleveref,multirow,courier,url,textcomp,pifont,array,makecell,diagbox,xspace}
\usepackage{bm}
\usepackage{tabularx}
\usepackage{enumerate}
\crefname{equation}{Eq.}{Eqs.}
\crefname{figure}{Fig.}{Figs.}
\crefname{algocf}{alg.}{algs.}
\Crefname{algocf}{Alg.}{Algorithms}

\usepackage[cmex10]{amsmath}
\usepackage{algorithm}
\usepackage{algorithmicx}
\usepackage{algpseudocode}
\usepackage[tight,footnotesize]{subfigure}
\usepackage{fixltx2e}

\usepackage{bbding}
\usepackage[usenames,dvipsnames]{color}

\newcolumntype{L}[1]{>{\raggedright\arraybackslash}p{#1}}  
\newcolumntype{C}[1]{>{\centering\arraybackslash}p{#1}}  
\newcolumntype{R}[1]{>{\raggedleft\arraybackslash}p{#1}}

\newcommand{\system}{{\sc TextBugger}\xspace}

\begin{document}
\bstctlcite{IEEEexample:BSTcontrol}

\title{\textsc{TextBugger}: Generating Adversarial Text Against Real-world Applications}

\author{\IEEEauthorblockN{Jinfeng Li\IEEEauthorrefmark{1},
Shouling Ji\thanks{\Envelope \quad Shouling Ji is the corresponding author.}\IEEEauthorrefmark{1}\IEEEauthorrefmark{2} \Envelope,
Tianyu Du\IEEEauthorrefmark{1}, 
Bo Li\IEEEauthorrefmark{3} and
Ting Wang\IEEEauthorrefmark{4}}
\IEEEauthorblockA{\IEEEauthorrefmark{1} Institute of Cyberspace Research and College of Computer Science and Technology, Zhejiang University \\Email: \{lijinfeng0713, sji, zjradty\}@zju.edu.cn}
\IEEEauthorblockA{\IEEEauthorrefmark{2} Alibaba-Zhejiang University Joint Research Institute of Frontier Technologies}
\IEEEauthorblockA{\IEEEauthorrefmark{3} University of Illinois Urbana-Champaign, Email: lxbosky@gmail.com}
\IEEEauthorblockA{\IEEEauthorrefmark{4} Lehigh University, Email: inbox.ting@gmail.com}}

\IEEEoverridecommandlockouts
\makeatletter\def\@IEEEpubidpullup{6.5\baselineskip}\makeatother
\IEEEpubid{\parbox{\columnwidth}{
    Network and Distributed Systems Security (NDSS) Symposium 2019\\
    24-27 February 2019, San Diego, CA, USA\\
    ISBN 1-891562-55-X\\
    https://dx.doi.org/10.14722/ndss.2019.23138\\
    www.ndss-symposium.org
}
\hspace{\columnsep}\makebox[\columnwidth]{}}

\maketitle

\begin{abstract}
Deep Learning-based Text Understanding (DLTU) is the backbone technique behind various applications, including question answering, machine translation, and text classification. Despite its tremendous popularity, the security vulnerabilities of DLTU are still largely unknown, which is highly concerning given its increasing use in security-sensitive applications such as sentiment analysis and toxic content detection. In this paper, we show that DLTU is inherently vulnerable to adversarial text attacks, in which maliciously crafted texts trigger target DLTU systems and services to misbehave. Specifically, we present \system, a general attack framework for generating adversarial texts. In contrast to prior works, \system differs in significant ways: (i) effective -- it outperforms state-of-the-art attacks 
in terms of attack success rate; (ii) evasive -- it preserves the utility of benign text, with 94.9\% of the adversarial text correctly recognized by human readers; and (iii) efficient -- it generates adversarial text with computational complexity sub-linear to the text length. We empirically evaluate \system on a set of real-world DLTU systems and services used for sentiment analysis and toxic content detection, demonstrating its effectiveness, evasiveness, and efficiency. 
For instance, \system achieves 100\% success rate on the IMDB dataset based on Amazon AWS Comprehend within 4.61 seconds and preserves 97\% semantic similarity. We further discuss possible defense mechanisms to mitigate such attack and the adversary's potential countermeasures, which leads to promising directions for further research.

\end{abstract}

\section{Introduction} \label{introduction}
Deep neural networks (DNNs) have been shown to achieve great success in various tasks such as classification, regression, and decision making. Such advances in DNNs have led to broad deployment of systems on important problems in physical world. However, though DNNs models have exhibited state-of-the-art performance in a lot of applications, recently they have been found to be vulnerable against adversarial examples which are carefully generated by adding small perturbations to the legitimate inputs to fool the targeted models~\cite{Goodfellow:2014tl,Szegedy:2014,nguyen2015deep,cheng2018seq2sick,lingdeepsec,xiao2018generating}. Such discovery has also raised serious concerns, especially when deploying such machine learning models to security-sensitive tasks.

In the meantime, DNNs-based text classification plays a more and more important role in information understanding and analysis nowadays. 
For instance, many online recommendation systems rely on the sentiment analysis of user reviews/comments \cite{medhat2014sentiment}.
Generally, such systems would classify the reviews/comments into two or three categories and then take the results into consideration when ranking movies/products.
Text classification is also important for enhancing the safety of online discussion environments, e.g., automatically detect online toxic content \cite{nobata2016abusive}, including irony, sarcasm, insults, harassment and abusive content.

Many studies have investigated the security of current machine learning models and proposed different attack methods, including causative attacks and exploratory attacks \cite{barreno2006can,barreno2010security, huang2011adversarial}.
Causative attacks aim to manipulate the training data thus misleading the classifier itself, and exploratory attacks craft malicious testing instances (adversarial examples) so as to evade a given classifier.
To defend against these attacks, several mechanisms have been proposed to obtain robust classifiers \cite{sculley2006spam,biggio2011design}.
Recently, adversarial attacks have been shown to be able to achieve a high attack success rate in image classification tasks \cite{Carlini:2017jb}, which has posed severe physical threats to many intelligent devices (e.g., self-driving cars)~\cite{evtimov2017robust}.

While existing works on adversarial examples mainly focus on the image domain, it is more challenging to deal with text data due to its discrete property,  which is hard to optimize.
Furthermore, in the image domain, the perturbation can often be made virtually imperceptible to human perception, causing humans and state-of-the-art models to disagree. 
However, in the text domain, small perturbations are usually clearly perceptible, and the replacement of a single word may drastically alter the semantics of the sentence.
In general, existing attack algorithms designed for images cannot be directly applied to text, and we need to study new attack techniques and corresponding defenses.



\begin{figure*}[tp]
    \centering
    \includegraphics[width=0.85\textwidth]{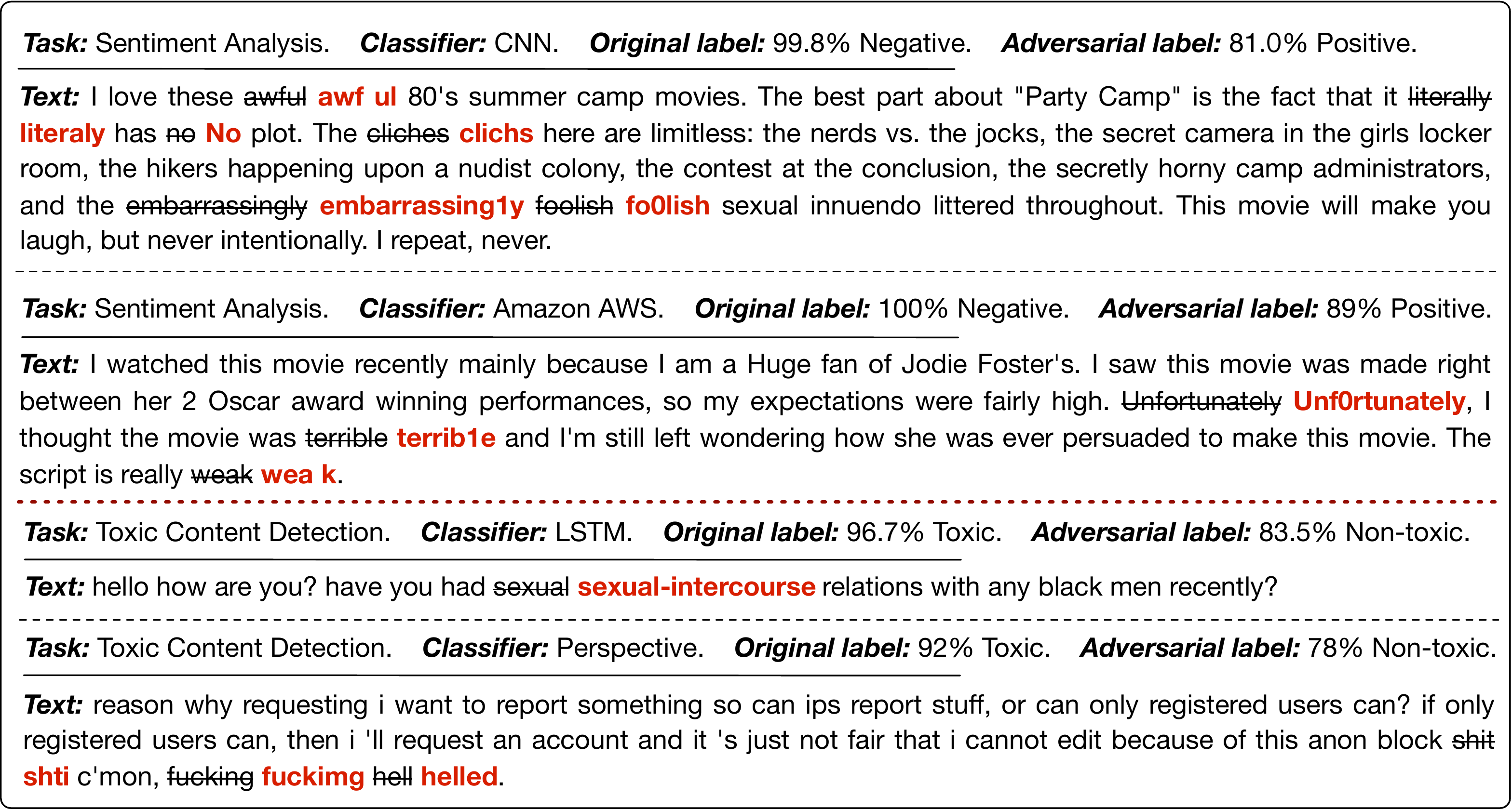}
    \caption{Adversarial examples against two natural language classification tasks. Replacing a fraction of the words in a document with adversarially-chosen bugs fools classifiers into predicting an incorrect label. The new document is classified correctly by humans and preserves most of the original meaning although it contains small perturbations. }
    \label{fig:examples}
    \vspace{-0.25cm} 
\end{figure*}

Recently, some mechanisms are proposed towards generating adversarial texts \cite{liang2017deep,samanta2017towards}. 
These work proposed to generate adversarial texts by replacing a word with an out-of-vocabulary one \cite{belinkov2017synthetic,hosseini2017deceiving,gao2018black}.
Although seminal, they are limited in practice due to the following reasons: (i) they are not computationally efficient, (ii) they are designed under the white-box setting, (iii) they require manual intervention, and/or (iv) they are designed against a particular NLP model and are not comprehensively evaluated.
Thus, the efficiency and effectiveness of current adversarial text generation techniques and the robustness of popular text classification models need to be studied.

In this paper, we propose \textsc{TextBugger}, a framework that can effectively and efficiently generate utility-preserving (i.e., keep its original meaning for human readers) adversarial texts against state-of-the-art text classification systems under both white-box and black-box settings.
In the white-box scenario, we first find important words by computing the Jacobian matrix of the classifier and then choose an optimal perturbation from the generated five kinds of perturbations.
In the black-box scenario, we first find the important sentences, and then use a scoring function to find important words to manipulate. 
Through extensive experiments under both settings, we show that an adversary can deceive multiple real-world online DLTU systems with the generated adversarial texts\footnote{We have reported our findings to their companies, and they replied that they would fix these bugs in the next version.}, including Google Cloud NLP, Microsoft Azure Text Analytics, IBM Watson Natural Language Understanding and Amazon AWS Comprehend, etc.
Several adversarial examples are shown in \Cref{fig:examples}.
The existence of such adversarial examples causes a lot of concerns for text classification systems and seriously undermines their usability.

\textbf{Our Contribution.}
Our main contributions can be summarized as follows.
\begin{itemize}
	\item We propose \textsc{TextBugger}, a framework that can effectively and efficiently generate utility-preserving adversarial texts under both white-box and black-box settings.
    \item We evaluate \system on a group of state-of-the-art machine learning models and popular real-world online DLTU applications, including sentiment analysis and toxic content detection. Experimental results show that \system is very effective and efficient. For instance, \system achieves 100\% attack success rate on the IMDB dataset when targeting the Amazon AWS and Microsoft Azure platforms under black-box settings. 
    We shows that transferability also exists in the text domain and the adversarial texts generated against offline models can be successfully transferred to multiple popular online DLTU systems.
    \item We conduct a user study on our generated adversarial texts and show that \system has little impact on human understanding.
    \item We further discuss two potential defense strategies to defend against the above attacks along with preliminary evaluations. Our results can encourage building more robust DLTU systems in the future.
\end{itemize}

\section{Attack Design} \label{sec:methodology}

\subsection{Problem Formulation}
Given a pre-trained text classification model $\mathcal{F}:\mathcal{X} \to \mathcal{Y}$, which maps from feature space $\mathcal{X}$ to a set of classes $\mathcal{Y}$, an adversary aims to generate an adversarial document $\bm{x_{adv}}$ from a legitimate document $\bm{x} \in \mathcal{X}$ whose ground truth label is $y \in \mathcal{Y}$, so that $\mathcal{F}(\bm{x_{adv}}) = t$ $(t\neq y)$.
The adversary also requires $S(\bm{x},\bm{x_{adv}}) \geq \epsilon$ for a domain-specific similarity function $S:\mathcal{X} \times \mathcal{X} \to \mathbb{R}_+$, where the bound $\epsilon \in \mathbb{R}$ captures the notion of utility-preserving alteration. For instance, in the context of text classification tasks, we may use $S$ to capture the semantic similarity between $\bm{x}$ and $\bm{x_{adv}}$.

\subsection{Threat Model}

We consider both white-box and black-box settings to evaluate different adversarial abilities.

\textbf{White-box Setting.}
We assume that attackers have complete knowledge about the targeted model including the model architecture parameters. 
White-box attacks find or approximate the worst-case attack for a particular model and input based on the kerckhoff's principle~\cite{shannon1949communication}. Therefore,  white-box attacks can expose a model's worst case vulnerabilities.

\begin{figure}[tp]
    \centering
    \includegraphics[width=0.48\textwidth]{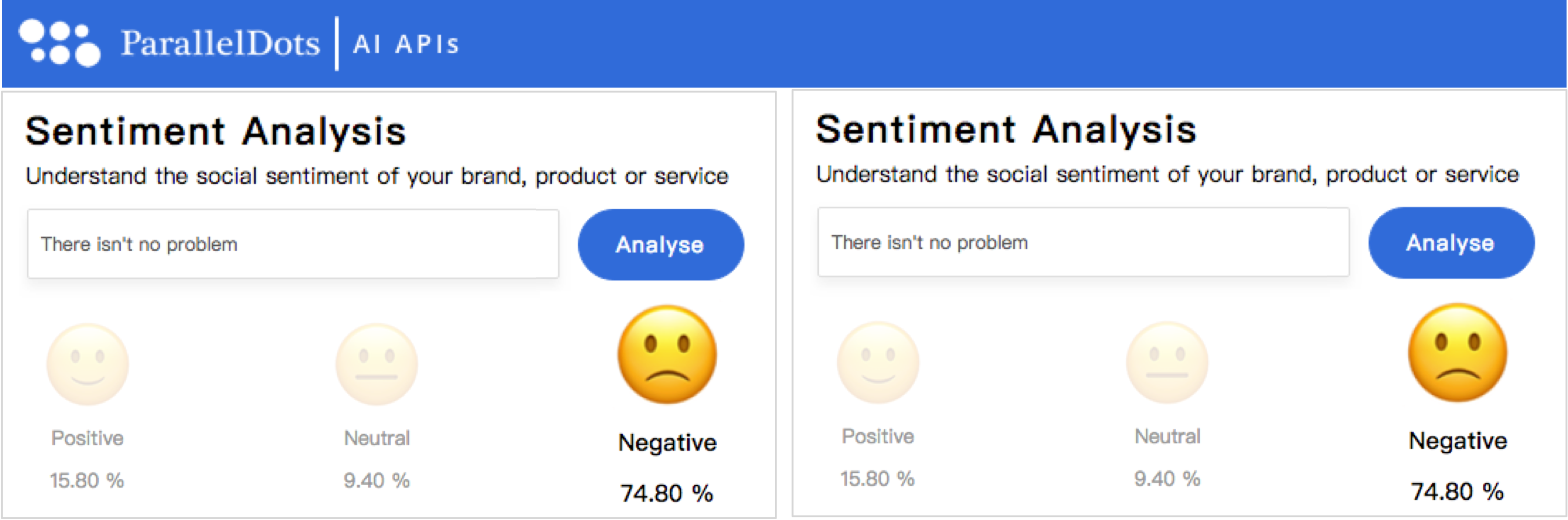}
    \caption{ParallelDots API: An example of deep learning text classification platform, which is a black-box scenario.}
    \label{fig:online_text_classification_platform}
    \vspace{-0.25cm} 
\end{figure}

\textbf{Black-box Setting.}
With the development of machine learning, many companies have launched their own Machine-Learning-as-a-Service (MLaaS) for DLTU tasks such as text classification.
Generally, MLaaS platforms have similar system design: the model is deployed on the cloud servers, and users can only access the model via an API. 
In such cases, we assume that the attacker is not aware of the model architecture, parameters or training data, and is only capable of querying the target model with output as the prediction or confidence scores.
Note that the free usage of the API is limited among these platforms.
Therefore, if the attackers want to conduct practical attacks against these platforms, they must take such limitation and cost into consideration.
Specifically, we take the ParallelDots\footnote{\url{https://www.paralleldots.com/}} as an example and show its sentiment analysis API and the abusive content classifier API in \Cref{fig:online_text_classification_platform}.
From \Cref{fig:online_text_classification_platform}, we can see that the sentiment analysis API would return the confidence value of three classes, i.e., ``positive'', ``neutral'' and ``negative''.
Similarly, the abusive content classifier would return the confidence value of two classes, i.e., ``abusive'' and ``non abusive''.
For both APIs, the sum of confidence values of an instance equal to 1, and the class with the highest confidence value is considered as the input's class.

\subsection{\textsc{TextBugger}}

We propose efficient strategies to change a word slightly, which is sufficient for creating adversarial texts in both white-box settings and black-box settings.
Specifically, we call the slightly changed words ``bugs''.


\subsubsection{White-box Attack}
We first find important words by computing the Jacobian matrix of the classifier $\mathcal{F}$, and generate five kinds of bugs. Then we choose an optimal bug in terms of the change of the confidence value. 
The algorithm of white-box attack is shown in \Cref{alg:textbugger_whitebox}.

\textbf{Step 1: Find Important Words (line 2-5).}
The first step is to compute the Jacobian matrix for the given input text $\bm{x}=(x_1, x_2, \cdots, x_N)$ (line 2-4), where $x_i$ is the $i^{th}$ word, and $N$ represents the total number of words within the input text. 
For a text classification task, the output of $\mathcal{F}$ is more than one dimension. Therefore the matrix is as follows:
\begin{equation} 
J_{\mathcal{F}}(\bm{x}) = \frac{\partial \mathcal{F}(\bm{x})}{\partial \bm{x}} = \left[\frac{\partial \mathcal{F}_j(\bm{x})}{\partial x_i}\right]_{i \in 1..N, j\in 1..K}
\end{equation}
where $K$ represents the total number of classes in $\mathcal{Y}$, and  $\mathcal{F}_j(\cdot)$ represents the confidence value of the $j^{th}$ class.
The importance of word $x_i$ is defined as:
\begin{equation} \label{eq:jacobian_score}
C_{x_i} = J_{\mathcal{F}(i,y)} = \frac{\partial \mathcal{F}_y(\bm{x})}{\partial x_i}
\end{equation}
i.e., the partial derivative of the confidence value based on the predicted class $y$ regarding to the input word $x_i$.
This allows us to find the important words that have significant impact on the classifier's outputs.
Once we have calculated the importance score of each word within the input sequences, we sort these words in inverse order according to the importance value (line 5).

\begin{algorithm}[tp]  
\caption{\textsc{TextBugger} under white-box settings}  
\begin{algorithmic}[1]  \label{alg:textbugger_whitebox}
\Require legitimate document $\bm{x}$ and its ground truth label $y$, classifier $\mathcal{F(\cdot)}$, threshould $\epsilon$
\Ensure adversarial document $\bm{x_{adv}}$
\State Inititialize: $\bm{x'} \gets \bm{x}$

\For {word $x_i$ in $\bm{x}$}
\State Compute $C_{x_i}$ according to Eq.\ref{eq:jacobian_score};\
\EndFor

\State $W_{ordered} \gets Sort(x_1, x_2, \cdots, x_m)$ according to $C_{x_{i}}$;\
\For {$x_i$ in $W_{ordered}$}
\State $bug = SelectBug(x_i, \bm{x'}, y, \mathcal{F}(\cdot))$;\
\State $\bm{x'} \gets$ replace $x_i$ with $bug$ in $\bm{x'}$
\If {$S(\bm{x}, \bm{x'})\leq \epsilon$}
		\State Return None.
\ElsIf {$\mathcal{F}_l(\bm{x'}) \neq y$}
	\State Solution found. Return $\bm{x'}$.
\EndIf
\EndFor 
\State \Return None 
\end{algorithmic}  
\end{algorithm}
\vspace{-0.2cm}

\textbf{Step 2: Bugs Generation (line 6-14).}
To generate bugs, many operations can be used.
However, we prefer small changes to the original words as we require the generated adversarial sentence is visually and semantically similar to the original one for human understanding. 
Therefore, we consider two kinds of perturbations, i.e., character-level perturbation and word-level perturbation.

For character-level perturbation, one key observation is that words are symbolic, and learning-based DLTU systems usually use a dictionary to represent a finite set of possible words. 
The size of the typical word dictionary is much smaller than the possible combinations of characters at a similar length (e.g., about $26^n$ for the English case, where $n$ is the length of the word). 
This means if we deliberately misspell important words, we can easily convert those important words to ``unknown'' (i.e., words not in the dictionary). 
The unknown words will be mapped to the ``unknown'' embedding vector in deep learning modeling. 
Our results strongly indicate that such simple strategy can effectively force text classification models to behave incorrectly.

For word-level perturbation, we expect that the classifier can be fooled after replacing a few words, which are obtained by nearest neighbor searching in the embedding space, without changing the original meaning.
However, we found that in some word embedding models (e.g., word2vec), semantically opposite words such as ``worst'' and ``better'' are highly syntactically similar in texts, thus ``better'' would be considered as the nearest neighbor of ``worst''. 
However, changing ``worst'' to ``better'' would completely change the sentiment of the input text.
Therefore, we make use of a semantic-preserving technique, i.e., replace the word with its $top_k$ nearest neighbors in a context-aware word vector space. 
Specifically, we use the pre-trained GloVe model \cite{pennington2014glove} provided by Stanford for word embedding and set $top_k=5$ in the experiment. 
Thus, the neighbors are guaranteed to be semantically similar to the original one.

According to previous studies, the meaning of the text is very likely to be preserved or inferred by the reader after a few character changes \cite{rawlinson2007significance}.
Meanwhile, replacing words with semantically and syntactically similar words can ensure that the examples are perceptibly similar \cite{alzantot2018generating}.
Based on these observations, we propose five bug generation methods for \textsc{TextBugger}:
(1) \textbf{Insert}: Insert a space into the word\footnote{Considering the usability of text, we apply this method only when the length of the word is shorter than 6 characters since long words might be split into two legitimate words.}. Generally, words are segmented by spaces in English. Therefore, we can deceive classifiers by inserting spaces into words.
(2) \textbf{Delete}: Delete a random character of the word except for the first and the last character.
(3) \textbf{Swap}: Swap random two adjacent letters in the word but do not alter the first or last letter\footnote{For this reason, this method is only applied to words longer than 4 letters.}. This is a common occurrence when typing quickly and is easy to implement. 
(4) \textbf{Substitute-C (Sub-C)}: Replace characters with visually similar characters (e.g., replacing ``o'' with ``0'', ``l'' with ``1'', ``a'' with ``@'') or adjacent characters in the keyboard (e.g., replacing ``m'' with ``n'').
(5) \textbf{Substitute-W (Sub-W)}: Replace a word with its $top_k$ nearest neighbors in a context-aware word vector space. 
Several substitute examples are shown in \Cref{tab:transformer_functions}.

As shown in \Cref{alg:bug_selection}, after generating five bugs, we choose the optimal bug according to the change of the confidence value, i.e., choosing the bug that decreases the confidence value of the ground truth class the most.
Then we will replace the word with the optimal bug to obtain a new text $\bm{x'}$ (line 8).
If the classifier gives the new text a different label (i.e., $\mathcal{F}_l(\bm{x'}) \neq y$) while preserving the semantic similarity (which is detailed in \Cref{sec:evaluation_metric}) above the threshold (i.e., $S(\bm{x}, \bm{x'})\geq \epsilon$), the adversarial text is found (line 9-13).
If not, we repeat above steps to replace the next word in $W_{ordered}$ until we find the solution or fail to find a semantic-preserving adversarial example.

\begin{algorithm}[h]  
\caption{Bug Selection algorithm}  
\begin{algorithmic}[1] \label{alg:bug_selection}
\Function {SelectBug}{$w, \bm{x}, y, \mathcal{F}(\cdot)$}
\State $bugs=BugGenerator(w)$;\
\For {$b_k$ in $bugs$}
\State $\bm{candidate(k)}$ = replace $w$ with $b_k$ in $\bm{x}$;\
\State $score(k) = \mathcal{F}_y(\bm{x}) - \mathcal{F}_y(\bm{candidate(k)})$;\
\EndFor
\State $bug_{best} = \mathop{\arg\max}_{b_k} score(k)$;\
\State \Return {$bug_{best}$};\
\EndFunction
\end{algorithmic}  
\end{algorithm} 

\begin{table}[tp]
    \caption{Examples for five bug generation methods.}
    \label{tab:transformer_functions}
    \centering
    \scalebox{1.2}{
    \begin{tabular}{cccccc}
    \toprule 
    \textbf{Original} & \textbf{Insert} & \textbf{Delete} & \textbf{Swap} & \textbf{Sub-C} & \textbf{Sub-W} \\
     \midrule
     foolish & f oolish & folish & fooilsh & fo0lish & silly \\
     awfully & awfull y & awfuly & awfluly & awfu1ly & terribly \\
     cliches & clich es & clichs & clcihes & c1iches & cliche \\
    \bottomrule
    \end{tabular}}
\end{table}


\subsubsection{Black-box Attack}
Under the black-box setting, gradients of the model are not directly available, and we need to change the input sequences directly without the guidance of gradients.
Therefore different from white-box attacks, where we can directly select important words based on gradient information, in black-box attacks, we will first find important sentences and then the important words within them.
Briefly, the process of generating word-based adversarial examples on text under black-box setting contains three steps: 
(1) Find the important sentences.
(2) Use a scoring function to determine the importance of each word regarding to the classification result, and rank the words based on their scores. 
(3) Use the bug selection algorithm to change the selected words. 
The black-box adversarial text generation algorithm is shown in \Cref{alg:blackbox_textbugger}.

\begin{algorithm}[tp] 
\caption{\textsc{TextBugger} under black-box settings}  
\begin{algorithmic}[1]  \label{alg:blackbox_textbugger}
\Require legitimate document $\bm{x}$ and its ground truth label $y$, classifier $\mathcal{F(\cdot)}$, threshould $\epsilon$
\Ensure adversarial document $\bm{x_{adv}}$
\State Inititialize: $\bm{x'} \gets \bm{x}$

\For{$\bm{s_i}$ in document $\bm{x}$}  
\State $C_{sentence}(i)=\mathcal{F}_y(\bm{s_i})$;\   
\EndFor 

\State $S_{ordered} \gets Sort(sentences)$ according to $C_{sentence}(i)$;\
\State Delete sentences in $S_{ordered}$ if $\mathcal{F}_l(\bm{s_i}) \neq y$;\

\For {$\bm{s_i}$ in $S_{ordered}$}  
	\For {$w_j$ in $\bm{s_i}$}
	\State Compute $C_{w_j}$ according to Eq.\ref{eq:word_contribution};\
	\EndFor
	\State $W_{ordered} \gets Sort(words)$ according to $C_{w_j}$;\
	\For {$w_j$ in $W_{ordered}$}
	\State $bug = SelectBug(w_j, \bm{x'}, y, \mathcal{F}(\cdot))$;\
	\State $\bm{x'} \gets$ replace $w_j$ with $bug$ in $\bm{x'}$
	\If {$S(\bm{x}, \bm{x'})\leq \epsilon$}
		\State Return None.
	\ElsIf {$\mathcal{F}_l(\bm{x'}) \neq y$}
		\State Solution found. Return $\bm{x'}$.
	\EndIf
	\EndFor 
\EndFor 
\State \Return None 
\end{algorithmic}  
\end{algorithm} 

\textbf{Step 1: Find Important Sentences (line 2-6).}
Generally, when people express their opinions, most of the sentences are describing facts and the main opinions usually depend on only a few of sentences which have a greater impact on the classification results.
Therefore, to improve the efficiency of \system, we first find the important sentences that contribute to the final prediction results most and then prioritize to manipulate them.

Suppose the input document $\bm{x}=(\bm{s_1}, \bm{s_2}, \cdots, \bm{s_n})$, where $\bm{s_i}$ represents the sentence at the $i^{th}$ position. 
First, we use the spaCy library\footnote{http://spacy.io} to segment each document into sentences.
Then we filter out the sentences that have different predicted labels with the original document label (i.e., filter out $\mathcal{F}_l(\bm{s_i}) \neq y$).
Then, we sort the important sentences in an inverse order according to their importance score.
The importance score of a sentence $\bm{s_i}$ is represented with the confidence value of the predicted class $\mathcal{F}_y$, i.e., $C_{s_i}=\mathcal{F}_y(\bm{s_i})$.

\textbf{Step 2: Find Important Words (line 8-11).}
Considering the vast search space of possible changes, we should first find the most important words that contribute the most to the original prediction results, and then modify them slightly by controlling the semantic similarity. 

\begin{figure}[tp]
\centering
\includegraphics[width=0.4\textwidth]{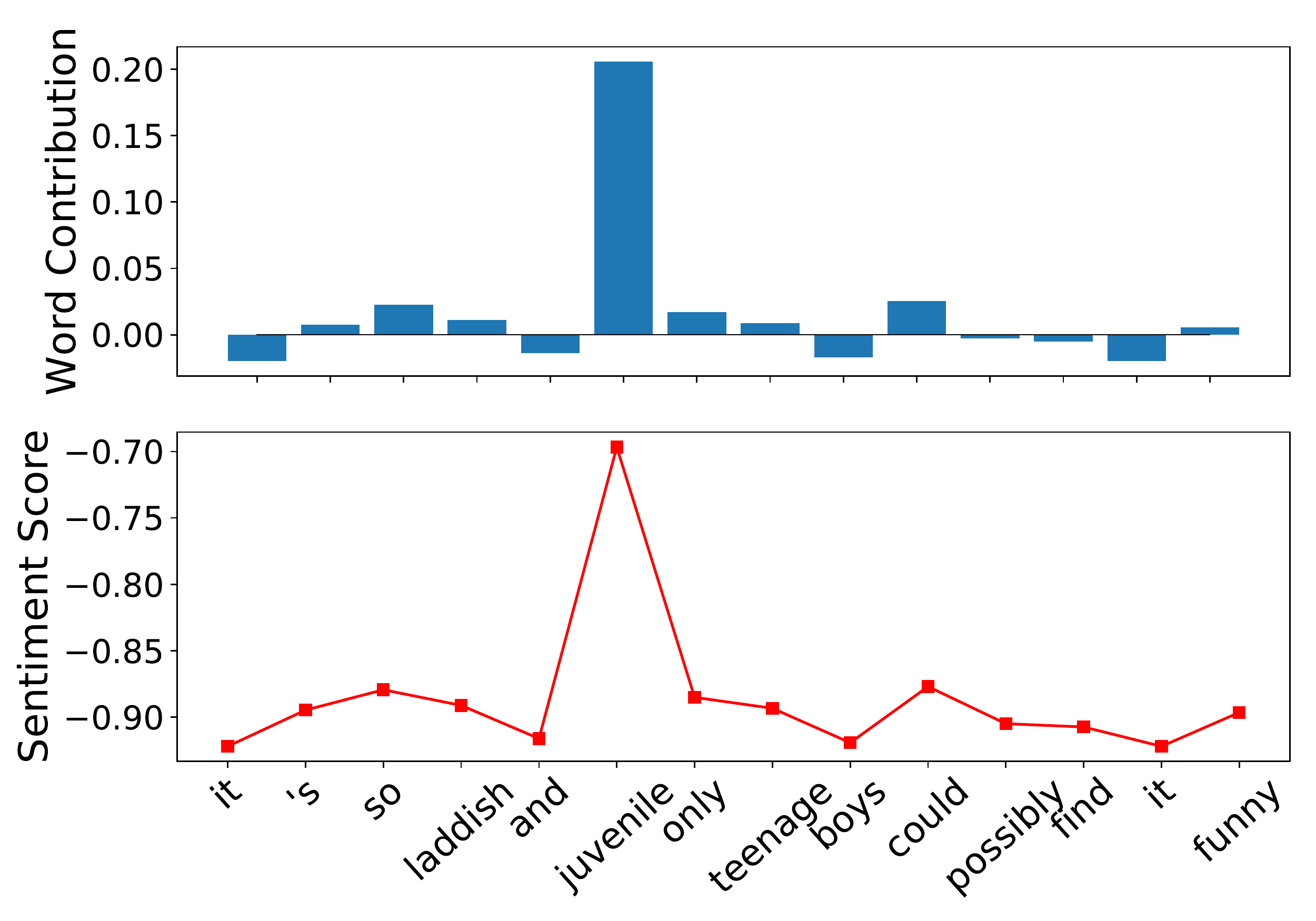}
\caption{Illustration of how to select important words to apply perturbations for the input sentence ``\textit{It is so laddish and juvenile, only teenage boys could possibly find it funny}''. The sentiment score of each word is the classification result's confidence value of the new text that deleting the word from the original text. The contribution of each word is the difference between the new confidence score and the original confidence score.}
\label{fig:textBugger_word_contribution}
\vspace{-0.25cm}
\end{figure}

One reasonable choice is to directly measure the effect of removing the $i^{th}$ word, since comparing the prediction before and after removing a word reflects how the word influences the classification result as shown in \Cref{fig:textBugger_word_contribution}.
Therefore, we introduce a scoring fuction that determine the importance of the $j^{th}$ word in $\bm{x}$ as:
\begin{small}
\begin{equation} \label{eq:word_contribution}
C_{w_j}\!=\! \mathcal{F}_y(w_1, w_2, \cdots \!, w_m) -\! \mathcal{F}_y(w_1, \cdots \!, w_{j\!-\! 1}, w_{j\!+\! 1}, \cdots \!, w_m)
\end{equation}
\end{small}
The proposed scoring function has the following properties:
(1) It is able to correctly reflect the importance of words for the prediction,  
(2) it calculates word scores without the knowledge of the parameters and structure of the classification model, and 
(3) it is efficient to calculate.

\textbf{Step 3: Bugs Generation (line 12-20).} 
This step is similar as that in white-box setting.

\section{Attack Evaluation: Sentiment Analysis} \label{sec:experiments_white}

Sentiment analysis refers to the use of NLP, statistics, or machine learning methods to extract, identify or characterize the sentiment content of a text unit.
It is widely applied to helping a business understand the social sentiment of their products or services by monitoring online conversations.

In this section, we investigate the practical performance of the proposed method for generating adversarial texts for sentiment analysis.
We start with introducing the datasets, targeted models, baseline algorithms, evaluation metrics and implementation details.
Then we will analyze the results and discuss potential reasons for the observed performance.


\subsection{Datasets} 
We study adversarial examples of text on two popular public benchmark datasets for sentiment analysis.
The final adversarial examples are generated and evaluated on the test set.

\textbf{IMDB \cite{maas-EtAl:2011:ACL-HLT2011}.} 
This dataset contains 50,000 positive and negative movie reviews that crawled from online sources, with 215.63 words as average length for each sample.
It has been divided into two parts, i.e., 25,000 reviews for training and 25,000 reviews for testing.
Specifically, we held out 20\% of the training set as a validation set and all parameters are tuned based on it.

\textbf{Rotten Tomatoes Movie Reviews (MR) \cite{pang2005seeing}.}
This dataset is a collection of movie reviews collected by Pang and Lee in \cite{pang2005seeing}.
It contains 5,331 positive and 5,331 negative processed sentences/snippets and has an average length of 32 words.
In our experiment, we divide this dataset into three parts, i.e., 80\%, 10\%, 10\% as training, validation and testing, respectively. 

\subsection{Targeted Models}
For white-box attacks, we evaluated \textsc{TextBugger} on LR, Kim's CNN \cite{kim2014convolutional} and the LSTM used in \cite{zhang2015character}.
In our implementation, the model's parameters are fine-tuned according to the sensitivity analysis on model performance conducted by Zhang \textit{et al.} \cite{zhang2017sensitivity}.
Meanwhile, all models were trained in a hold-out test strategy, and hyper-parameters were tuned only on the validation set.

For black-box attacks, we evaluated the \textsc{TextBugger} on ten sentiment analysis platforms/models, i.e., Google Cloud NLP, IBM Waston Natural Language Understanding (IBM Watson), Microsoft Azure Text Analytics (Microsoft Azure), Amazon AWS Comprehend (Amazon AWS), Facebook fastText (fastText), ParallelDots, TheySay Sentiment, Aylien Sentiment, TextProcessing, and Mashape Sentiment. For fastText, we used a pre-trained model\footnote{\url{https://s3-us-west-1.amazonaws.com/fasttext-vectors/supervised_models/amazon_review_polarity.bin}} provided by Facebook. This model is trained on the Amazon Review Polarity dataset and we do not have any information about  the models' parameters or architecture.

\subsection{Baseline Algorithms}
We implemented and compared the other three methods with our white-box attack method.
In total, the three methods are: 
(1) \textbf{Random:} Randomly selects words to modify. For each sentence, we select 10\% words to modify.
(2) \textbf{FGSM+Nearest Neighbor Search (NNS):} The FGSM method was first proposed in \cite{Goodfellow:2014tl} to generate adversarial images, which adds to the whole image the noise that is proportional to sign($\nabla(L_x)$), where $L$ represent the loss function and $x$ is the input data.
It was combined with NNS to generate adversarial texts as in \cite{gong2018adversarial}: first, generating adversarial embeddings by applying FGSM on the embedding vector of the texts, then reconstructing the adversarial texts via NNS.
(3) \textbf{DeepFool+NNS:} The DeepFool method is first proposed in \cite{moosavi2016deepfool} to generate adversarial images, which iteratively finds the optimal direction to search for the minimum distance to cross the decision boundary.
It was combined with NNS to generate adversarial texts as in \cite{gong2018adversarial}.

\subsection{Evaluation Metrics} \label{sec:evaluation_metric}
We use four metrics, i.e., edit distance, Jaccard similarity coefficient, Euclidean distance and semantic similarity, to evaluate the utility of the generated adversarial texts.
Specifically, the edit distance and Jaccard similarity coefficient are calculated on the raw texts, while the Euclidean distance and semantic similarity are calculated on word vectors.

\textbf{Edit Distance.}
Edit distance is a way of quantifying how dissimilar two strings (e.g., sentences) are by counting the minimum number of operations required to transform one string to the other. 
Specifically, different definitions of the edit distance use different sets of string operations.
In our experiment, we use the most common metrics, i.e., the Levenshtein distance, whose operations include removal, insertion, and substitution of characters in the string.

\textbf{Jaccard Similarity Coefficient.}
The Jaccard similarity coefficient is a statistic used for measuring the similarity and diversity of finite sample sets.
It is defined as the size of the intersection divided by the size of the union of the sample sets:
\begin{equation} \label{eq:jaccard_coef}
J(A,B) = \frac{|A \cap B|}{|A \cup B|} = \frac{|A \cap B|}{|A| + |B| - |A \cap B|}
\end{equation}
Larger Jaccard similarity coefficient means higher sample similarity.
In our experiment, one sample set consists of all the words in the sample.

\textbf{Euclidean Distance.}
Euclidean distance is a measure of the true straight line distance between two points in the Euclidean space.
If $\bm{p}=(p_1, p_2, \cdots, p_n)$ and $\bm{q}=(q_1, q_2, \cdots, q_n)$ are two samples in the word vector space, then the Euclidean distance between $\bm{p}$ and $\bm{q}$ is given by:
\begin{equation} \label{eq:euclidean_distance}
d(\bm{p}, \bm{q}) = \sqrt{(p_1-q_1)^2 + (p_2-q_2)^2 + \cdots + (p_n-q_n)^2}
\end{equation}
In our experiment, the Euclidean space is exactly the word vector space.

\textbf{Semantic Similarity.}
The above three metrics can only reflect the magnitude of the perturbation to some extent. 
They cannot guarantee that the generated adversarial texts will preserve semantic similarity from original texts. 
Therefore, we need a fine-grained metric that measures the degree to which two pieces of text carry the similar meaning so as to control the quality of the generated adversarial texts.

In our experiment, we first use the Universal Sentence Encoder \cite{cer2018universal}, a model trained on a number of natural language prediction tasks that require modeling the meaning of word sequences, to encode sentences into high dimensional vectors. Then, we use the cosine similarity to measure the semantic similarity between original texts and adversarial texts.
The cosine similarity of two \textit{n}-dimensional vectors $\bm{p}$ and $\bm{q}$ is defined as:
\begin{equation} \label{eq:cosine_similarity}
S(\bm{p},\bm{q})=\frac{\bm{p} \cdot \bm{q}}{||\bm{p}||\cdot ||\bm{q}||} = \frac{\sum_{i=1}^n p_i \times q_i}{\sqrt{\sum_{i=1}^n (p_i)^2} \times \sqrt{\sum_{i=1}^n (q_i)^2}}
\end{equation}
Generally, it works better than other distance measures because the norm of the vector is related to the overall frequency of which words occur in the training corpus. 
The direction of a vector and the cosine distance is unaffected by this, so a common word like ``frog'' will still be similar to a less frequent word like ``Anura'' which is its scientific name.

Since our main goal is to successfully generate adversarial texts, we only need to control the semantic similarity to be above a specific threshold.

\begin{table*}[tp]
    \caption{Results of the white-box attacks on IMDB and MR datasets.}
    \label{tab:sentiment_white_box_summary}
    \centering
    \scalebox{1.05}{
    \begin{tabular}{C{1.0cm}C{1.0cm}C{1.0cm}C{1.0cm}C{1.2cm}C{1.0cm}C{1.2cm}C{1.0cm}C{1.2cm}C{1.0cm}C{1.2cm}}
    \toprule 
    \multirowcell{2}[-1.5ex][c]{\centering \textbf{Model}} & \multirowcell{2}[-1.5ex][c]{\centering \textbf{Dataset}} & \multirowcell{2}[-1.5ex][c]{\centering \textbf{Accuracy}} & \multicolumn{2}{c}{\textbf{Random}} & \multicolumn{2}{c}{\textbf{FGSM+NNS \cite{gong2018adversarial}}} & \multicolumn{2}{c}{\textbf{DeepFool+NNS \cite{gong2018adversarial}}} & \multicolumn{2}{c}{\textbf{\textsc{TextBugger}}} \\
    \cmidrule(r){4-5} \cmidrule(r){6-7} \cmidrule(r){8-9} \cmidrule(r){10-11}
     & & & Success Rate & Perturbed Word & Success Rate & Perturbed Word & Success Rate & Perturbed Word & Success Rate & Perturbed Word \\
     \midrule
     \multirow{2}{*}{\textbf{LR}} 
     & MR & 73.7\% & 2.1\% & 10\% & 32.4\% & 4.3\% & 35.2\% & 4.9\% & \textbf{92.7\%} & 6.1\% \\
     & IMDB & 82.1\% & 2.7\% & 10\% & 41.1\% & 8.7\% & 30.0\% & 5.8\% & \textbf{95.2\%} & 4.9\% \\
     \midrule
     \multirow{2}{*}{\textbf{CNN}} 
     & MR & 78.1\% & 1.5\% & 10\% & 25.7\% & 7.5\% & 28.5\% & 5.4\% & \textbf{85.1\%} & 9.8\% \\
     & IMDB & 89.4\% & 1.3\% & 10\% & 36.2\% & 10.6\% & 23.9\% & 2.7\% & \textbf{90.5\%} & 4.2\% \\
     \midrule
     \multirow{2}{*}{\textbf{LSTM}}
     & MR & 80.1\% & 1.8\% & 10\% & 25.0\% & 6.6\% & 24.4\% & 11.3\% & \textbf{80.2\%} & 10.2\% \\
     & IMDB & 90.7\% & 0.8\% & 10\% & 31.5\% & 9.0\% & 26.3\% & 3.6\% & \textbf{86.7\%} & 6.9\% \\
    \bottomrule
    \end{tabular}}
\end{table*}

\begin{table*}[tp]
    \caption{Results of the black-box attack on IMDB.}
    \label{tab:sentiment_blackbox_IMDB}
    \centering
    \scalebox{1.15}{
    \begin{tabular}{ccccccccc}
    \toprule 
    \multirowcell{2}[-.5ex][c]{\centering \textbf{Targeted Model}} & \multirowcell{2}[-.5ex][c]{\centering \textbf{Original Accuracy}} & \multicolumn{3}{c}{\textbf{DeepWordBug \cite{gao2018black}}} & \multicolumn{3}{c}{\textbf{\textsc{TextBugger}}} \\
     \cmidrule(r){3-5} \cmidrule(r){6-8}
     & & Success Rate & Time (s) & Perturbed Word & Success Rate & Time (s) & Perturbed Word \\
     \midrule
     Google Cloud NLP & 85.3\% & 43.6\% & 266.69 & 10\% & \textbf{70.1\%} & 33.47 & 1.9\% \\
     IBM Waston & 89.6\% & 34.5\% & 690.59 & 10\% & \textbf{97.1\%} & 99.28 & 8.6\% \\
     Microsoft Azure & 89.6\% & 56.3\% & 182.08 & 10\% & \textbf{100.0\%} & 23.01 & 5.7\% \\
     Amazon AWS & 75.3\% & 68.1\% & 43.98 & 10\% & \textbf{100.0\%} & 4.61 & 1.2\% \\
     Facebook fastText & 86.7\% & 67.0\% & 0.14 & 10\% & \textbf{85.4\%} & 0.03 & 5.0\% \\
     ParallelDots & 63.5\% & 79.6\% & 812.82 & 10\% & \textbf{92.0\%} & 129.02 & 2.2\% \\
     TheySay & 86.0\% & 9.5\% & 888.95 & 10\% & \textbf{94.3\%} & 134.03 & 4.1\% \\
     Aylien Sentiment & 70.0\% & 63.8\% & 674.21 & 10\% & \textbf{90.0\%} & 44.96 & 1.4\% \\
     TextProcessing & 81.7\% & 57.3\% & 303.04 & 10\% & \textbf{97.2\%} & 59.42 & 8.9\% \\
     Mashape Sentiment & 88.0\% & 31.1\% & 585.72 & 10\% & \textbf{65.7\%} & 117.13 & 6.1\%  \\
    \bottomrule
    \end{tabular}}
\end{table*}

\begin{table*}[tp]
    \caption{Results of the black-box attack on MR.}
    \label{tab:sentiment_blackbox_MR}
    \centering
    \scalebox{1.15}{
    \begin{tabular}{ccccccccc}
    \toprule 
    \multirowcell{2}[-.5ex][c]{\centering \textbf{Targeted Model}} & \multirowcell{2}[-.5ex][c]{\centering \textbf{Original Accuracy}} & \multicolumn{3}{c}{\textbf{DeepWordBug \cite{gao2018black}}} & \multicolumn{3}{c}{\textbf{\textsc{TextBugger}}} \\
     \cmidrule(r){3-5} \cmidrule(r){6-8}
     & & Success Rate & Time (s) & Perturbed Word & Success Rate & Time (s) & Perturbed Word \\
     \midrule
     Google Cloud NLP & 76.7\% & 67.3\% & 34.64 & 10\% & \textbf{86.9\%} & 13.85 & 3.8\% \\
     IBM Waston & 84.0\% & 70.8\% & 150.45 & 10\% & \textbf{98.8\%} & 43.59 & 4.6\% \\
     Microsoft Azure & 67.5\% & 71.3\% & 43.98 & 10\% & \textbf{96.8\%} & 12.46 & 4.2\% \\
     Amazon AWS & 73.9\% & 69.1\% & 39.62 & 10\% & \textbf{95.7\%} & 3.25 & 4.8\% \\
     Facebook fastText & 89.5\% & 37.0\% & 0.02 & 10\% & \textbf{65.5\%} & 0.01 & 3.9\% \\
     ParallelDots & 54.5\% & 76.6\% & 150.89 & 10\% & \textbf{91.7\%} & 70.56 & 4.2\% \\
     TheySay & 72.3\% & 56.3\% & 69.61 & 10\% & \textbf{90.2\%} & 30.12 & 3.1\% \\
     Aylien Sentiment & 65.3\% & 65.2 & 83.63 & 10\% & \textbf{94.1\%} & 13.71 & 3.5\% \\
     TextProcessing & 77.6\% & 38.1\% & 59.44 & 10\% & \textbf{87.0\%} & 12.36 & 5.7\% \\
     Mashape Sentiment & 72.0\% & 73.6\% & 113.54 & 10\% & \textbf{94.8\%} & 18.24 & 5.1\% \\
    \bottomrule
    \end{tabular}}
    \vspace{-0.2cm} 
\end{table*}

\subsection{Implementation}
We conducted the experiments on a server with two Intel Xeon E5-2640 v4 CPUs running at 2.40GHz, 64 GB memory, 4TB HDD and a GeForce GTX 1080 Ti GPU card.
We repeated each experiment 5 times and report the mean value. 
This replication is important because training is stochastic and thus introduces variance in performance \cite{zhang2017sensitivity}.

In our experiment, we did not filter out stop-words before feature extraction as most NLP tasks do.
This is because we observe that the stop-words also have impact on the prediction results.
In particular, our experiments utilize the 300-dimension GloVe embeddings\footnote{\url{http://nlp.stanford.edu/projects/glove/}} trained on 840 billion tokens of Common Crawl. 
Words not present in the set of pre-trained words are initialized by randomly sampling from the uniform distribution in [-0.1, 0.1].
Furthermore, the semantic similarity threshold $\epsilon$ is set as 0.8 to guarantee a good trade-off between quality and strength of the generated adversarial text.

\subsection{Attack Performance}

\textbf{Effectiveness and Efficiency.}
The main results of white-box attacks on the IMDB and MR datasets and comparison of the performance of baseline methods are summarized in \Cref{tab:sentiment_white_box_summary}, where the third column of \Cref{tab:sentiment_white_box_summary} shows the original model accuracy in non-adversarial setting.
We do not give the average time of generating one adversarial example under white-box settings since the models are offline and the attack is very efficient (e.g., generating hundreds of adversarial texts in one second).
From \Cref{tab:sentiment_white_box_summary}, we can see that randomly choosing words to change (i.e., Random in \Cref{tab:sentiment_white_box_summary}) has hardly any influence on the final result.
This implies randomly changing words would not fool classifiers and choosing important words to modify is necessary for successful attack.
From \Cref{tab:sentiment_white_box_summary}, we can also see that the targeted models all perform quite well in non-adversarial setting.
However, the adversarial texts generated by \textsc{TextBugger} still has high attack success rate on these models.
In addition, the linear model is more susceptible to adversarial texts than deep learning models.
Specifically, \textsc{TextBugger} only perturbs a few words to achieve a high attack success rate and performs much better than baseline algorithms against all models as shown in \Cref{tab:sentiment_white_box_summary}.
For instance, it only perturbs 4.9\% words of one sample when achieving 95.2\% success rate on the IMDB dataset against the LR model, while all baselines achieve no more than 42\% success rate in this case.
As the IMDB dataset has an average length of 215.63 words, \textsc{TextBugger} only perturbed about 10 words for one sample to conduct successful attacks.
This means that \textsc{TextBugger} can successfully mislead the classifiers into assigning significantly higher positive scores to the negative reviews via subtle manipulation. 

The main results of black-box attacks on the IMDB and MR datasets and comparison of the performance of different methods are summarized in \Cref{tab:sentiment_blackbox_IMDB,tab:sentiment_blackbox_MR} respectively, and the second column of which shows the original model accuracy in non-adversarial setting.
From \Cref{tab:sentiment_blackbox_IMDB,tab:sentiment_blackbox_MR}, we can see that \textsc{TextBugger} achieves high attack success rate and performs much better than DeepWordBug \cite{gao2018black} against all real-world online DLTU platforms.
For instance, it achieves 100\% success rate on the IMDB dataset when targeting Azure and AWS platforms, while DeepWordBug only achieves 56.3\% and 68.1\% success rate respectively.
Besides, \textsc{TextBugger} only perturbs a few words to achieve a high success rate as shown in \Cref{tab:sentiment_blackbox_IMDB,tab:sentiment_blackbox_MR}.
For instance, it only perturbs 7\% words of one sample when achieving 96.8\% success rate on the MR dataset targeting the Microsoft Azure platform.
As the MR dataset has an average length of 32 words, \textsc{TextBugger} only perturbed about 2 words for one sample to conduct successful attacks.
Again, that means an adversary can subtly modify highly negative reviews in a way that the classifier assigns significantly higher positive scores to them. 

\begin{figure*}[tp]
\centering
\subfigure[Success Rate]{
    \centering
    \includegraphics[width=0.31\textwidth]{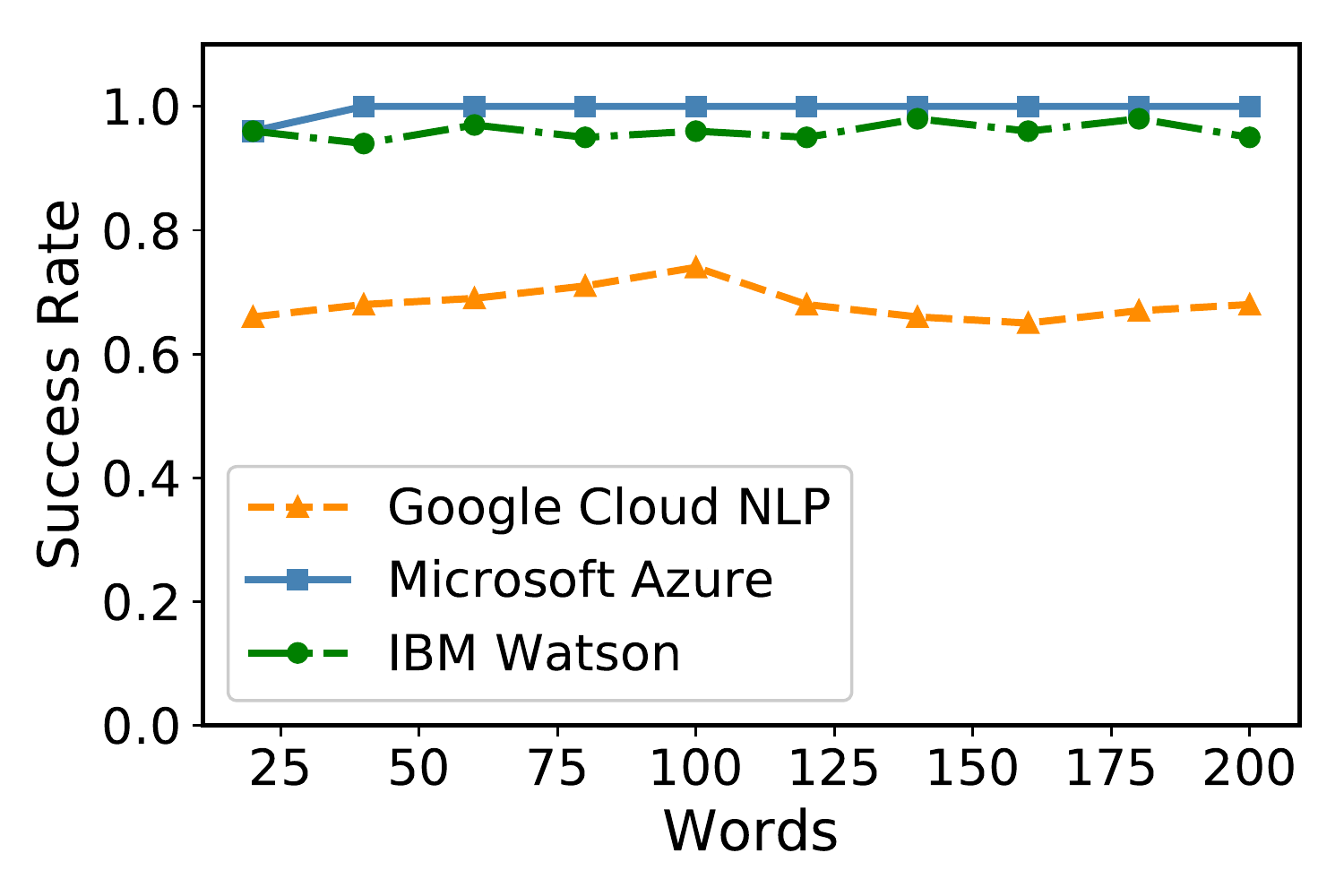}
    \label{fig:sentiment_wordlength_success_rate}
}
\subfigure[Score]{
    \centering
    \includegraphics[width=0.31\textwidth]{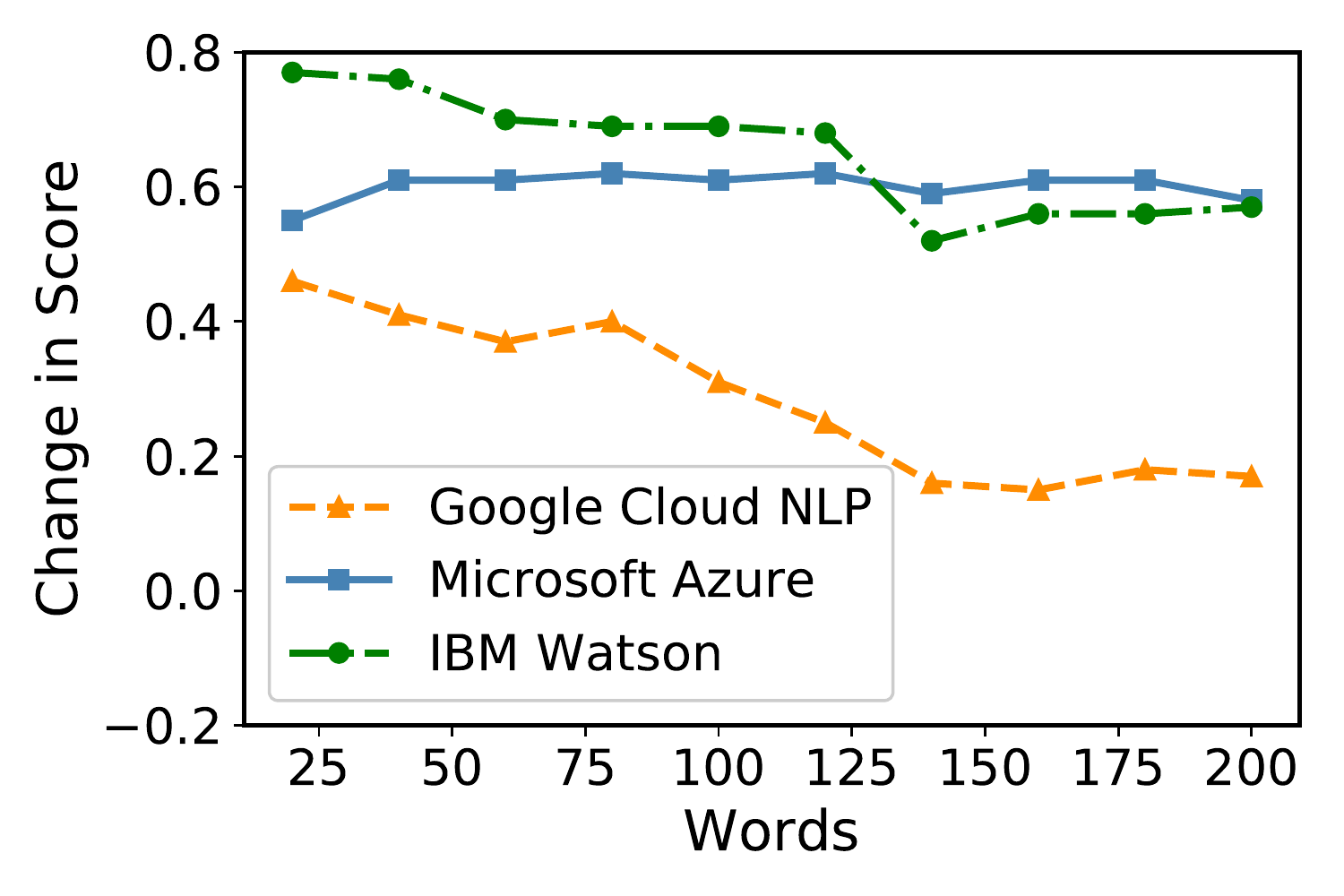}
    \label{fig:sentiment_wordlength_score}
}
\subfigure[Time]{
    \centering
    \includegraphics[width=0.31\textwidth]{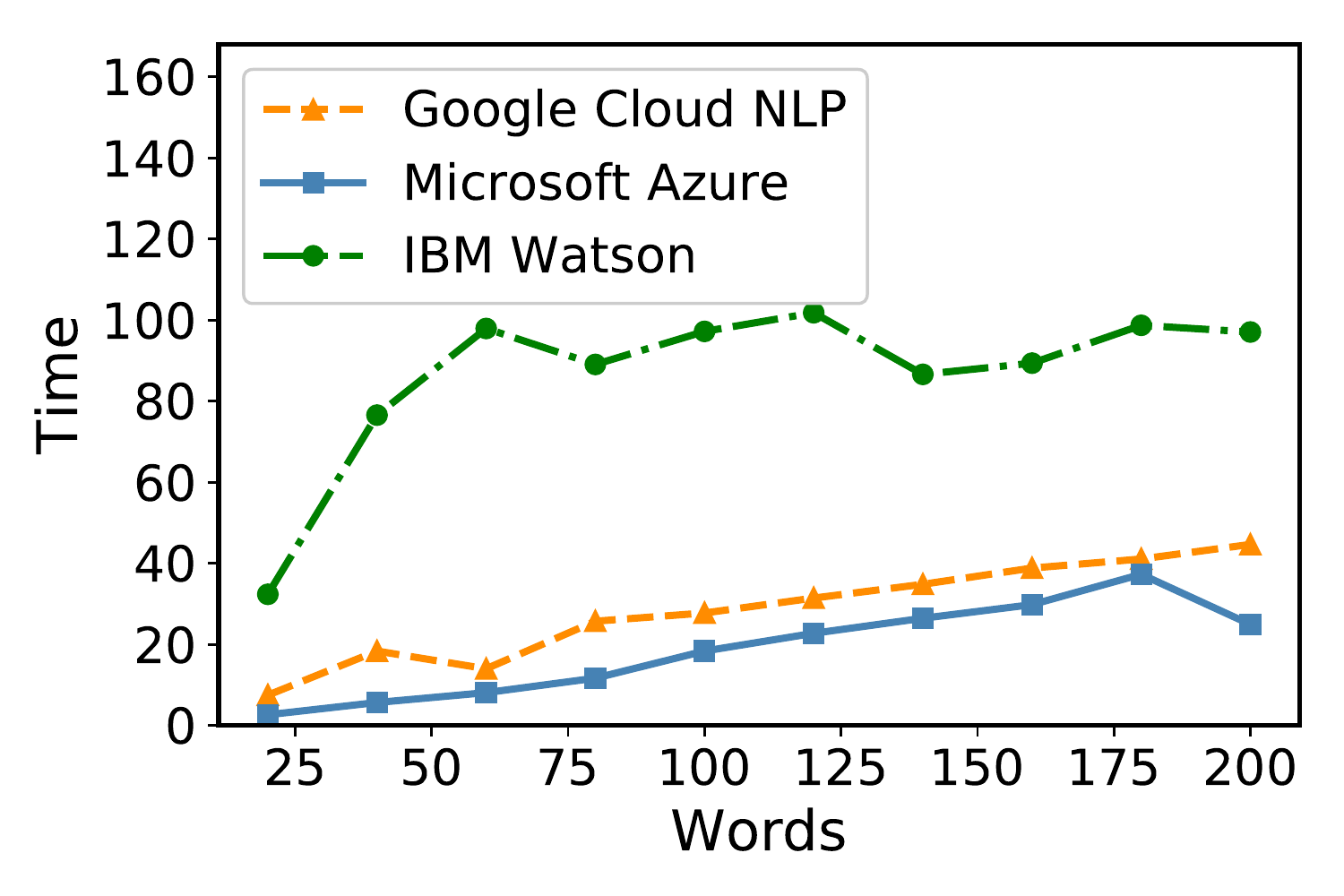}
    \label{fig:sentiment_wordlength_time}
}
\caption{The impact of document length (i.e. number of words in a document) on attack's performance against three online platforms: Google Cloud NLP, IBM Watson and Microsoft Azure.
The sub-figures are: (a) the success rate and document length, (b) the change of negative class's confidence value. For instance, the original text is classified as negative with 90\% confidence, while the adversarial text is classified as positive with 80\% confidence (20\% negative), the score changes 0.9-0.2=0.7. (c) the document length and the average time of generating an adversarial text.}
\label{fig:sentiment_wordlength_effectiveness_and_efficiency}
\vspace{-0.25cm}
\end{figure*}

\begin{figure*}[tp]
\centering
\subfigure[IMDB]{
    \centering
    \includegraphics[width=0.245\textwidth]{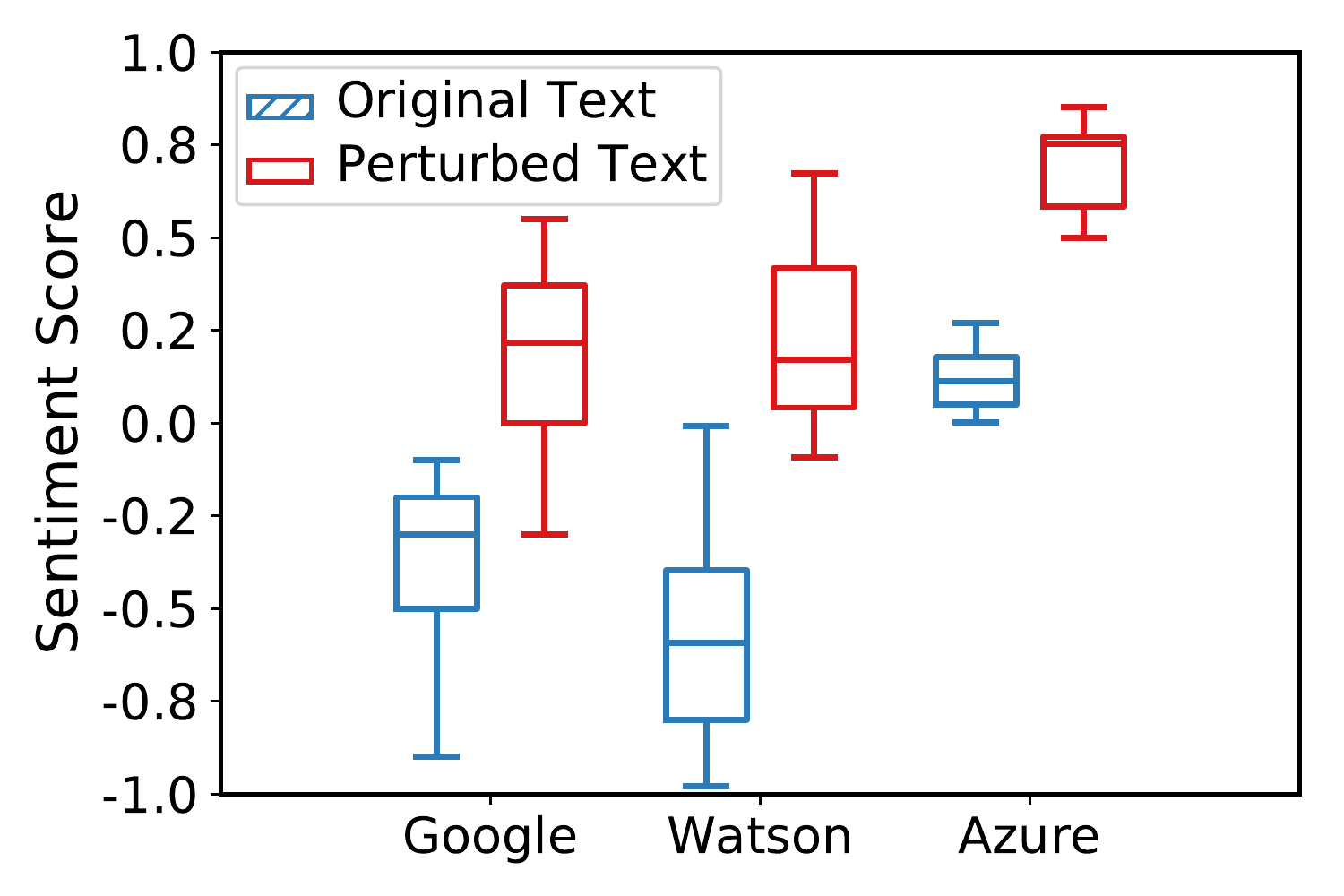}
    \label{fig:imdb_score_improvement_1}
    \hspace{-0.5cm}
}
\subfigure[IMDB]{
    \centering
    \includegraphics[width=0.245\textwidth]{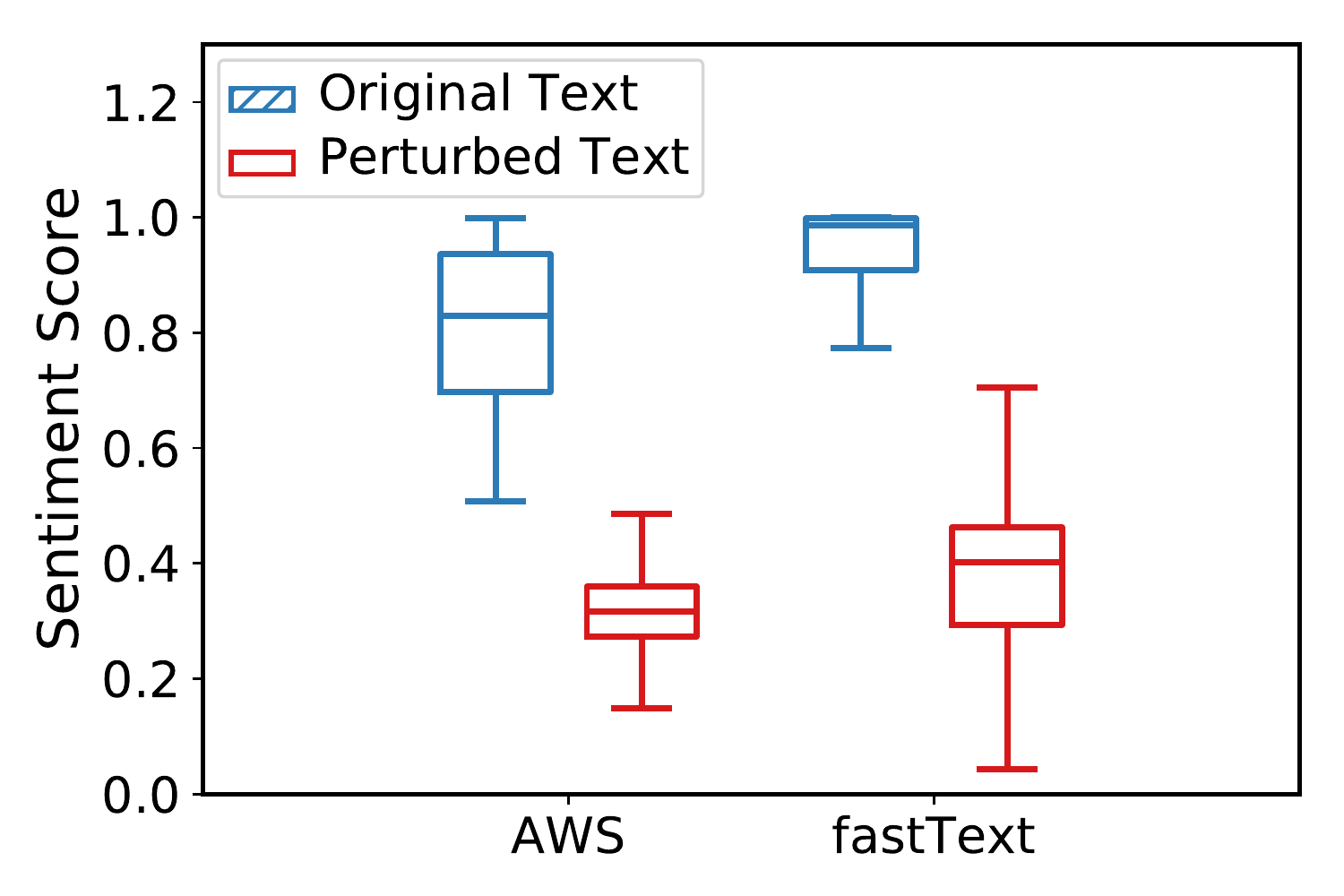}
    \label{fig:imdb_score_improvement_2}
    \hspace{-0.5cm}
}
\subfigure[MR]{
    \centering
    \includegraphics[width=0.245\textwidth]{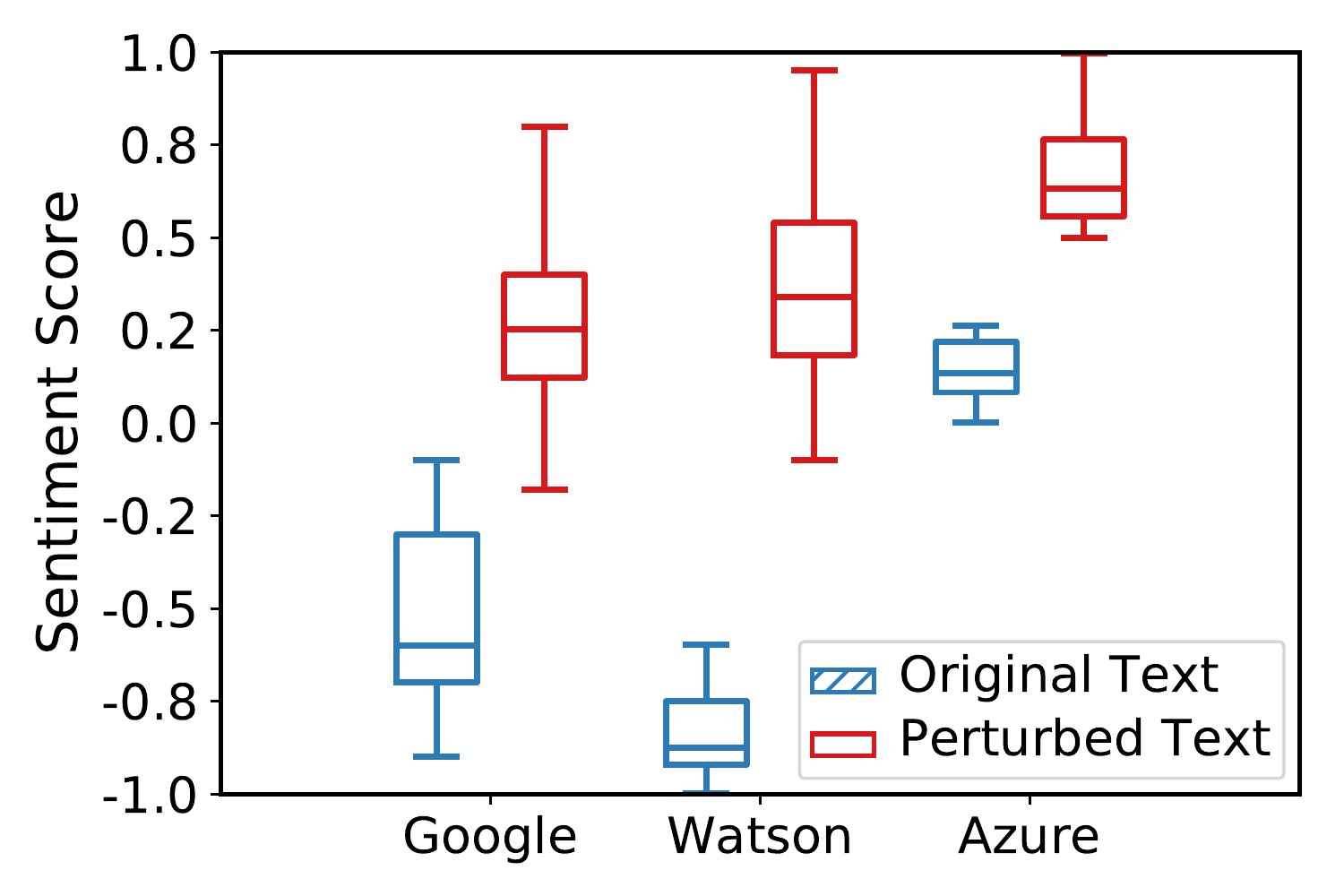}
    \label{fig:mr_score_improvement_1}
    \hspace{-0.5cm}
}
\subfigure[MR]{
    \centering
    \includegraphics[width=0.245\textwidth]{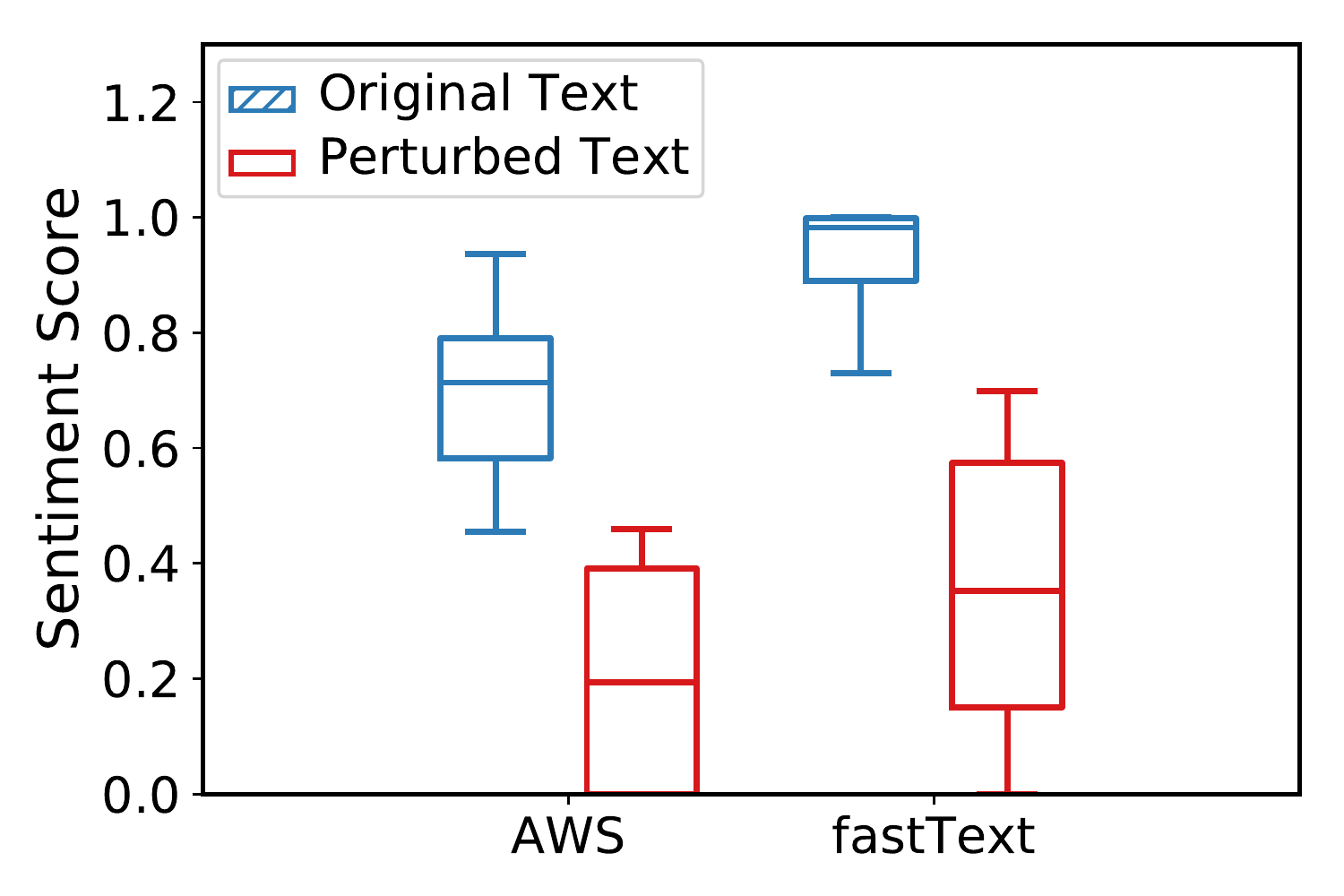}
    \label{fig:mr_score_improvement_2}
}
\caption{The change of sentiment score evaluated on IMDB and MR datasets for 5 black-box platforms/models. For Google Cloud NLP (Google), IBM Watson (Watson), the range of ``negative'' score is [-1, 0] and the range of ``positive'' score is [0, 1]. For Microsoft Azure (Azure), the range of ``negative'' score is [0, 0.5] and the range of ``positive'' score is [0.5, 1]. For Amazon AWS (AWS) and fastText, the range of ``negative'' score is [0.5, 1] and the range of ``positive'' score is [0, 0.5].}
\label{fig:sentiment_score_distribution}
\vspace{-0.25cm}
\end{figure*}

\textbf{The Impact of Document Length.}
We also study the impact of document length on the effectiveness and efficiency of the attacks and the corresponding results are shown in \Cref{fig:sentiment_wordlength_effectiveness_and_efficiency}.
From \Cref{fig:sentiment_wordlength_success_rate}, we can see that the document length has little impact on the attack success rate. 
This implies attackers can achieve high success rate no matter how long the sample is.
However, the confidence value of prediction results decrease for IBM Watson and Google Cloud NLP as shown in \Cref{fig:sentiment_wordlength_score}.
This means the attack on long documents would be a bit weaker than that on short documents.
From \Cref{fig:sentiment_wordlength_time}, we can see that the time required for generating one adversarial text and the average length of documents are positively correlated overall for Microsoft Azure and Google Cloud NLP. There is a very intuitive reason: the longer the length of the document is, the more information it contains that may need to be modified. Therefore, as the length of the document grows, the time required for generating one adversarial text increases slightly, since it takes more time to find important sentences.
For IBM Watson, the run time first increases before 60 words, then vibrates after that. We carefully analyzed the generated adversarial texts and found that when the document length is less than 60 words, the total length of the perturbed sentences increases sharply with the growth of document length. 
However, when the document length exceeds 60 words,  the total length of the perturbed sentences changes negligibly.
In general, generating one adversarial text only needs no more than 100 seconds for all the three platforms while the maximum length of a document is limited to 200 words. This means \system method is very efficient in practice.

\textbf{Adversarial Text Examples.} Two successful examples for sentiment analysis are shown in \Cref{fig:examples}.
The first adversarial text for sentiment analysis in \Cref{fig:examples} contains six modifications, i.e., one insert operation (``awful'' to ``aw ful''), one Sub-W operation (``no'' to ``No''), two delete operations (``literally'' to ``literaly'', ``cliches'' to ``clichs''), and two Sub-C operations (``embarrassingly'' to ``embarrassing1y'', ``foolish'' to ``fo0lish'').
These modifications successfully convert the prediction result of the CNN model, i.e., from 99.8\% negative to 81.0\% positive.
Note that the modification from ``no'' to ``No'' only capitalizes the first letter but really affects the prediction result. 
After further analysis, we find capitalization operation is common for both offline models and online platforms. 
We guess the embedding model may be trained without changing uppercase letters to lowercase, thus causing the same word in different forms get two different word vectors. 
Furthermore, capitalization sometimes may cause out-of-vocabulary phenomenon.
The second adversarial text for sentiment analysis in \Cref{fig:examples} contains three modifications, i.e., one insert operation (``weak'' to ``wea k'') and two Sub-C operations (``Unfortunately'' to ``Unf0rtunately'', ``terrible'' to ``terrib1e'').
These modifications successfully convert the prediction result of the Amazon AWS sentiment analysis API.

\textbf{Score Distribution.}
Even though \textsc{TextBugger} fails to convert the negative reviews to positive reviews in some cases, it can still reduce the confidence value of the classification results.
Therefore, we computed the change of the confidence value over all the samples including the failed samples before and after modification and show the results in \Cref{fig:sentiment_score_distribution}.
From \Cref{fig:sentiment_score_distribution}, we can see that the overall score of the texts has been moved to the positive direction.

\subsection{Utility Analysis}
\begin{figure}[tp]
\centering
\subfigure[Edit Distance]{
    \centering
    \includegraphics[width=0.235\textwidth]{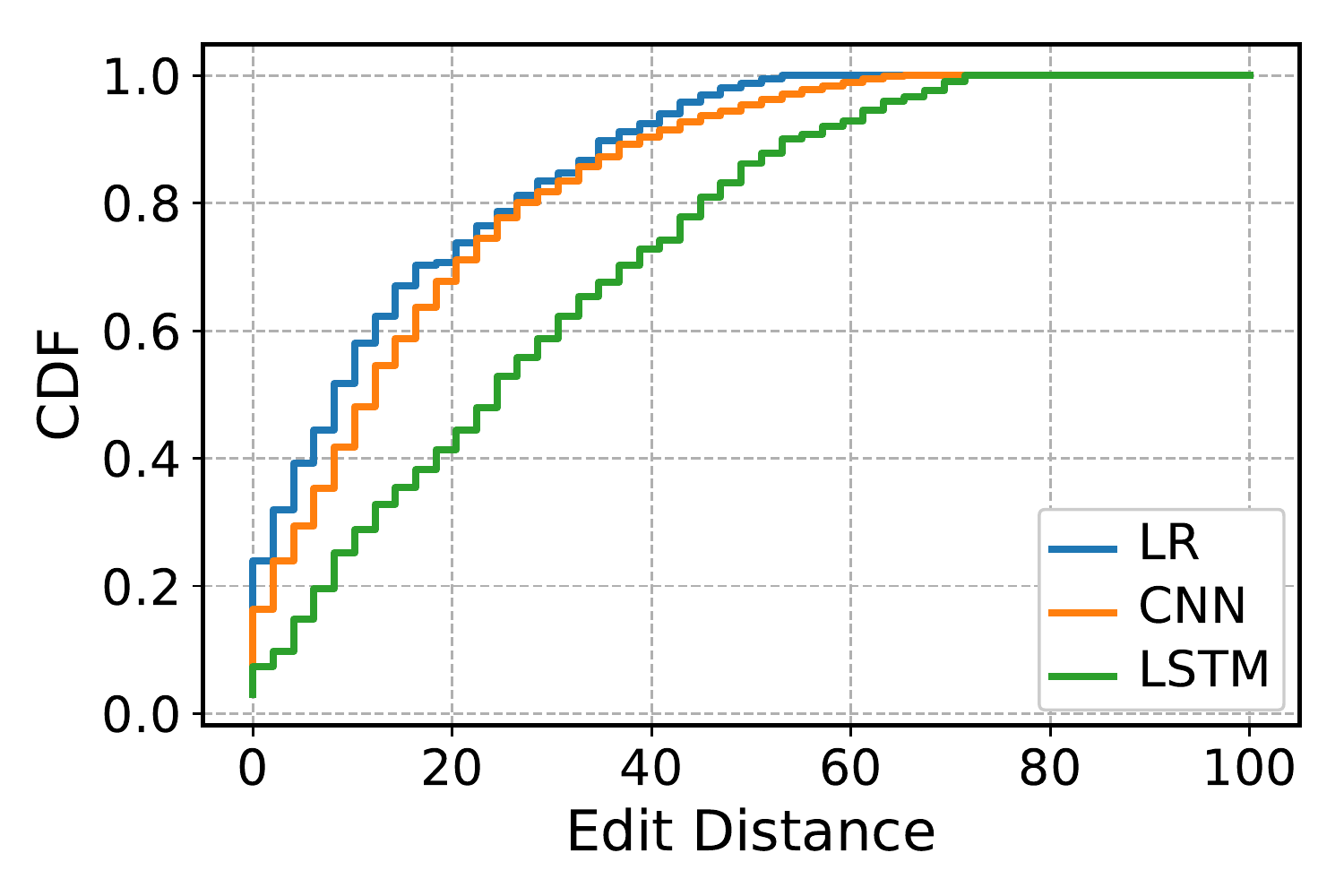}
    \label{fig:sentiment_whitebox_IMDB_edit_distance}
    \hspace{-0.5cm}
}
\subfigure[Jaccard Coefficient]{
    \centering
    \includegraphics[width=0.235\textwidth]{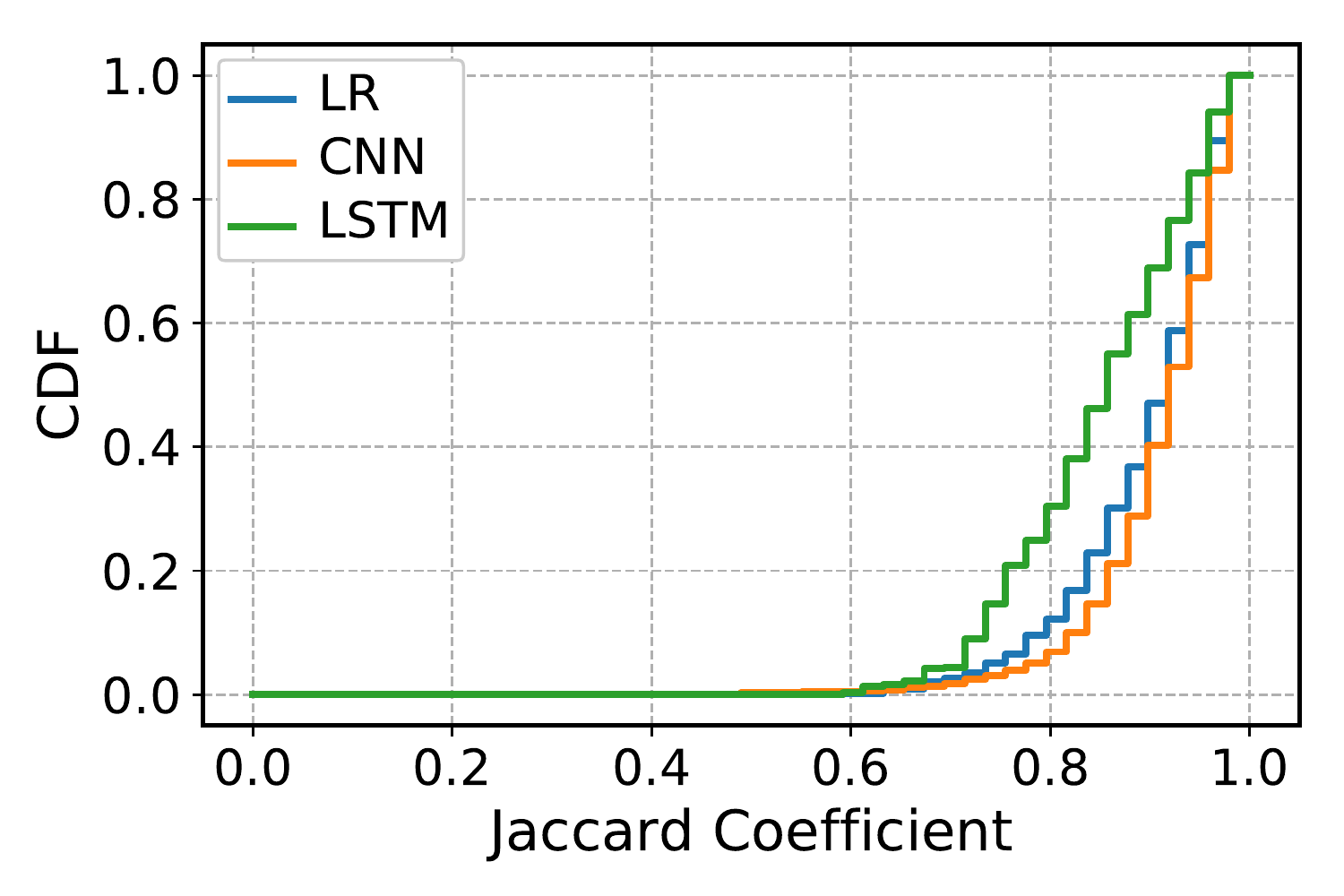}
    \label{fig:sentiment_whitebox_IMDB_jaccard_coefficient}
}
\subfigure[Euclidean Distance]{
    \centering
    \includegraphics[width=0.235\textwidth]{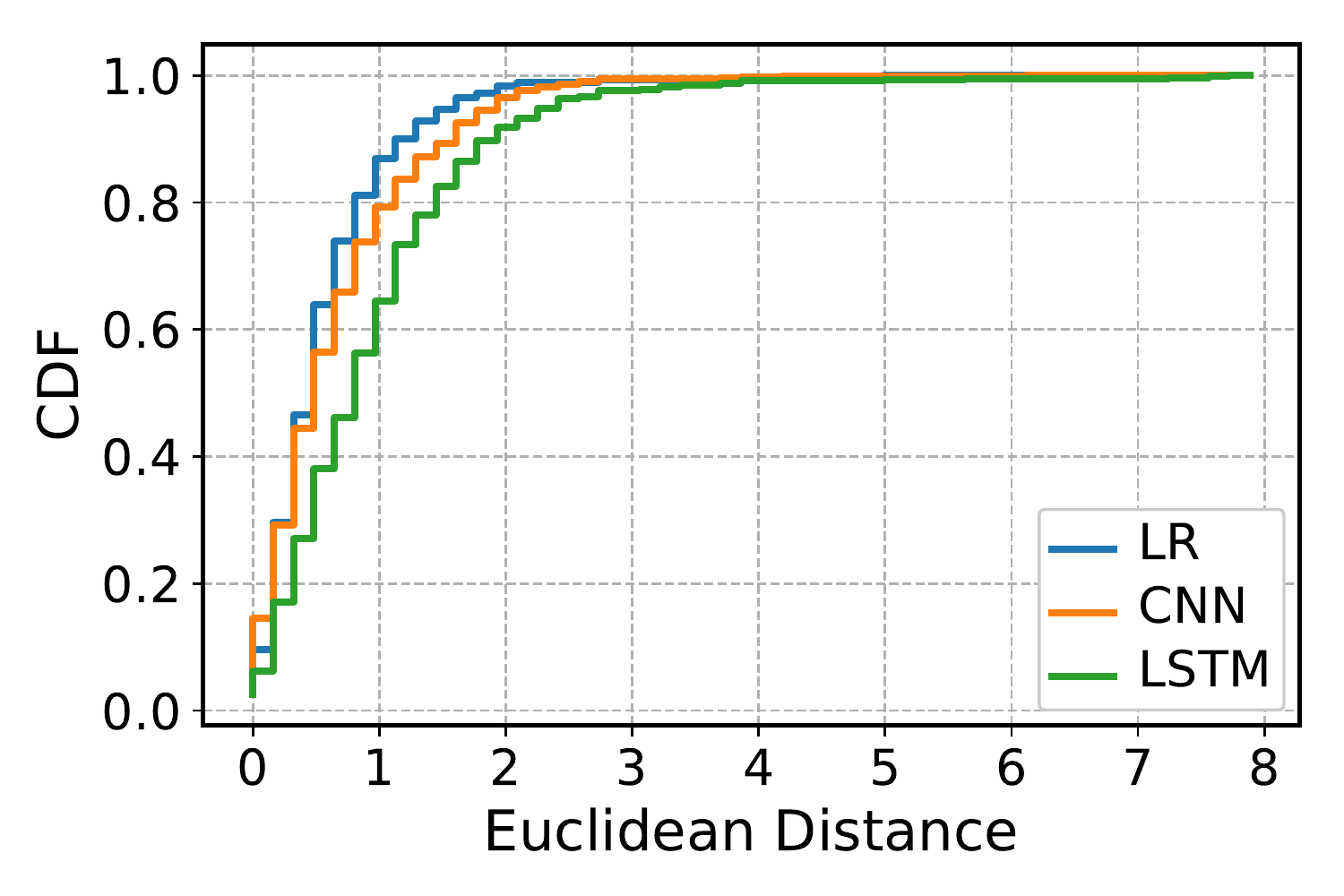}
    \label{fig:sentiment_whitebox_IMDB_euclidean_distance}
    \hspace{-0.5cm}
}
\subfigure[Semantic Similarity]{
    \centering
    \includegraphics[width=0.235\textwidth]{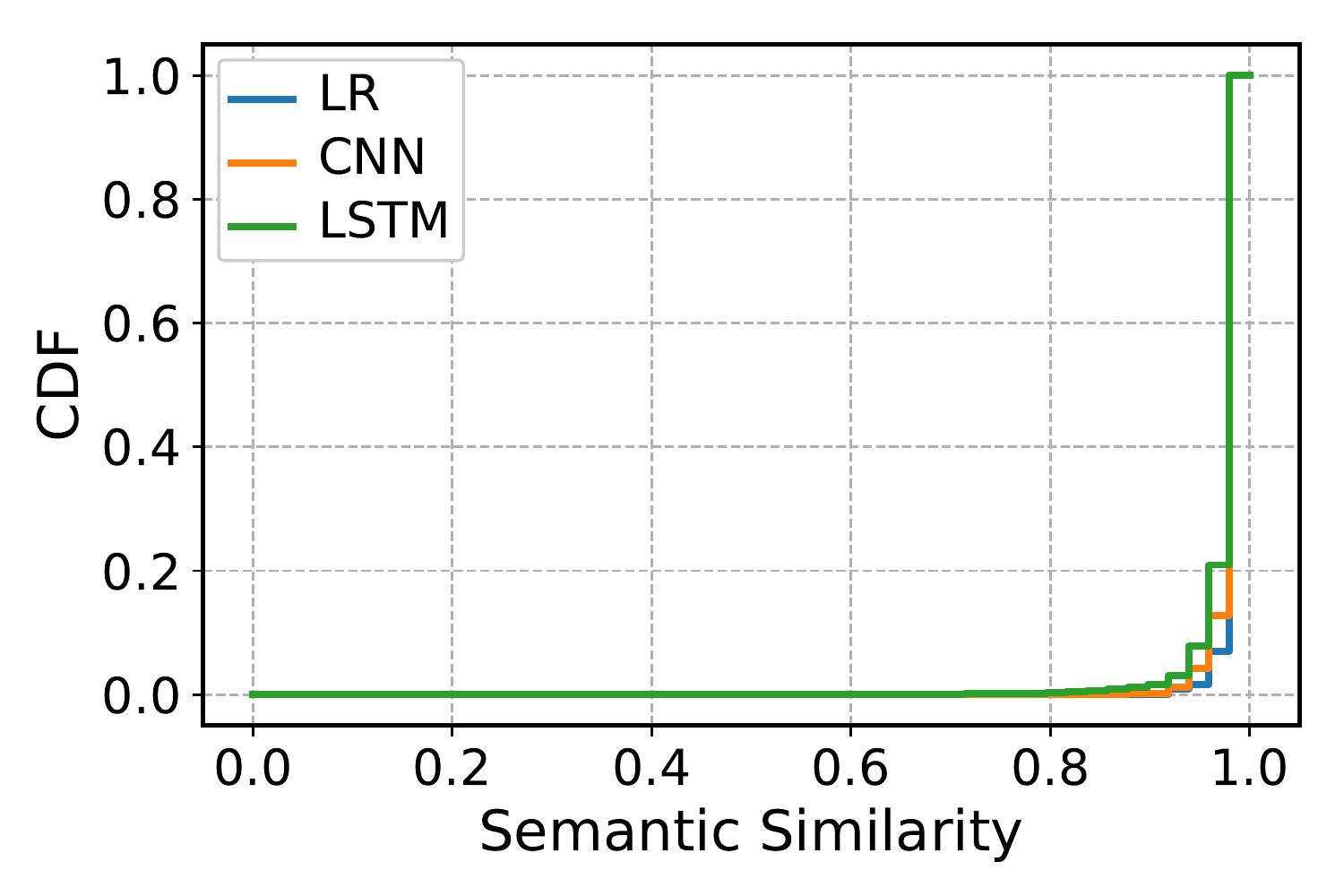}
    \label{fig:sentiment_whitebox_IMDB_semantic_similarity}
}
\caption{The utility of adversarial texts generated on IMDB dataset under white-box settings for LR, CNN and LSTM models.}
\label{fig:sentiment_whitebox_utility_IMDB}
\vspace{-0.25cm} 
\end{figure}

\begin{figure}[tp]
\centering
\subfigure[Edit Distance]{
    \centering
    \includegraphics[width=0.235\textwidth]{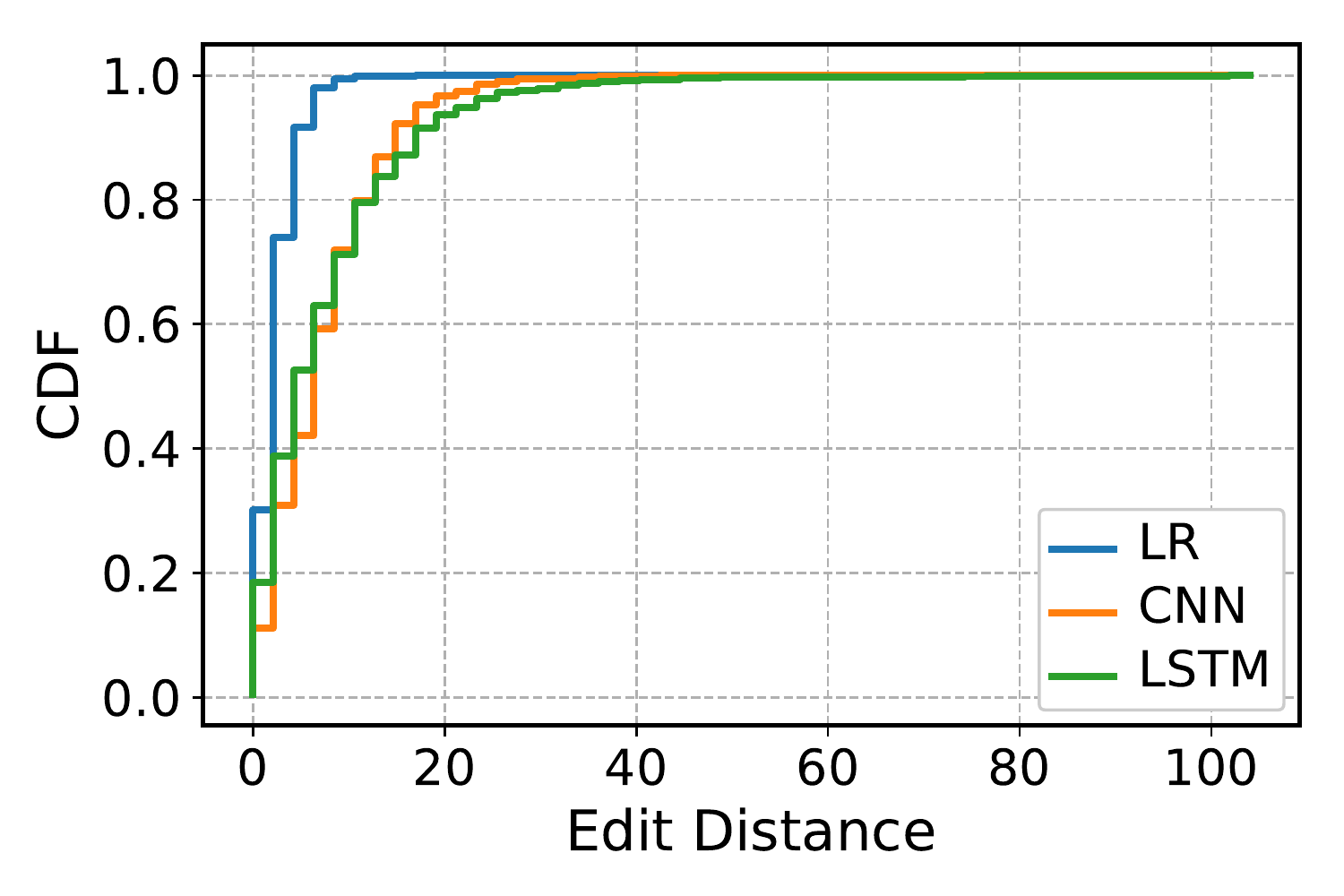}
    \label{fig:sentiment_whitebox_mr_edit_distance}
    \hspace{-0.5cm}
}
\subfigure[Jaccard Coefficient]{
    \centering
    \includegraphics[width=0.235\textwidth]{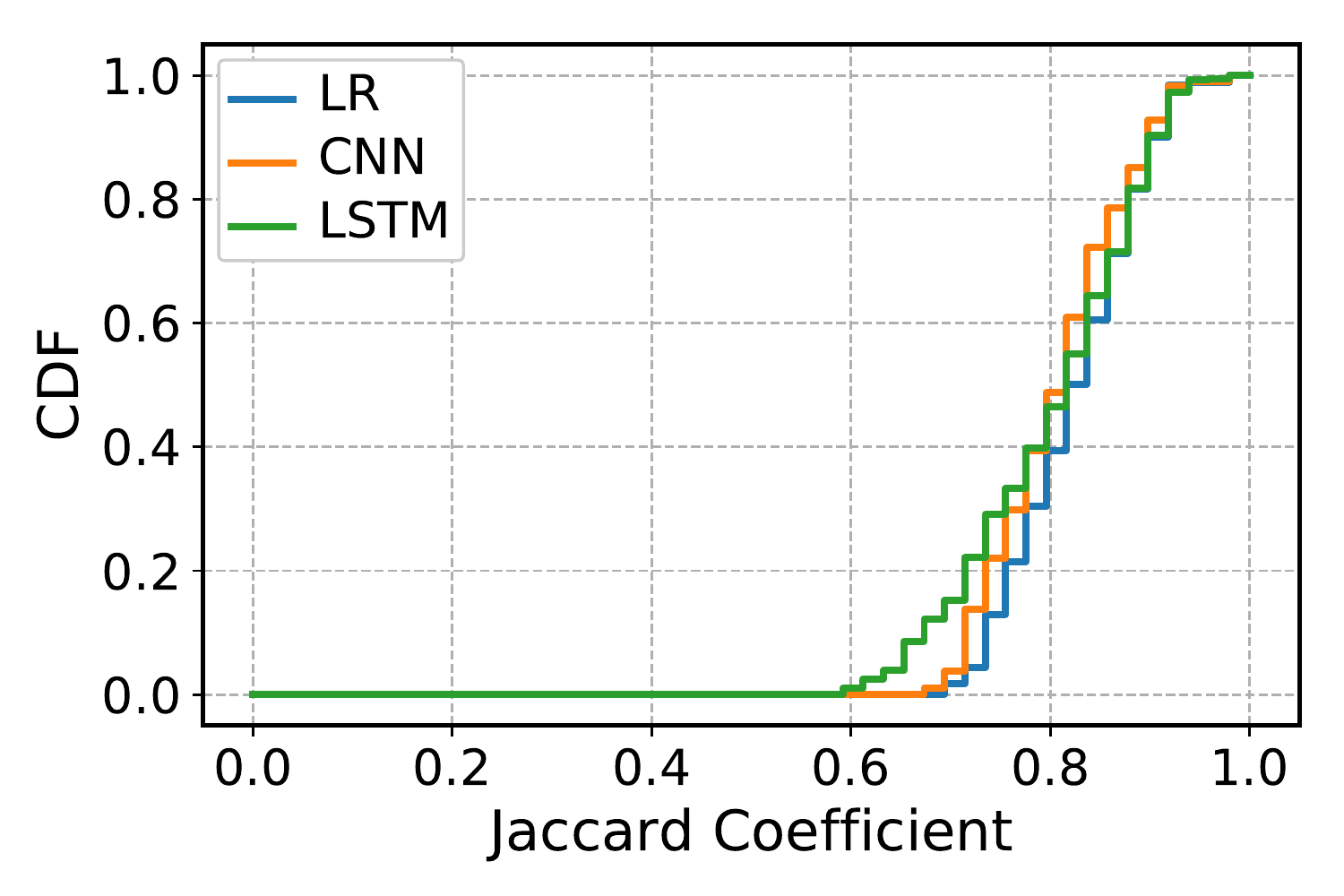}
    \label{fig:sentiment_whitebox_mr_jaccard_coefficient}
}
\subfigure[Euclidean Distance]{
    \centering
    \includegraphics[width=0.235\textwidth]{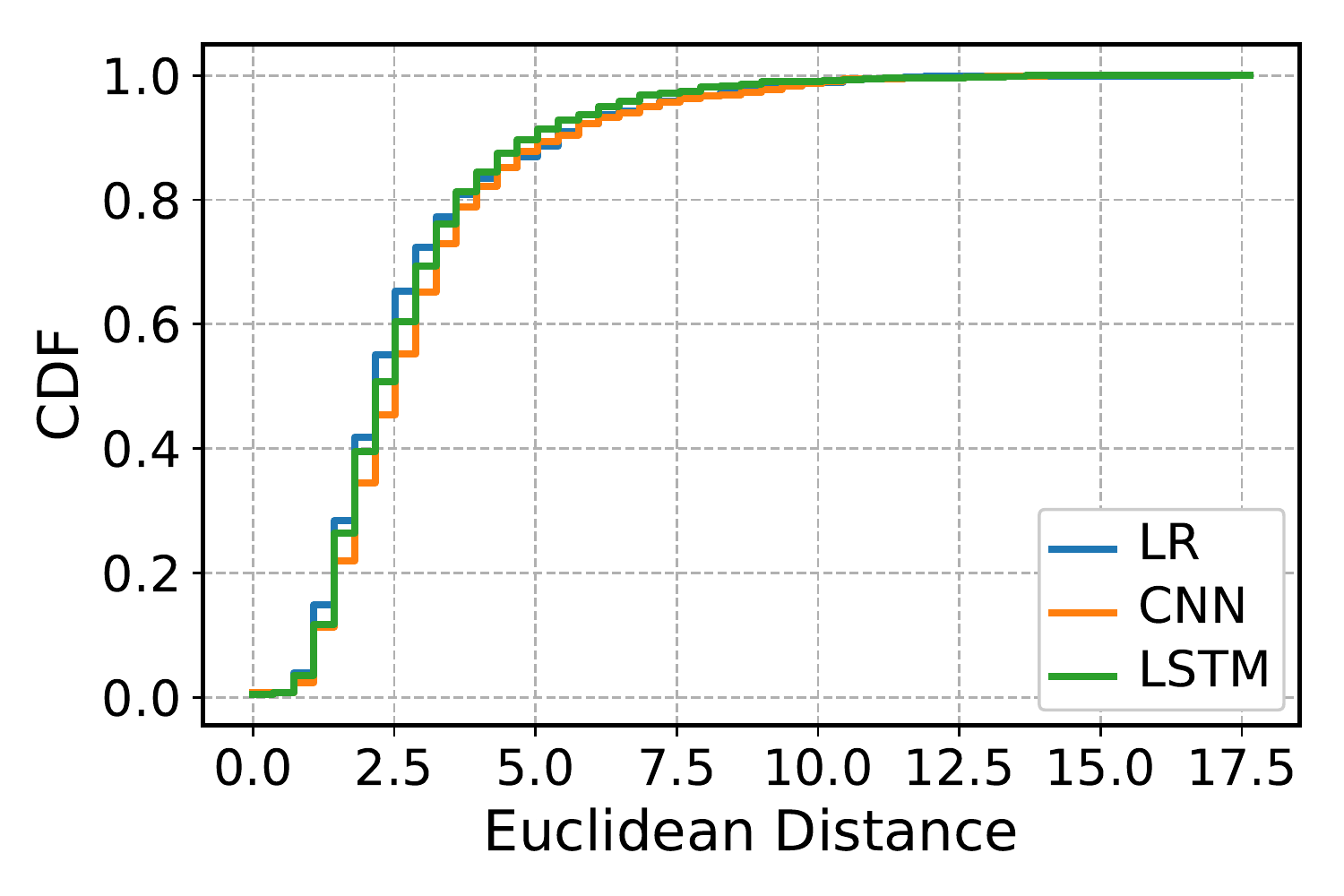}
    \label{fig:sentiment_whitebox_mr_euclidean_distance}
    \hspace{-0.5cm}
}
\subfigure[Semantic Similarity]{
    \centering
    \includegraphics[width=0.235\textwidth]{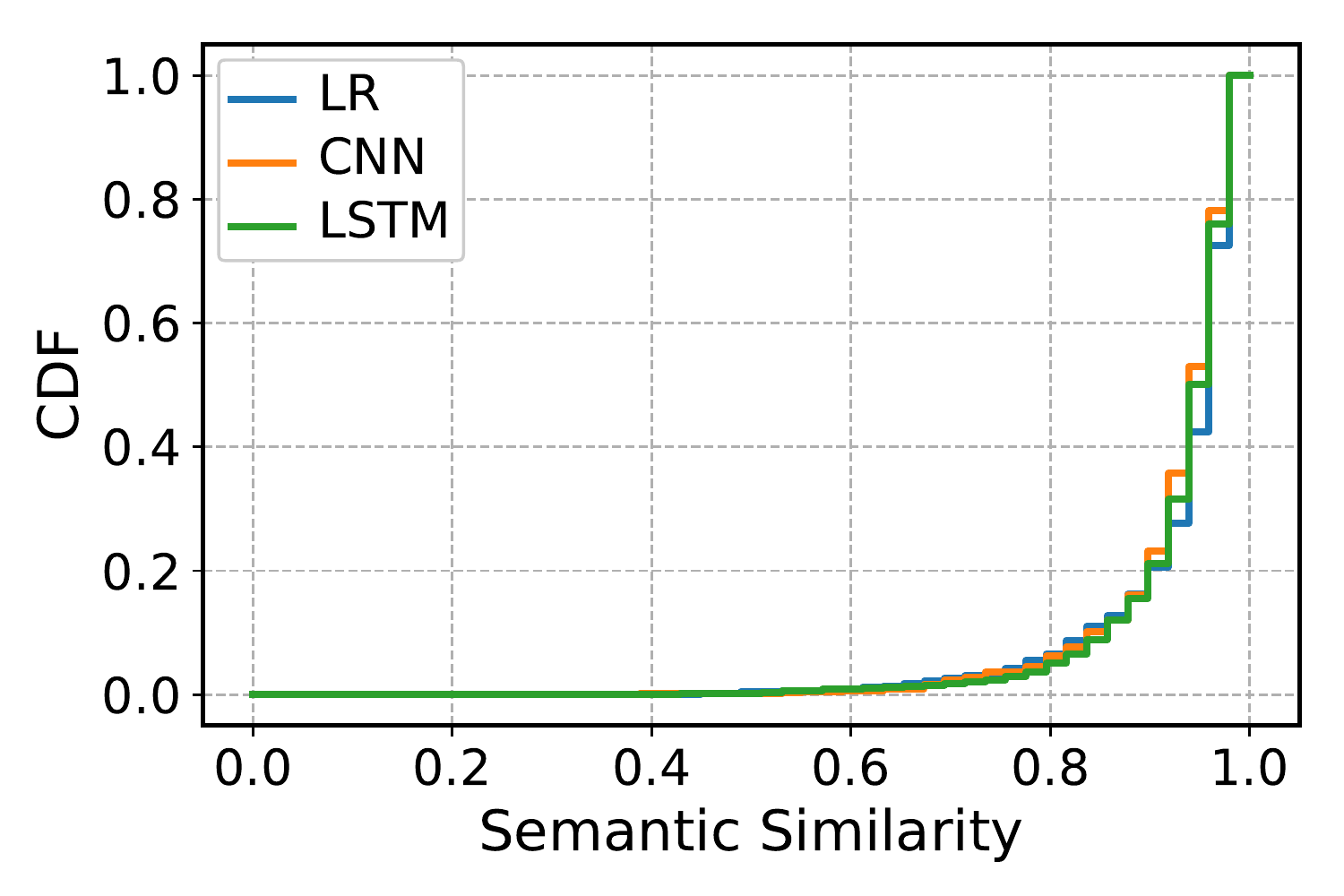}
    \label{fig:sentiment_whitebox_mr_semantic_similarity}
}
\caption{The utility of adversarial texts generated on MR dataset under white-box settings for LR, CNN and LSTM models.}
\label{fig:sentiment_whitebox_utility_MR}
\vspace{-0.25cm} 
\end{figure}

For white-box attacks, the similarity between original texts and adversarial texts against LR, CNN and LSTM models are shown in \Cref{fig:sentiment_whitebox_utility_IMDB,fig:sentiment_whitebox_utility_MR}.
We do not compare \textsc{TextBugger} with baselines in terms of utility since baselines only achieve low success rate as shown in \Cref{tab:toxic_whitebox_summary}.
From \Cref{fig:sentiment_whitebox_IMDB_edit_distance,fig:sentiment_whitebox_IMDB_jaccard_coefficient,fig:sentiment_whitebox_mr_edit_distance,fig:sentiment_whitebox_mr_jaccard_coefficient}, we can see that adversarial texts preserve good utility in terms of word-level.
Specifically, \Cref{fig:sentiment_whitebox_IMDB_edit_distance} shows that almost 80\% adversarial texts have no more than 25 edit distance comparing with original texts for LR and CNN models.
Meanwhile, \Cref{fig:sentiment_whitebox_IMDB_euclidean_distance,fig:sentiment_whitebox_IMDB_semantic_similarity,fig:sentiment_whitebox_mr_euclidean_distance,fig:sentiment_whitebox_mr_semantic_similarity} show that adversarial texts preserve good utility in terms of vector-level.
Specifically, from \Cref{fig:sentiment_whitebox_IMDB_semantic_similarity}, we can see that almost 90\% adversarial texts preserve at least 0.9 semantic similarity of the original texts.
This indicates that \textsc{TextBugger} can generate utility-preserving adversarial texts which fool the classifiers with high success rate.

\begin{figure}[tp]
\centering
\subfigure[Edit Distance]{
    \centering
    \includegraphics[width=0.235\textwidth]{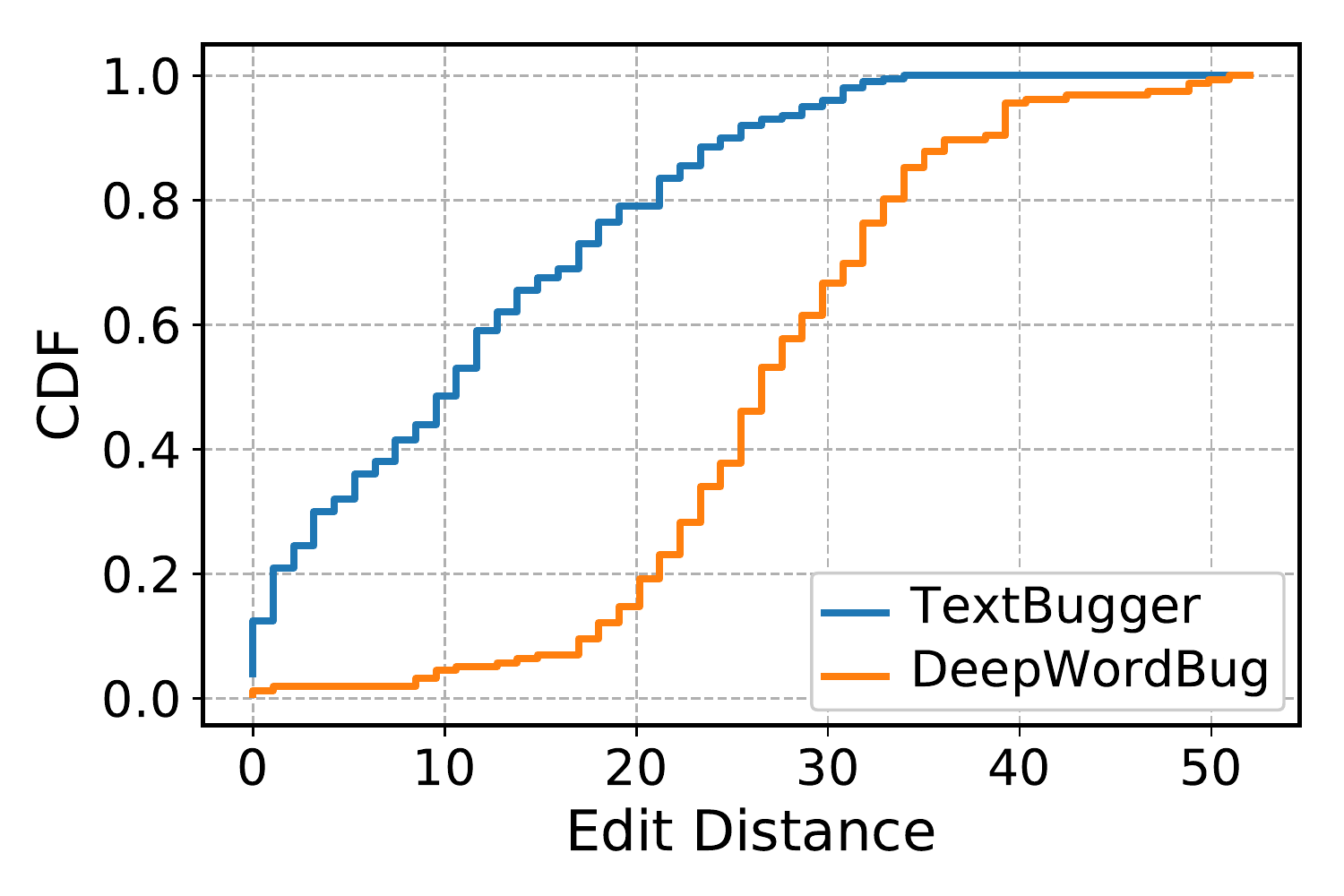}
    \label{fig:sentiment_blackbox_IMDB_edit_distance}
    \hspace{-0.5cm}
}
\subfigure[Jaccard Coefficient]{
    \centering
    \includegraphics[width=0.235\textwidth]{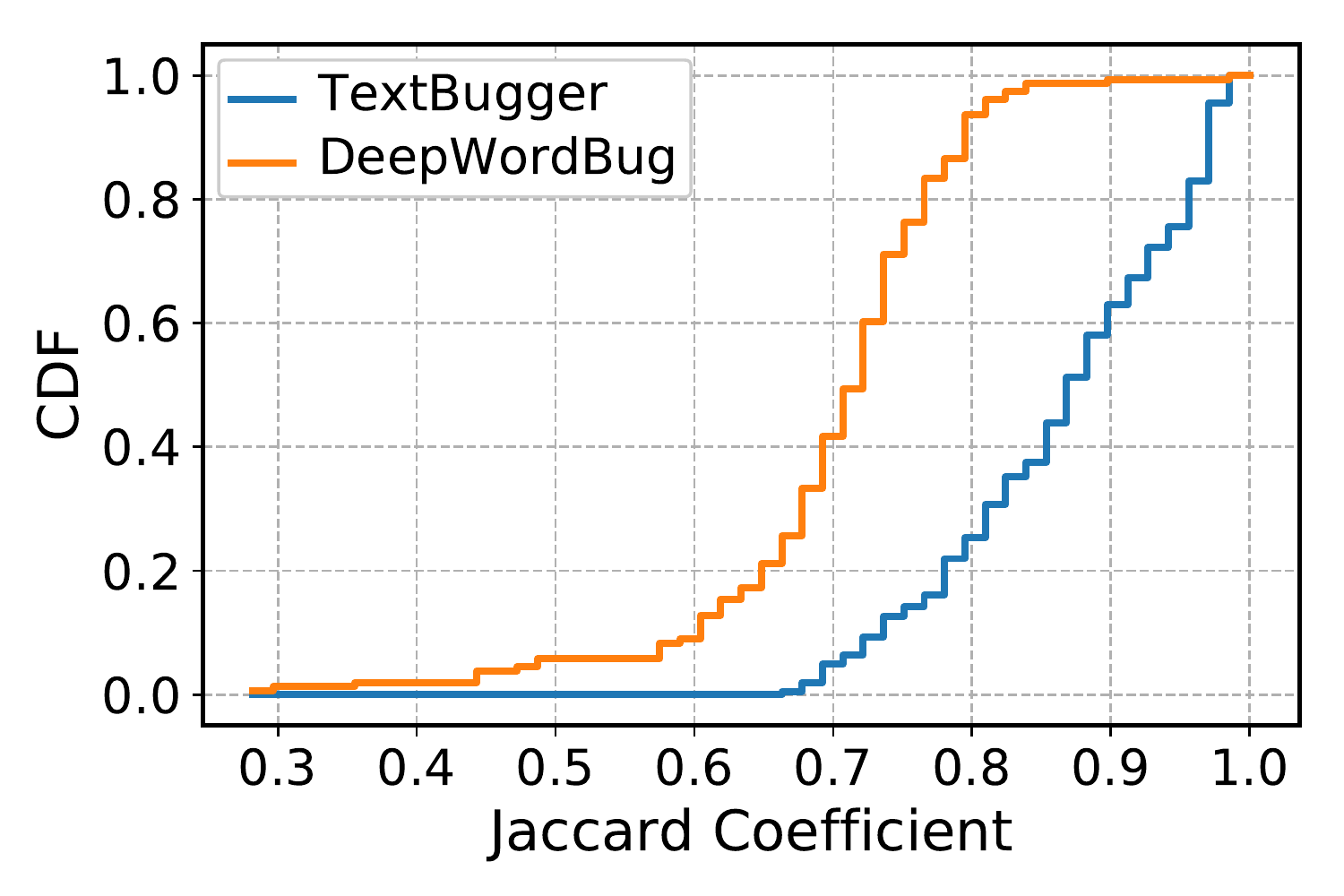}
    \label{fig:sentiment_blackbox_IMDB_jaccard_coefficient}
}
\subfigure[Euclidean Distance]{
    \centering
    \includegraphics[width=0.235\textwidth]{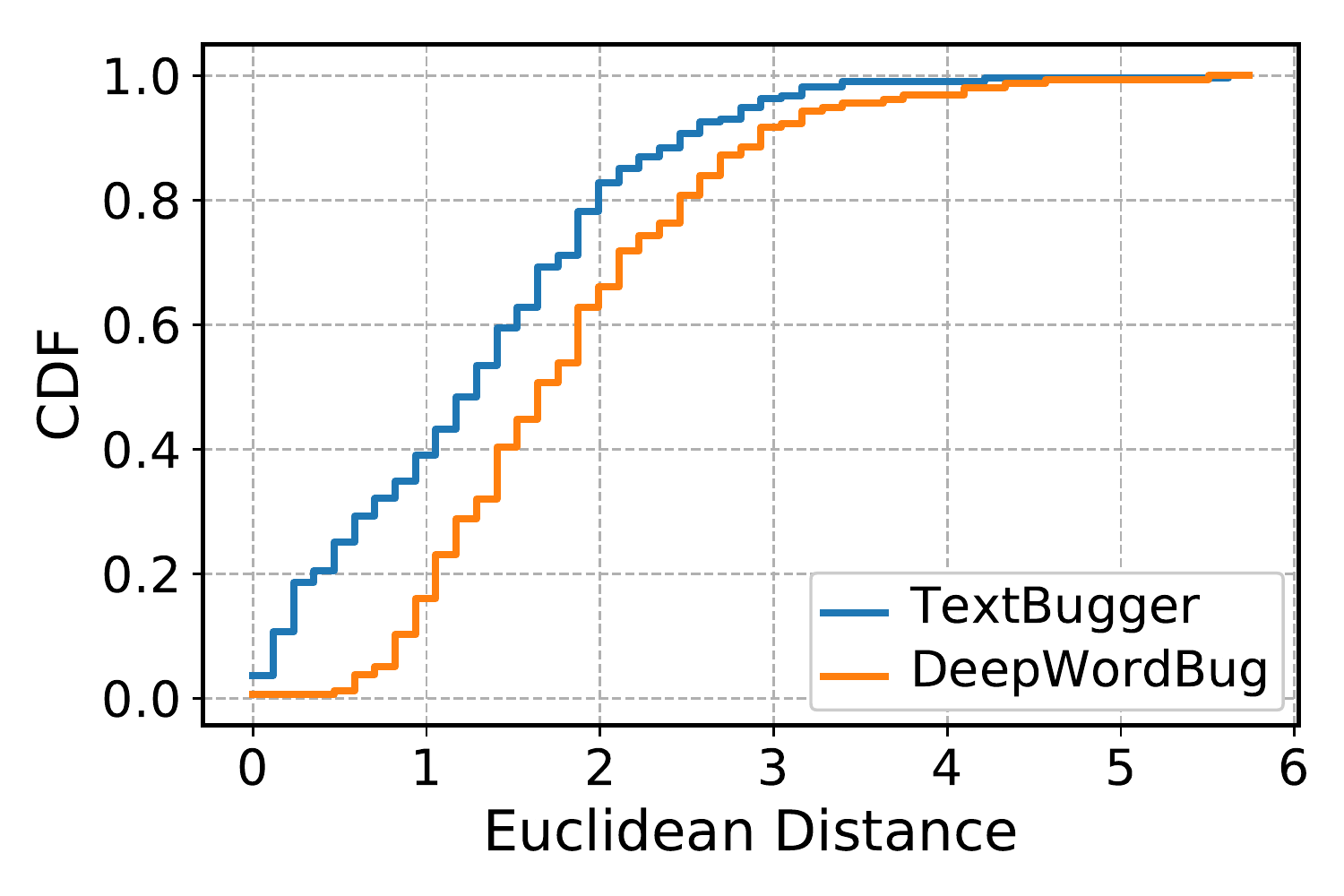}
    \label{fig:sentiment_blackbox_IMDB_euclidean_distance}
    \hspace{-0.5cm}
}
\subfigure[Semantic Similarity]{
    \centering
    \includegraphics[width=0.235\textwidth]{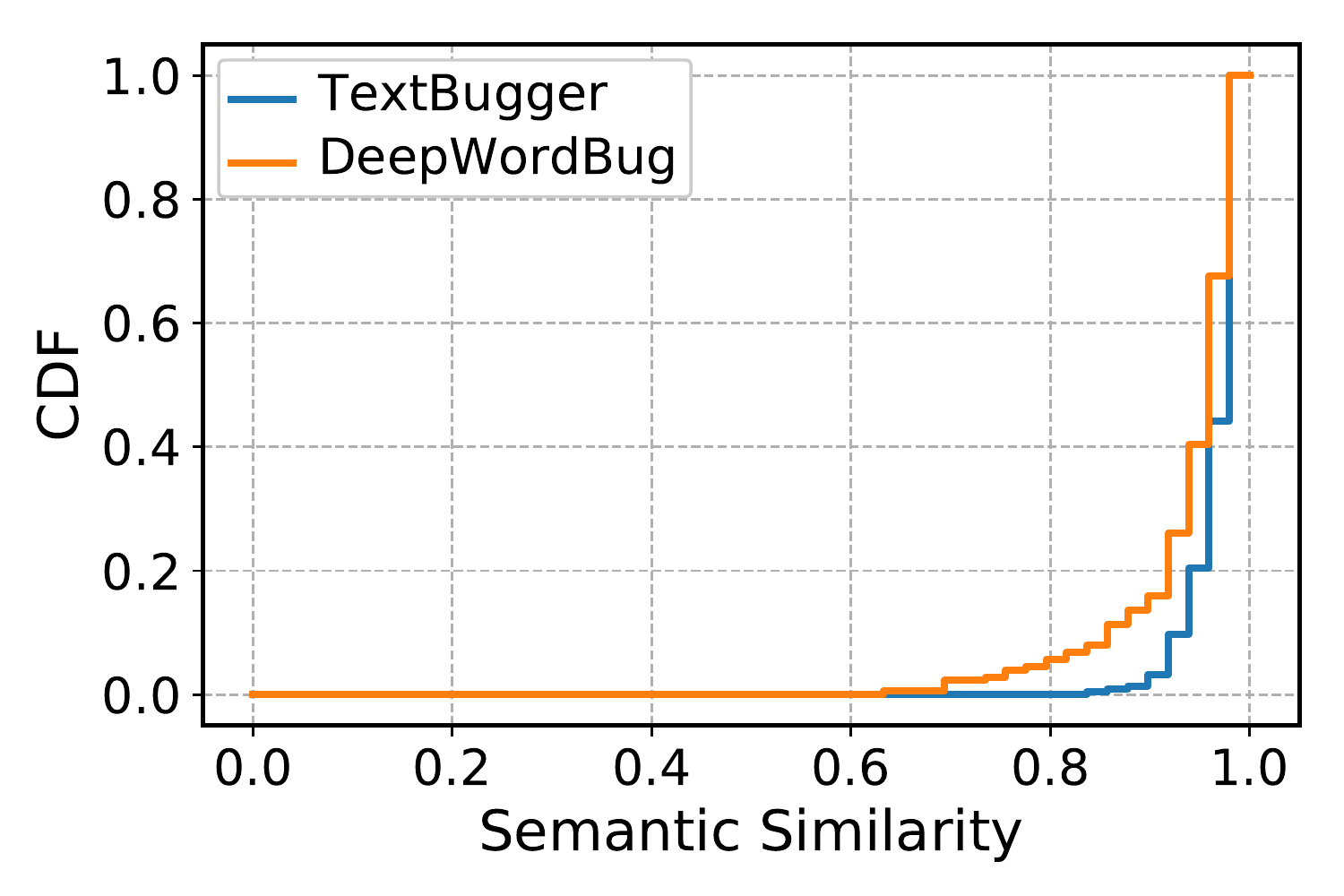}
    \label{fig:sentiment_blackbox_IMDB_semantic_similarity}
}
\caption{The average utility of adversarial texts generated on IMDB dataset under black-box settings for 10 platforms.}
\label{fig:sentiment_blackbox_utility_IMDB}
\vspace{-0.25cm} 
\end{figure}

\begin{figure}[tp]
\centering
\subfigure[Edit Distance]{
    \centering
    \includegraphics[width=0.235\textwidth]{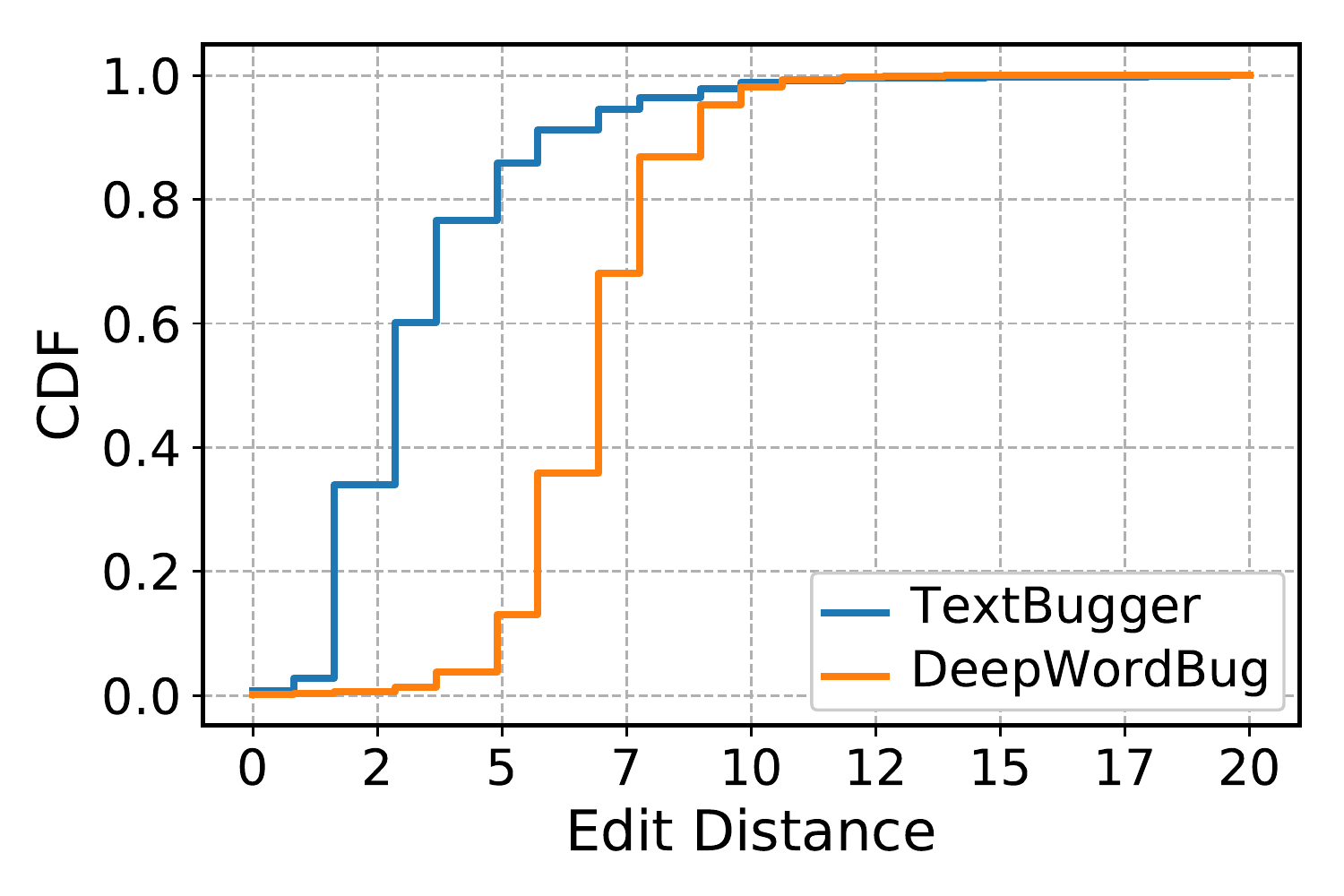}
    \label{fig:sentiment_blackbox_MR_edit_distance}
    \hspace{-0.5cm}
}
\subfigure[Jaccard Coefficient]{
    \centering
    \includegraphics[width=0.235\textwidth]{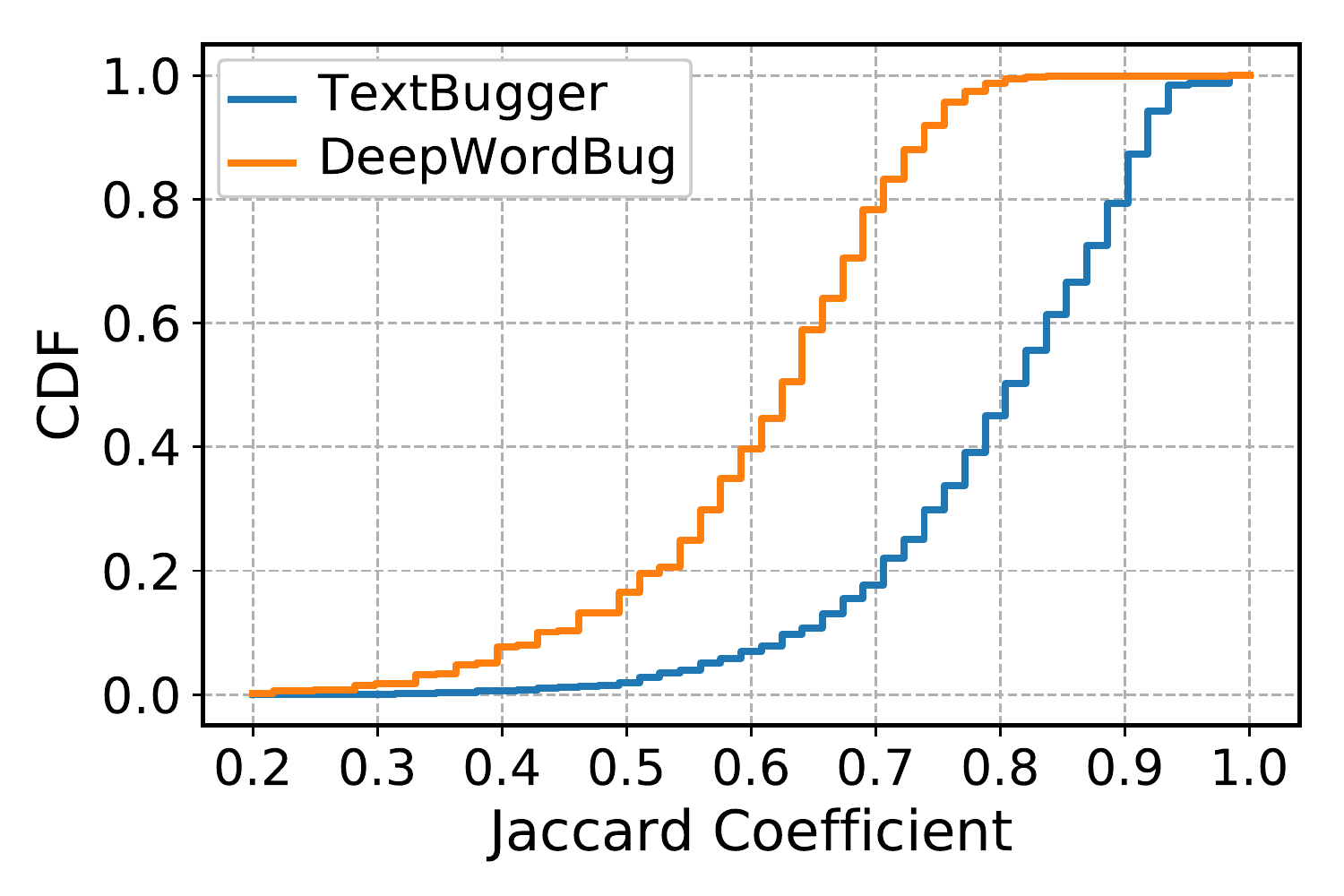}
    \label{fig:sentiment_blackbox_MR_jaccard_coefficient}
}
\subfigure[Euclidean Distance]{
    \centering
    \includegraphics[width=0.235\textwidth]{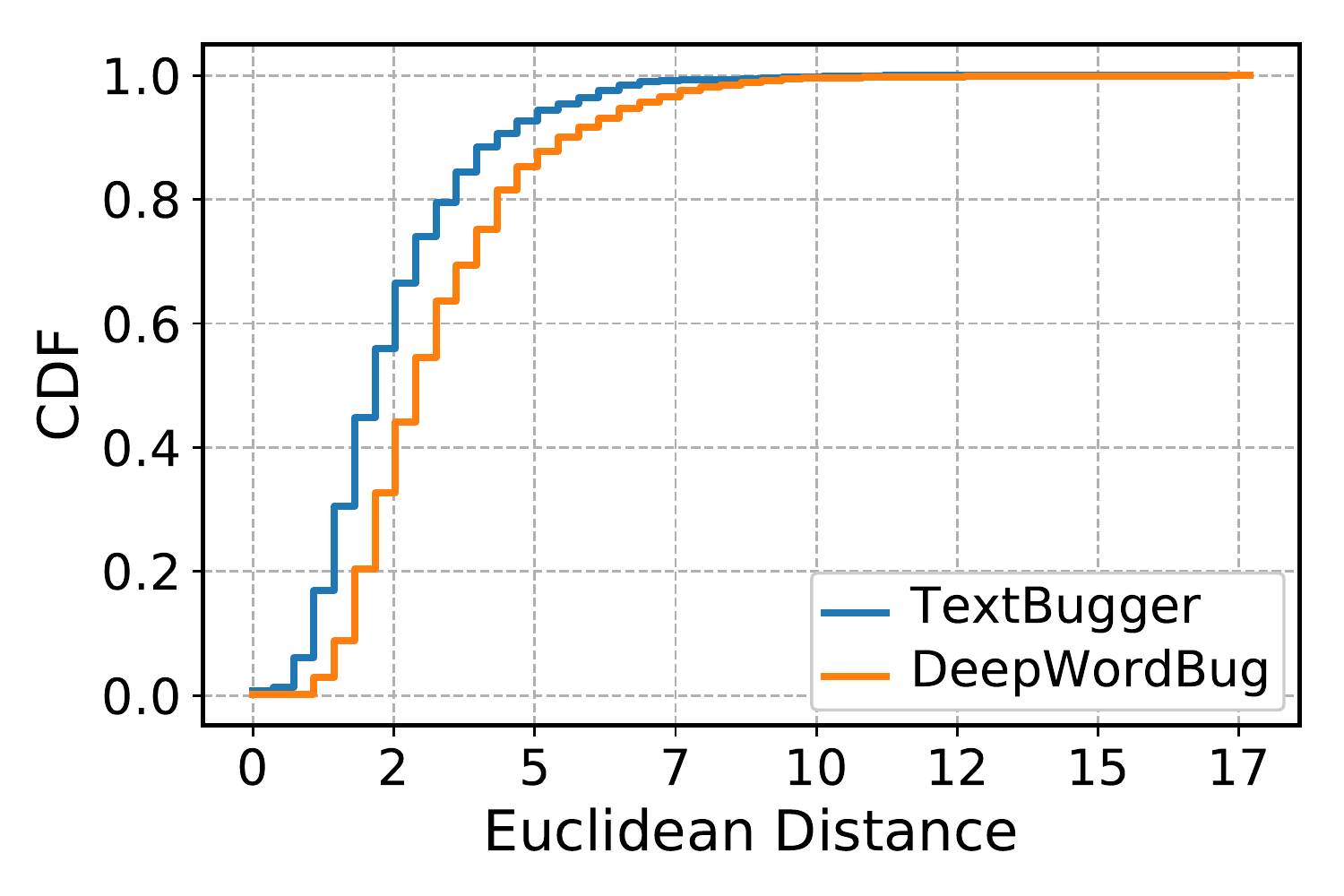}
    \label{fig:sentiment_blackbox_MR_euclidean_distance}
    \hspace{-0.5cm}
}
\subfigure[Semantic Similarity]{
    \centering
    \includegraphics[width=0.235\textwidth]{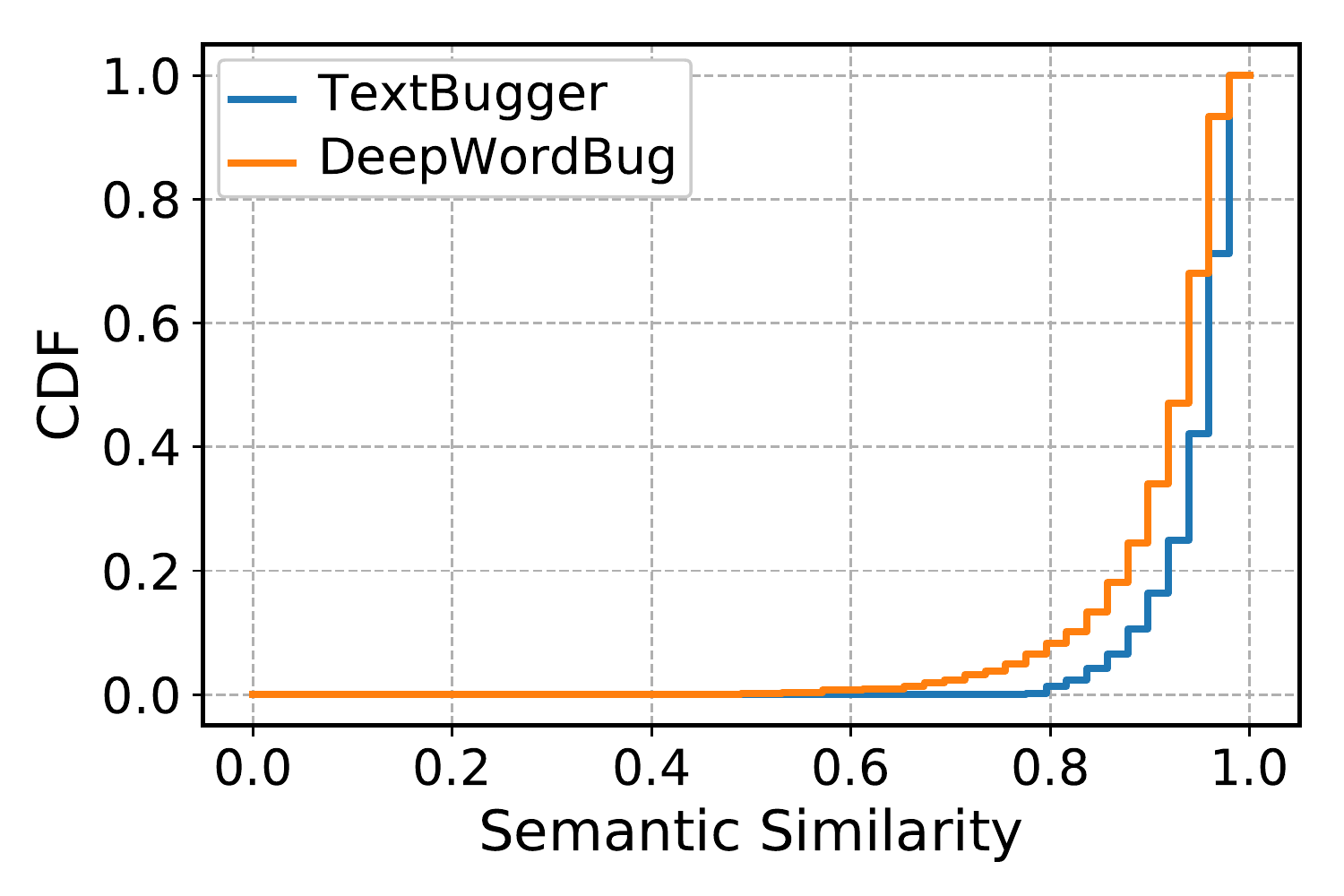}
    \label{fig:sentiment_blackbox_MR_semantic_similarity}
}
\caption{The average utility of adversarial texts generated on MR dataset under black-box settings for 10 platforms.}
\label{fig:sentiment_blackbox_utility_MR}
\vspace{-0.25cm} 
\end{figure}

For black-box attacks, the average similarity between original texts and adversarial texts against 10 platforms/models are shown in \Cref{fig:sentiment_blackbox_utility_IMDB,fig:sentiment_blackbox_utility_MR}.
From \Cref{fig:sentiment_blackbox_IMDB_edit_distance,fig:sentiment_blackbox_IMDB_jaccard_coefficient,fig:sentiment_blackbox_MR_edit_distance,fig:sentiment_blackbox_MR_jaccard_coefficient}, we can see that the adversarial texts generated by \textsc{TextBugger} are more similar to original texts than that generated by DeepWordBug in word-level.
From \Cref{fig:sentiment_blackbox_IMDB_euclidean_distance,fig:sentiment_blackbox_IMDB_semantic_similarity,fig:sentiment_blackbox_MR_euclidean_distance,fig:sentiment_blackbox_MR_semantic_similarity} we can see that the adversarial texts generated by \textsc{TextBugger} are more similar to original texts than that generated by DeepWordBug in the word vector space.
These results implies that the adversarial texts generated by \textsc{TextBugger} preserve more utility than that generated by DeepWordBug.
One reason is that DeepWordBug randomly chooses a bug from generated bugs, while \textsc{TextBugger} chooses the optimal bug that can change the prediction score most.
Therefore, DeepWordBug needs to manipulate more words than \textsc{TextBugger} to achieve successful attack.

\begin{figure}[tp]
\centering
\subfigure[Number of Perturbed Words]{
    \centering
    \includegraphics[width=0.235\textwidth]{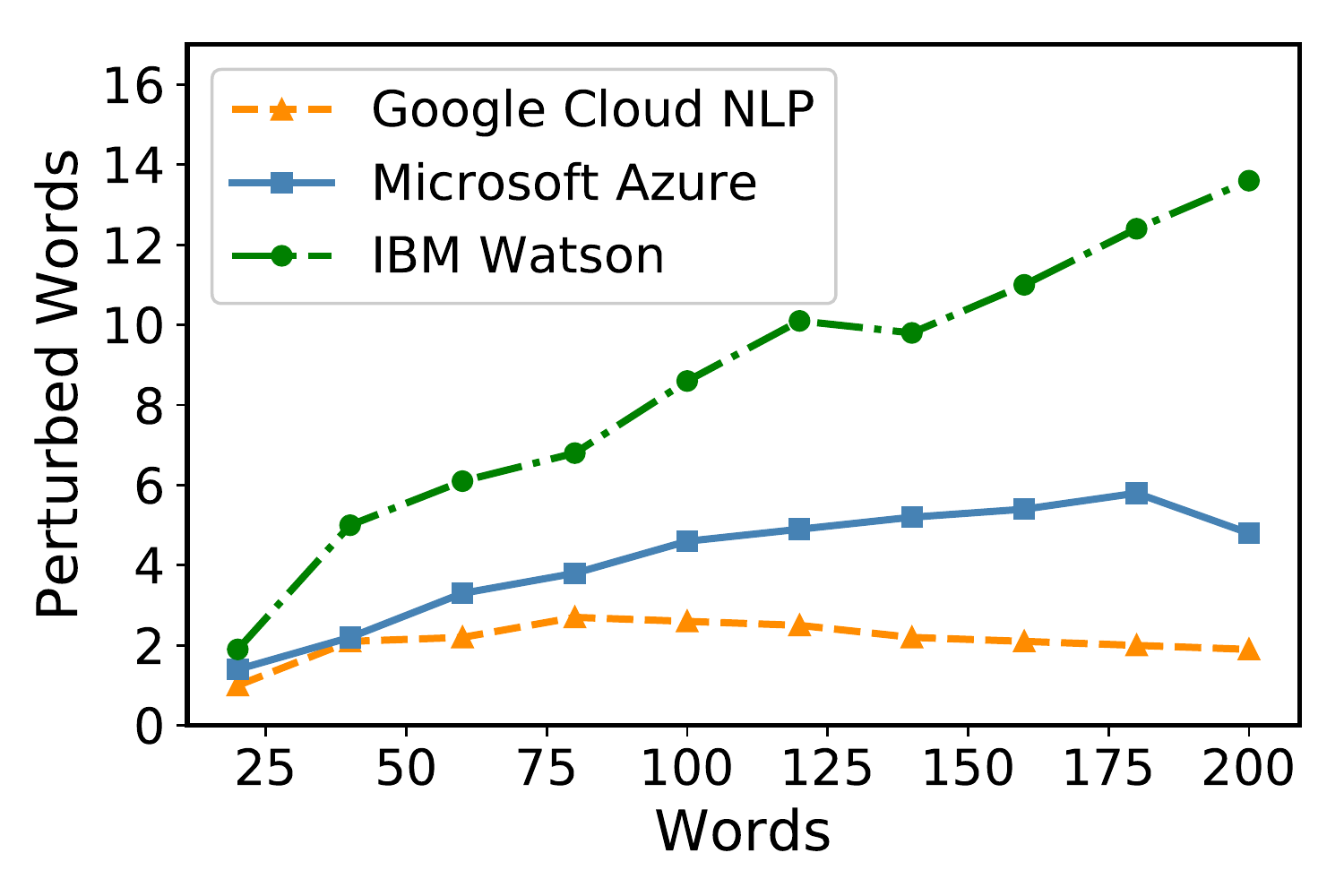}
    \label{fig:sentiment_wordlength_utility_perturbed_words}
    \hspace{-0.5cm}
}
\subfigure[Semantic Similarity]{
    \centering
    \includegraphics[width=0.235\textwidth]{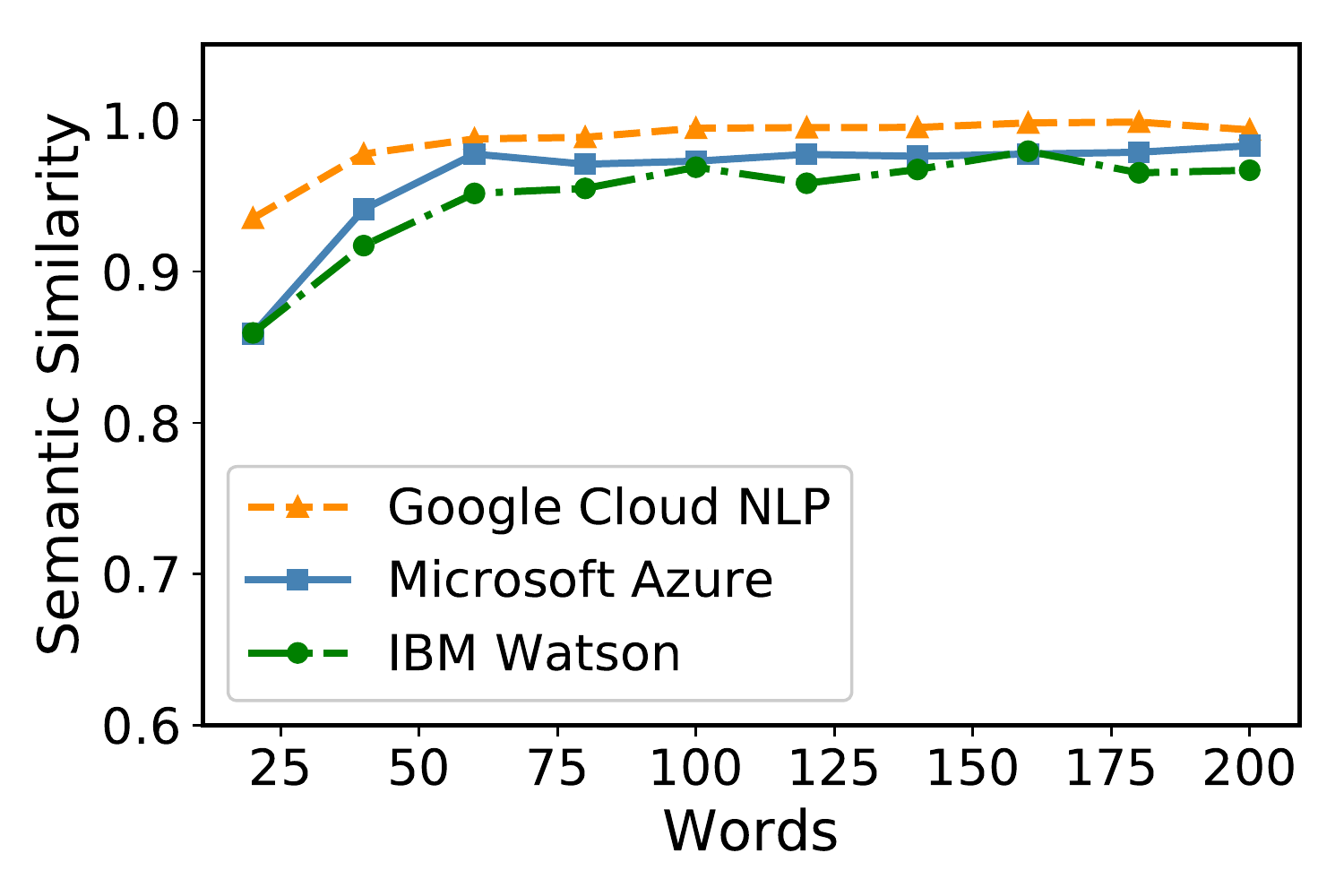}
    \label{fig:sentiment_wordlength_utility_semantic_similarity}
}
\caption{The impact of document length on the utility of generated adversarial texts in three online platforms: Google Cloud NLP, IBM Watson and Microsoft Azure.
The subfigures are: (a) the number of perturbed words and document length, (b) the document length and the semantic similarity between generated adversarial texts and original texts.}
\label{fig:sentiment_wordlength_utility}
\vspace{-0.25cm}
\end{figure}

\textbf{The Impact of Document Length.}
We also study the impact of word length on the utility of generated adversarial texts and show the results in \Cref{fig:sentiment_wordlength_utility}.
From \Cref{fig:sentiment_wordlength_utility_perturbed_words}, for IBM Watson and Microsoft Azure, we can see that the number of perturbed words roughly has a positive correlation with the average length of texts; for Google Cloud NLP, the number of perturbed words changes little with the increasing length of texts.
However, as shown in \Cref{fig:sentiment_wordlength_utility_semantic_similarity}, the increasing perturbed words do not decrease the semantic similarity of the adversarial texts.
This is because longer text would have richer semantic information, while the proportion of the perturbed words is always controlled within a small range by \textsc{TextBugger}.
Therefore, with the length of input text increasing, the perturbed words have smaller impact on the semantic similarity between original and adversarial texts.

\subsection{Discussion}

\begin{figure}[tp]
\centering
\subfigure[Word Cloud]{
    \centering
    \includegraphics[width=0.235\textwidth]{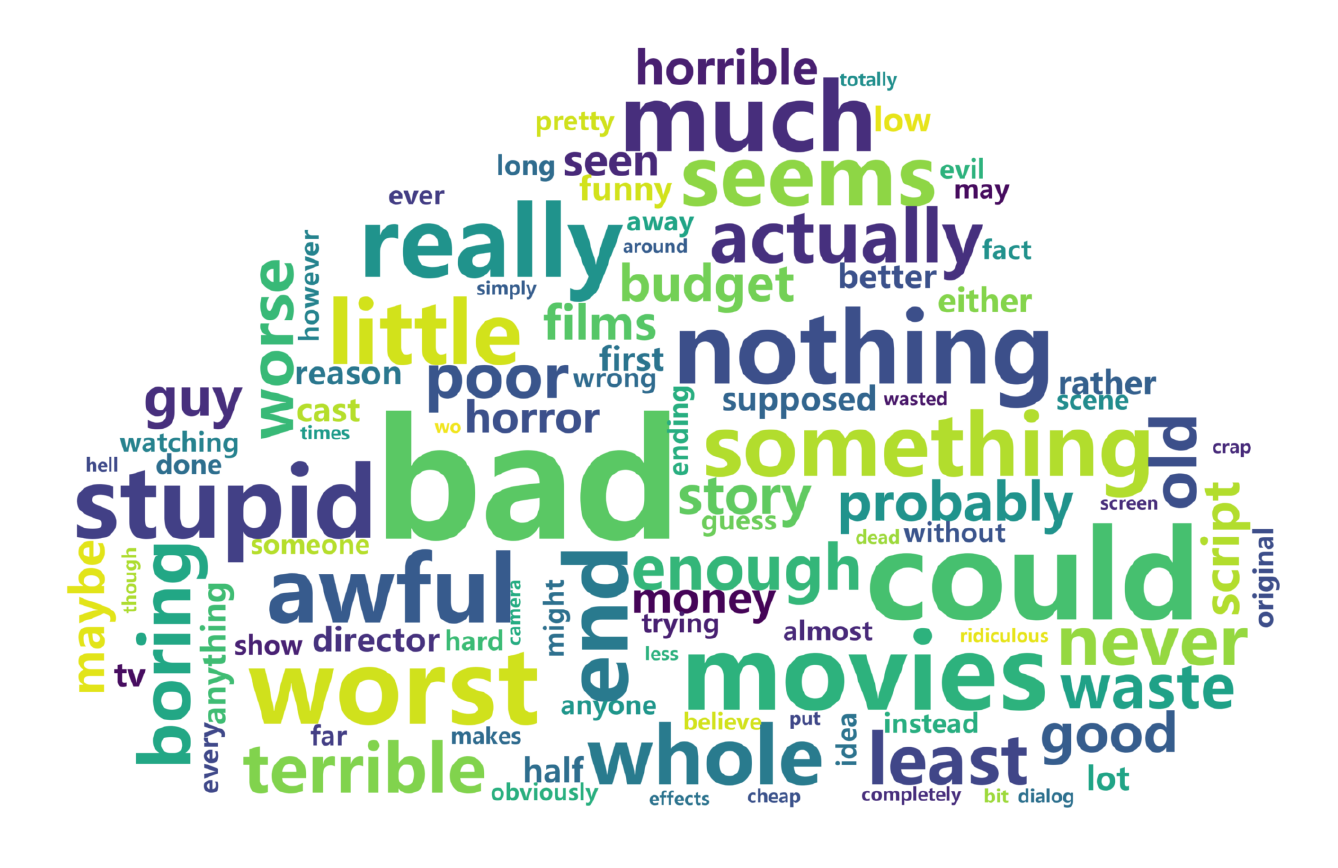}
    \label{fig:sentiment_whitebox_imdb_wordcloud}
    \hspace{-0.5cm}
}
\subfigure[Bug Distribution]{
    \centering
    \includegraphics[width=0.235\textwidth]{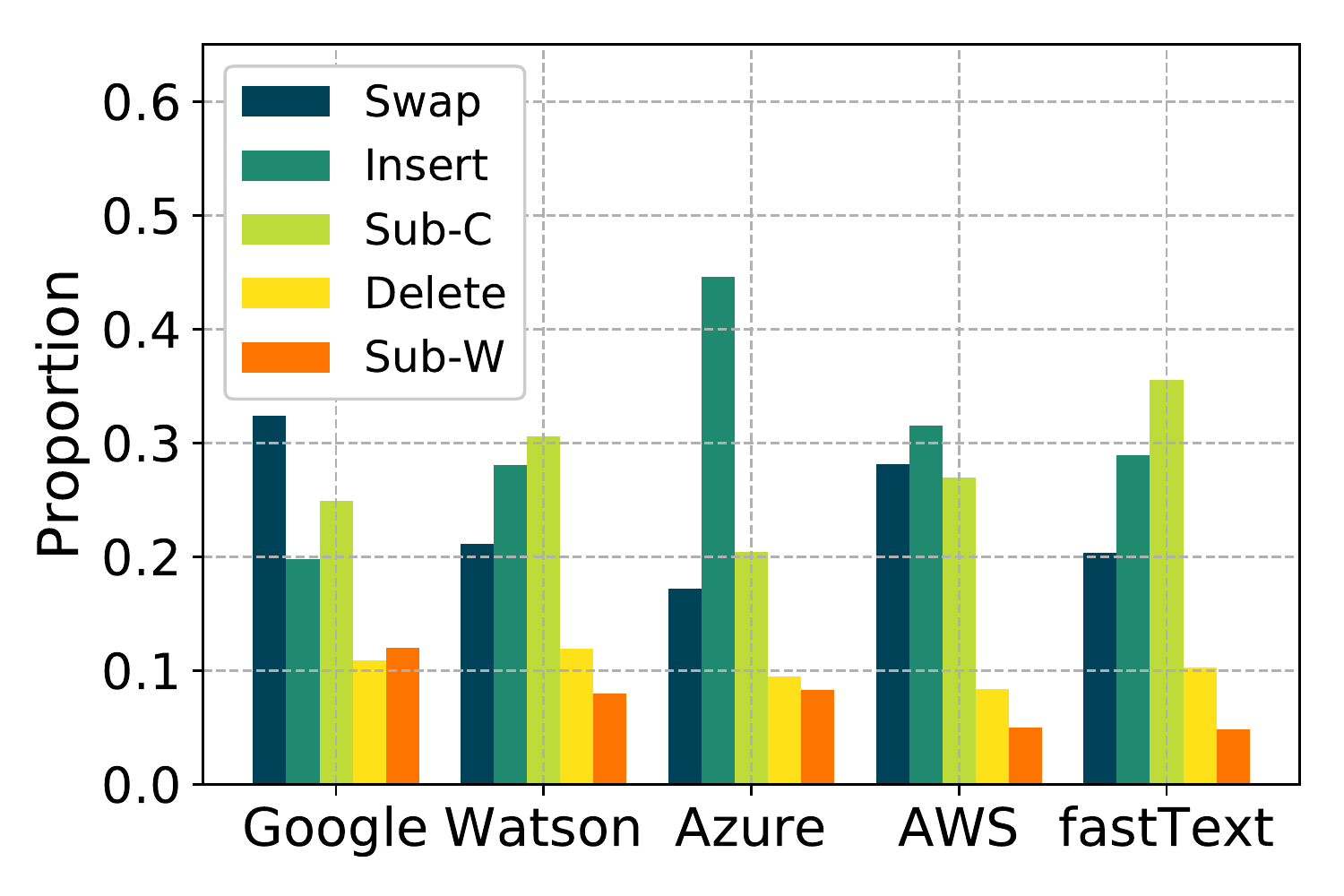}
    \label{fig:sentiment_blackbox_imdb_bug_distribution}
}
\caption{(a) The word cloud is generated from IMDB dataset against the CNN model. (b) The bug distribution of the adversarial texts is generated from IMDB dataset against the online platforms.}
\label{fig:sentiment_wordcloud_and_bug_distribution}
\vspace{-0.25cm} 
\end{figure}

\textbf{Toxic Words Distribution.}
To demonstrate the effectiveness of our method, we visualize the found important words according to their frequency in \Cref{fig:sentiment_whitebox_imdb_wordcloud}, in which the words higher frequency will be represented with larger font.
From \Cref{fig:sentiment_whitebox_imdb_wordcloud}, we can see that the found important words are indeed negative words, e.g., ``bad'', ``awful'', ``stupid'', ``worst'', ``terrible'', etc for negative texts.
Slight modification on these negative words would decrease the negative extent of input texts.
This is why \textsc{TextBugger} can generate adversarial texts whose only difference to the original texts are few character-level modifications. 

\textbf{Types of Perturbations.}
The proportion of each operation chosen by the adversary for the experiments are shown in \Cref{fig:sentiment_blackbox_imdb_bug_distribution}.
We can see that insert is the dominant operation for Microsoft Azure and Amazon AWS, while Sub-C is the dominant operation for IBM Watson and fastText.
One reason could be that Sub-C is deliberately designed for creating visually similar adversarial texts, while swap, insert and delete are common in typo errors.
Therefore, the bugs generated by Sub-C are less likely to be found in the large-scale word vector space, thus causing the ``out-of-vocabulary'' phenomenon.
Meanwhile, delete and Sub-W are used less than the others.
One reason is that Sub-W should satisfy two conditions: substituting with semantic similar words while changing the score largely in the five types of bugs.
Therefore, the proportion of Sub-W is less than other operations.

\section{Attack Evaluation: Toxic Content Detection} \label{sec:experiments_black}

Toxic content detection aims to apply NLP, statistics, and machine learning methods to detect illegal or toxic-related (e.g., irony, sarcasm, insults, harassment, racism, pornography, terrorism, and riots, etc.) content for online systems.
Such toxic content detection can help moderators to improve the online conversation environment.

In this section, we investigate practical performance of the proposed method for generating adversarial texts against real-world toxic content detection systems.
We start with introducing the datasets, targeted models and implementation details.
Then we will analyze the results and discuss potential reasons for the observed performance.

\subsection{Dataset}
We apply the dataset provided by the Kaggle Toxic Comment Classification competition\footnote{https://www.kaggle.com/c/jigsaw-toxic-comment-classification-challenge}.
This dataset contains a large number of Wikipedia comments which have been labeled by human raters for toxic behavior. 
There are six types of indicated toxicity, i.e., ``toxic'', ``severe toxic'', ``obscene'', ``threat'', ``insult'', and ``identity hate'' in the original dataset. We consider these categories as toxic and perform binary classification for toxic content detection. 
For more coherent comparisons, a balanced subset of this dataset is constructed for evaluation. 
This is achieved by random sampling of the non-toxic texts, obtaining a subset with equal number of samples with the toxic texts.
Further, we removed some abnormal texts (i.e., containing multiple repeated characters) and select the samples that have no more than 200 words for our experiment, due to the fact that some APIs limit the maximum length of input sentences. 
We obtained 12,630 toxic texts and non-toxic texts respectively.

\subsection{Targeted Model \& Implementation}
For white-box experiments, we evaluated the \textsc{TextBugger} on self-trained LR, CNN and LSTM models as we do in \Cref{sec:experiments_white}.
All models are trained in a hold-out test strategy, i.e., 80\%, 10\%, 10\% of the data was used for training, validation and test, respectively. 
Hyper-parameters were tuned only on the validation set, and the final adversarial examples are generated and evaluated on the test set.

For black-box experiments, we evaluated the \textsc{TextBugger} on five toxic content detection platforms/models, including Google Perspective, IBM Natural Language Classifier, Facebook fastText, ParallelDots AI, and Aylien Offensive Detector.
Since the IBM Natural Language Classifier and the Facebook fastText need to be trained by ourselves\footnote{We do not know the models' parameters or architechtures because they only provide training and predicting interfaces.}, we selected 80\% of the Kaggle dataset for training and the rest for testing. 
Note that we do not selected samples for validation since these two models only require training and testing set.

The implementation details of our toxic content attack are similar with that in the sentiment analysis attack, including the baselines. 

\begin{table*}[tp]
    \caption{Results of the White-box Attack on Kaggle Dataset.}
    \label{tab:toxic_whitebox_summary}
    \centering
    \scalebox{1.05}{
    \begin{tabular}{C{1.7cm}C{2.3cm}C{1.0cm}C{1.2cm}C{1.0cm}C{1.2cm}C{1.0cm}C{1.2cm}C{1.0cm}C{1.2cm}}
    \toprule 
    \multirowcell{2}[-1.6ex][c]{\centering \textbf{Targeted Model}} & \multirowcell{2}[-1.6ex][c]{\centering \textbf{Original Accuracy}} & \multicolumn{2}{c}{\textbf{Random}} & \multicolumn{2}{c}{\textbf{FGSM+NNS \cite{gong2018adversarial}}} & \multicolumn{2}{c}{\textbf{DeepFool+NNS \cite{gong2018adversarial}}} & \multicolumn{2}{c}{\textbf{\textsc{TextBugger}}} \\
    \cmidrule(r){3-4} \cmidrule(r){5-6} \cmidrule(r){7-8} \cmidrule(r){9-10}
     & & Success Rate & Perturbed Word & Success Rate & Perturbed Word & Success Rate & Perturbed Word & Success Rate & Perturbed Word \\
     \midrule
     \textbf{LR} & 88.5\% & 1.4\% & 10\% & 33.9\% & 5.4\% & 29.7\% & 7.3\% & \textbf{92.3\%} & 10.3\% \\
	 \textbf{CNN} & 93.5\% & 0.5\% & 10\% & 26.3\% & 6.2\% & 27.0\% & 9.9\% & \textbf{82.5\%} & 10.8\% \\
     \textbf{LSTM} & 90.7\% & 0.9\% & 10\% & 28.6\% & 8.8\% & 30.3\% & 10.3\% & \textbf{94.8\%} & 9.5\% \\
    \bottomrule
    \end{tabular}}
\end{table*}

\begin{table*}[tp]
    \caption{Results of the black-box attack on Kaggle dataset.}
    \label{tab:toxic_blackbox_summary}
    \centering
    \scalebox{1.1}{
    \begin{tabular}{C{3.0cm}C{2.3cm}ccccccc}
    \toprule 
    \multirowcell{2}[-.5ex][c]{\centering \textbf{Targeted Platform/Model}} & \multirowcell{2}[-.5ex][c]{\centering \textbf{Original Accuracy}} & \multicolumn{3}{c}{\textbf{DeepWordBug \cite{gao2018black}}} & \multicolumn{3}{c}{\textbf{\textsc{TextBugger}}} \\
     \cmidrule(r){3-5} \cmidrule(r){6-8}
     & & Success Rate & Time (s) & Perturbed Word & Success Rate & Time (s) & Perturbed Word \\
     \midrule
     Google Perspective & 98.7\% & 33.5\% & 400.20 & 10\% & \textbf{60.1\%} & 102.71 & 5.6\% \\
     IBM Classifier & 85.3\% & 9.1\% & 75.36 & 10\% & \textbf{61.8\%} & 21.53 & 7.0\% \\
     Facebook fastText & 84.3\% & 31.8\% & 0.05 & 10\% & \textbf{58.2\%} & 0.03 & 5.7\% \\
     ParallelDots & 72.4\% & 79.3\% & 148.67 & 10\% & \textbf{82.1\%} & 23.20 & 4.0\% \\
     Aylien Offensive Detector & 74.5\% & 53.1\% & 229.35 & 10\% & \textbf{68.4\%} & 37.06 & 32.0\% \\
    \bottomrule
    \end{tabular}}
    \vspace{-0.25cm} 
\end{table*}

\subsection{Attack Performance}

\textbf{Effectiveness and Efficiency.} 
\Cref{tab:toxic_whitebox_summary,tab:toxic_blackbox_summary} summarize the main results of the white-box and black-box attacks on the Kaggle dataset.
We can observe that under white-box settings, the Random strategy has minor influence on the final results in \Cref{tab:toxic_whitebox_summary}.
On the contrary, \textsc{TextBugger} only perturbs a few words to achieve high attack success rate and performs much better than baseline algorithms against all models/platforms.
For instance, as shown in \Cref{tab:toxic_whitebox_summary}, it only perturbs 10.3\% words of one sample to achieve 92.3\% success rate on the LR model, while all baselines achieve no more than 40\% attack success rate.
As the Kaggle dataset has an average length of 55 words, \textsc{TextBugger} only perturbed about 6 words for one sample to conduct successful attacks.
Furthermore, as shown in \Cref{tab:toxic_blackbox_summary}, it only perturbs 4.0\% words (i.e., about 3 words) of one sample when achieves 82.1\% attack success rate on the ParallelDots platform.
These results imply that an adversary can successfully mislead the system into assigning significantly different toxicity scores to the original sentences via modifying them slightly.

\textbf{Successful Attack Examples.} Two successful examples are shown in \Cref{fig:examples} as demonstration.
The first adversarial text for toxic content detection in \Cref{fig:examples} contains one Sub-W operation (``sexual'' to ``sexual-intercourse''), which successfully converts the prediction result of the LSTM model from 96.7\% toxic to 83.5\% non-toxic. 
The second adversarial text for toxic content detection in \Cref{fig:examples} contains three modifications, i.e., one swap operation (``shit'' to ``shti''), one Sub-C operation (``fucking'' to ``fuckimg'') and one Sub-W operation (``hell'' to ``helled'').
These modifications successfully convert the prediction result of the Perspective API from 92\% toxic to 78\% non-toxic\footnote{Since the Perspective API only returns the toxic score, we consider that 22\% toxic score is equal to 78\% non-toxic score.}.

\begin{figure}[tp]
    \centering
    \includegraphics[width=0.43\textwidth]{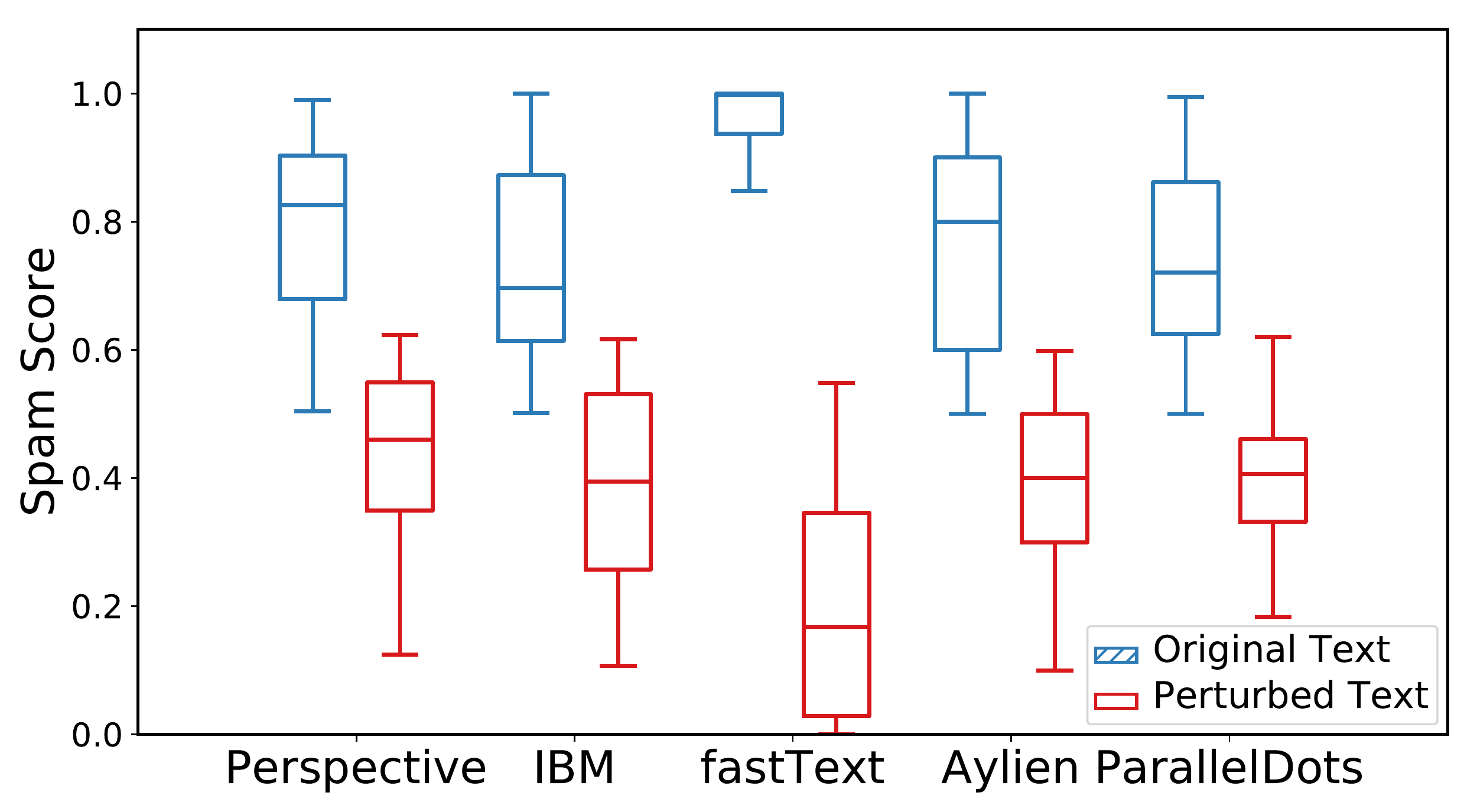}
    \caption{Score distribution of the after-modification texts. These texts are generated from Kaggle dataset against LR model.}
    \label{fig:toxic_score_distribution}
    \vspace{-0.25cm} 
\end{figure}

\textbf{Score Distribution.}
We also measured the change of the confidence value over all the samples including the failed samples before and after modifications.
The results are shown in \Cref{fig:toxic_score_distribution}, where the overall score of the after-modification texts has drifted to non-toxic for all platforms/models.

\subsection{Utility Analysis}
\begin{figure}[tp]
\centering
\subfigure[Edit Distance]{
    \centering
    \includegraphics[width=0.235\textwidth]{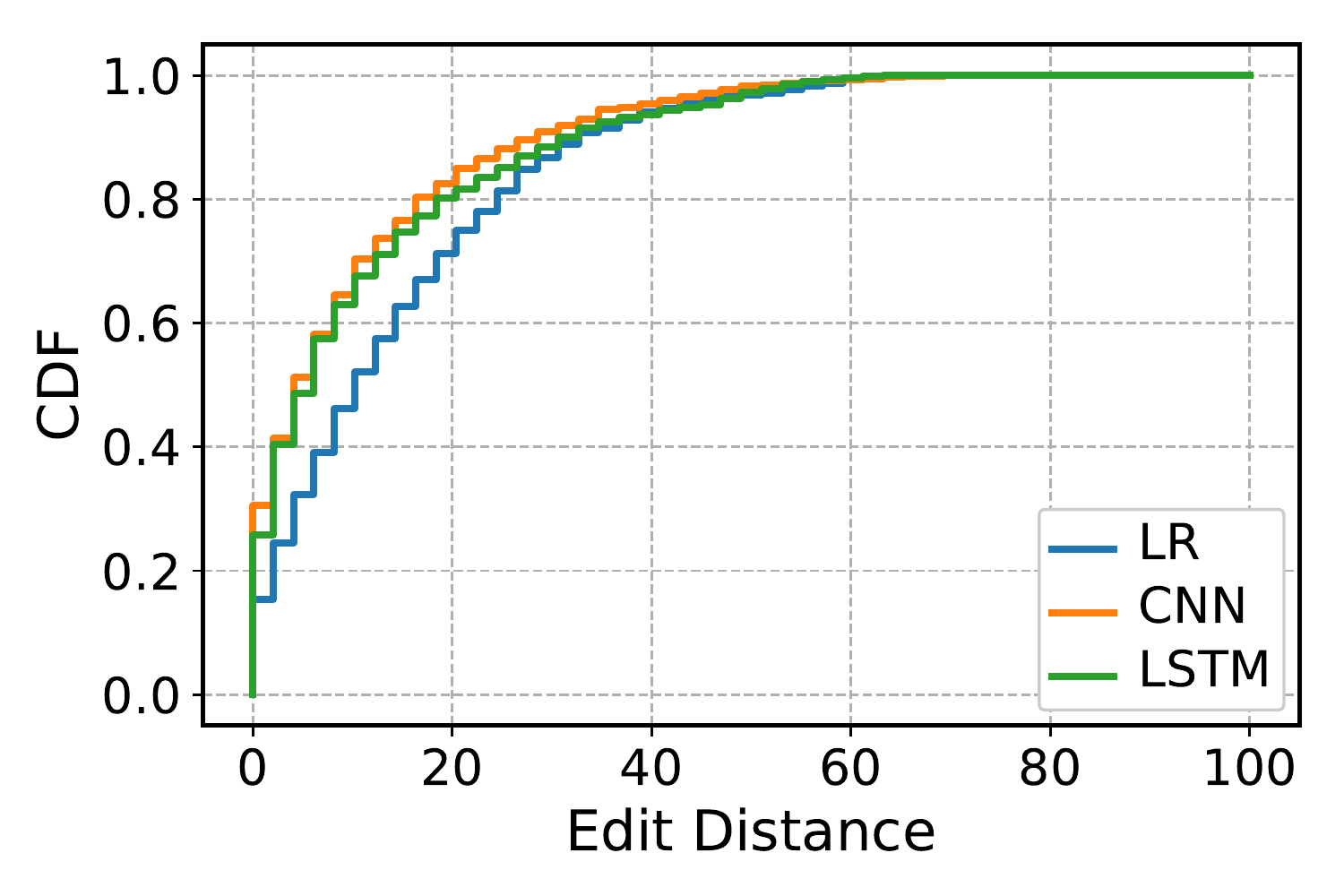}
    \label{fig:toxic_whitebox_utility_edit_distance}
    \hspace{-0.5cm}
}
\subfigure[Jaccard Coefficient]{
    \centering
    \includegraphics[width=0.235\textwidth]{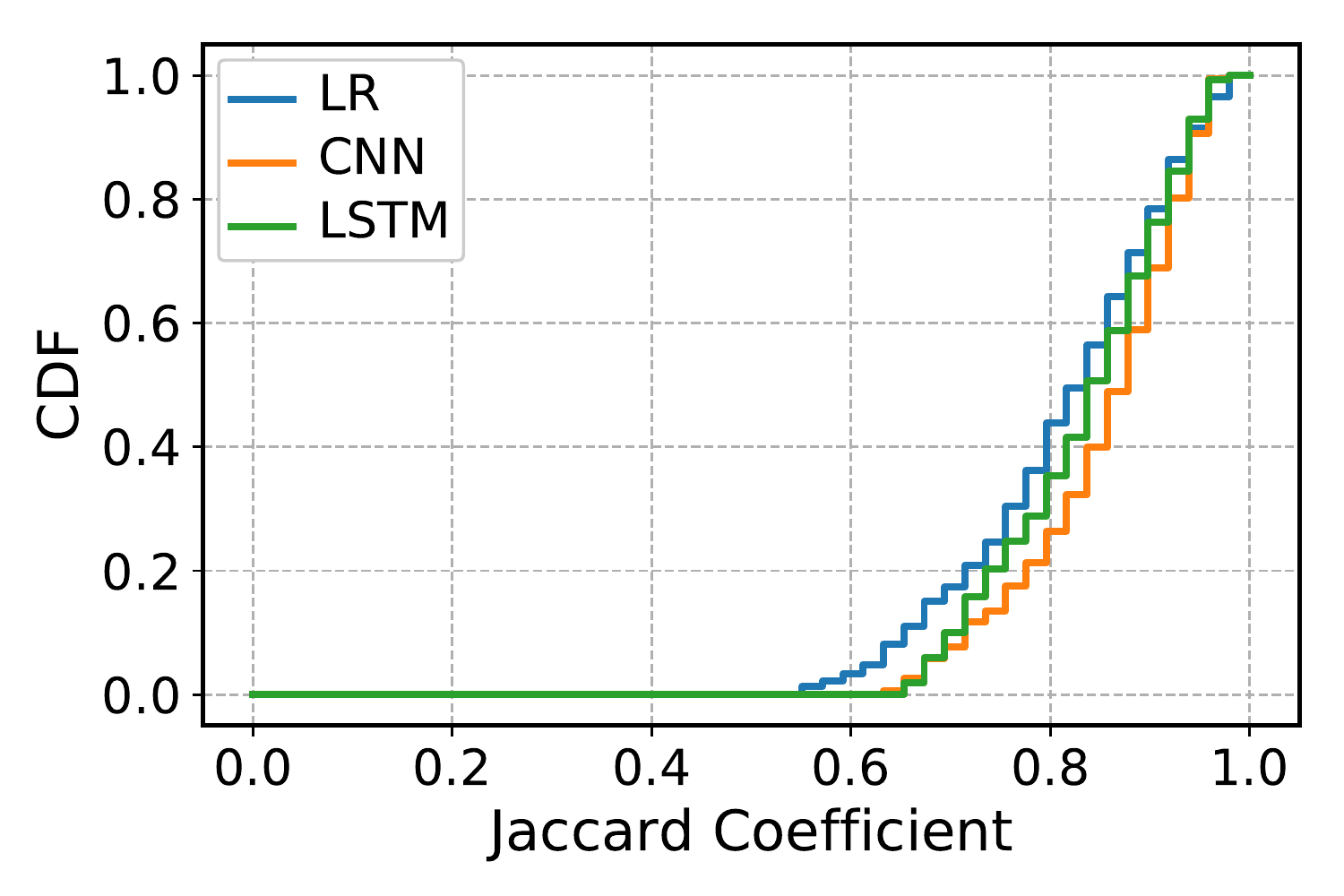}
    \label{fig:toxic_whitebox_utility_Jaccard_coef}
}
\subfigure[Euclidean Distance]{
    \centering
    \includegraphics[width=0.235\textwidth]{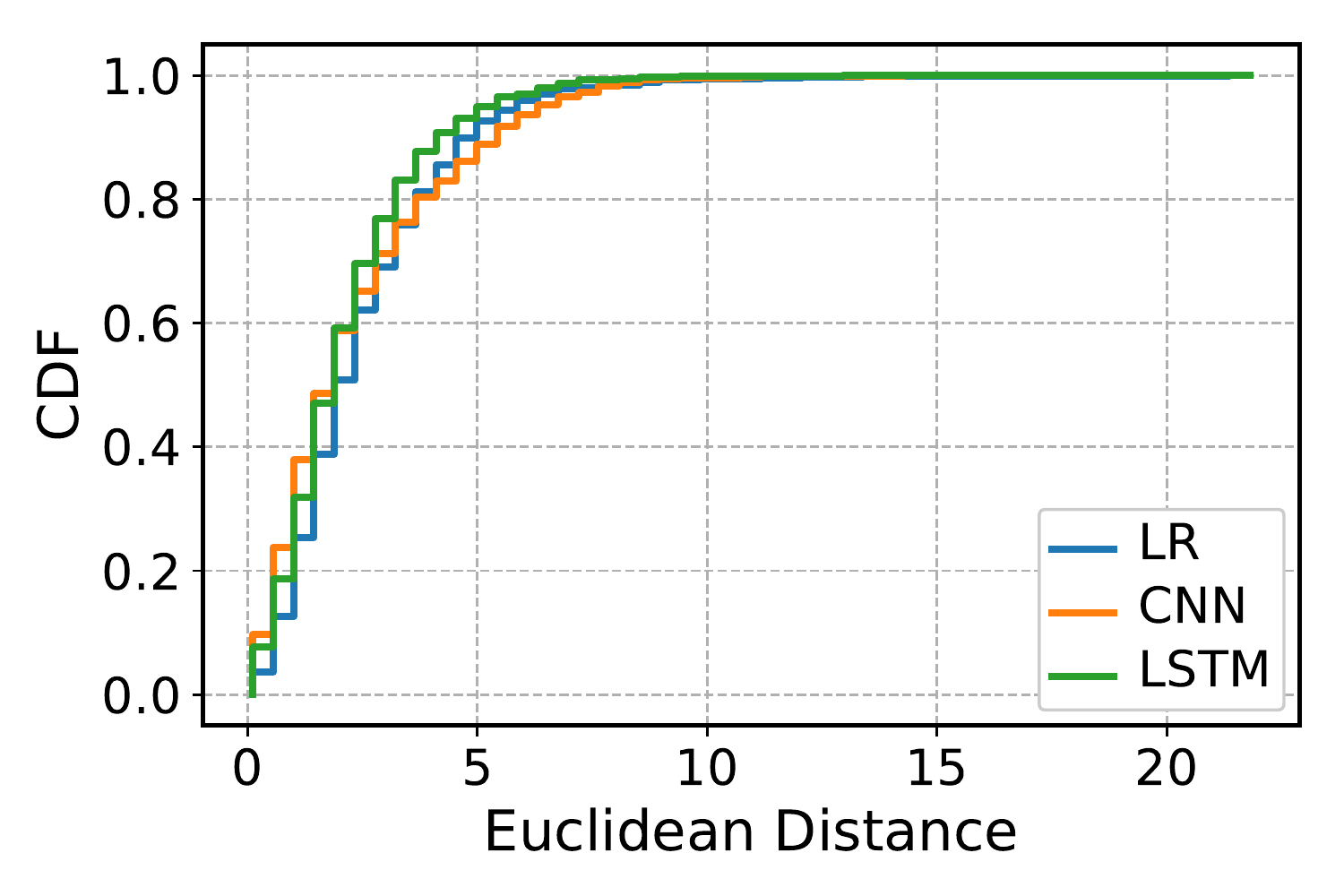}
    \label{fig:toxic_whitebox_utility_Euclidean_distance}
    \hspace{-0.5cm}
}
\subfigure[Semantic Similarity]{
    \centering
    \includegraphics[width=0.235\textwidth]{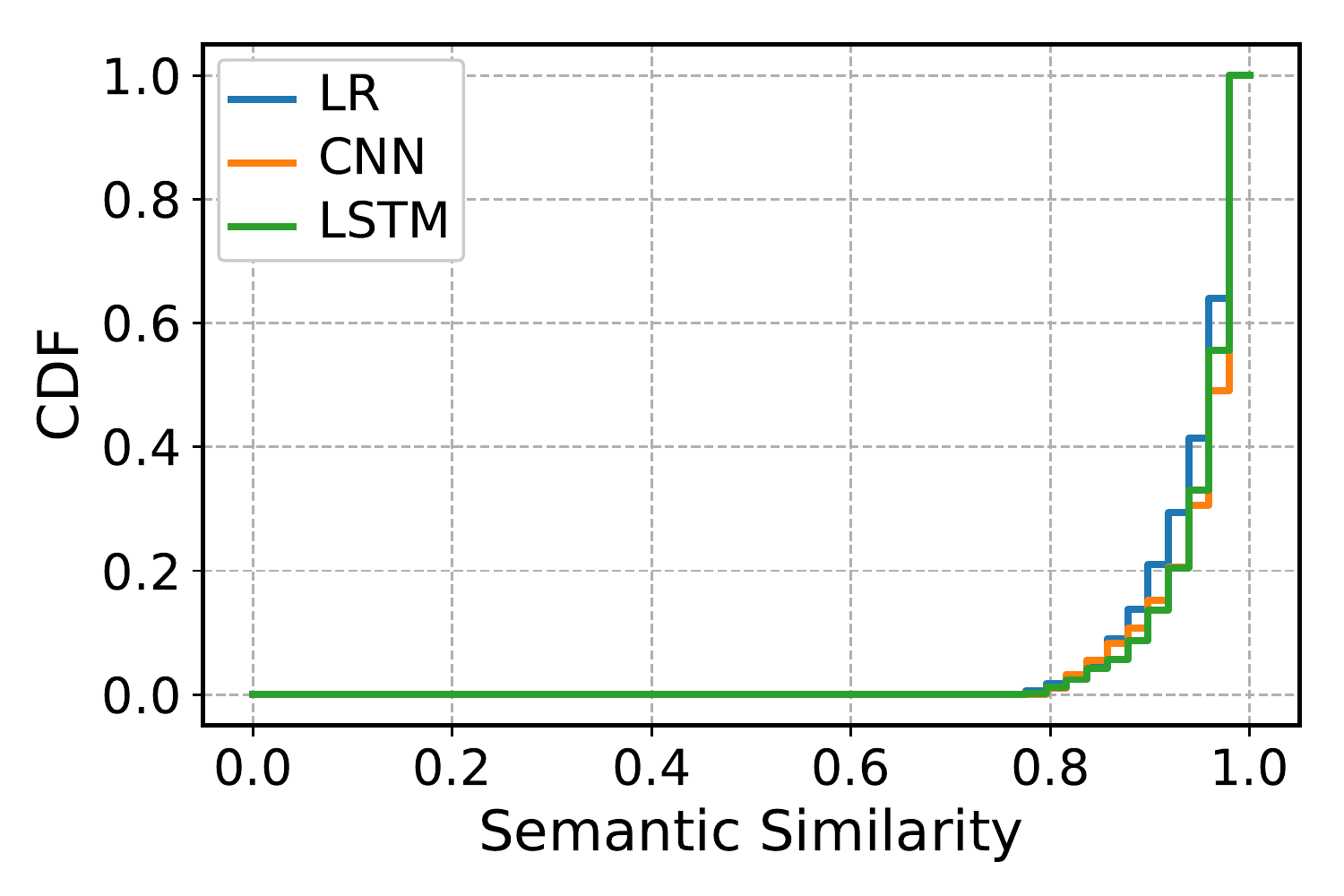}
    \label{fig:toxic_whitebox_utility_semantic_similarity}
}
\caption{The utility of adversarial texts generated on the Kaggle dataset under white-box settings for LR, CNN and LSTM models.}
\label{fig:toxic_whitebox_utility}
\vspace{-0.25cm} 
\end{figure}

\begin{figure}[tp]
\centering
\subfigure[Edit Distance]{
    \centering
    \includegraphics[width=0.235\textwidth]{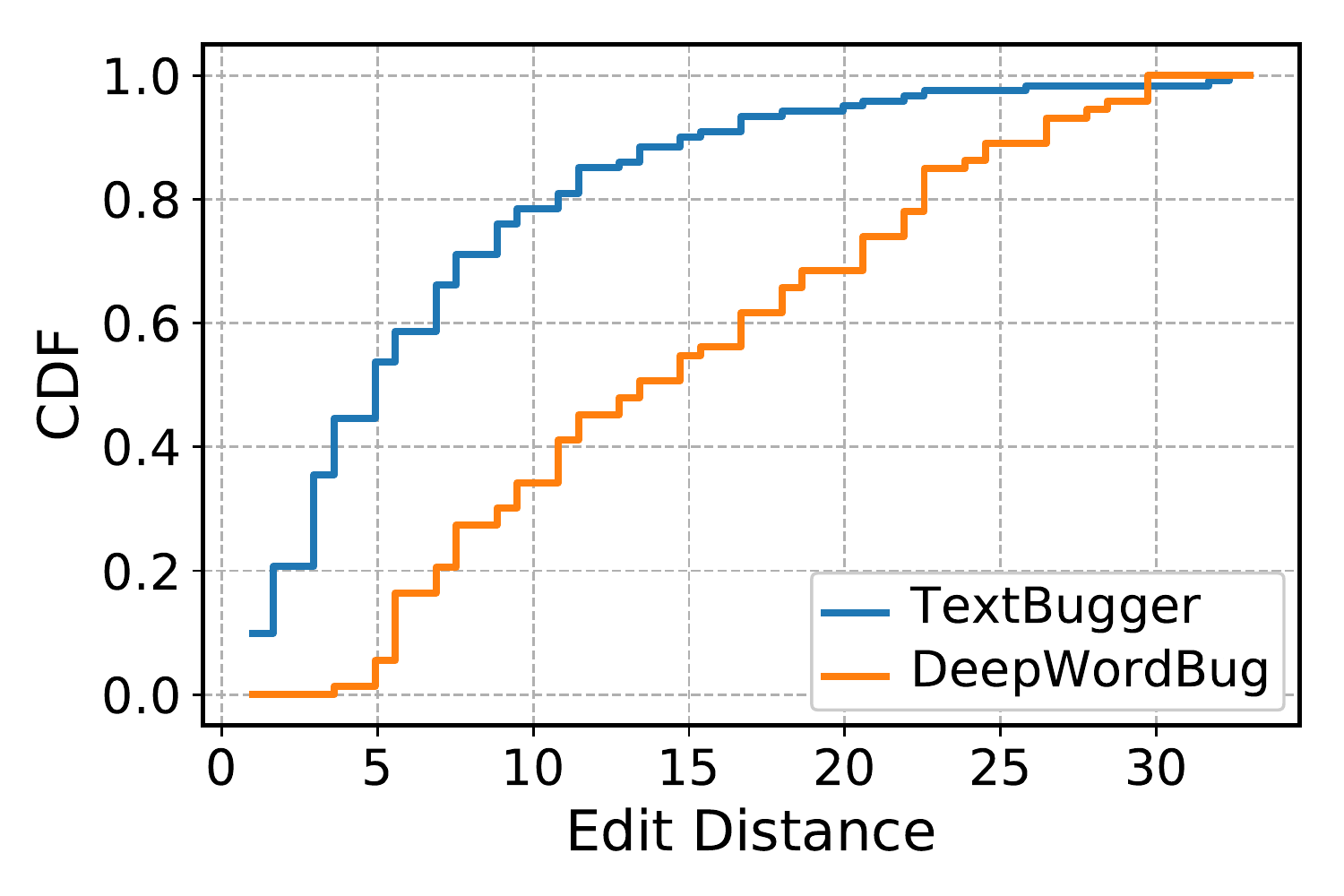}
    \label{fig:toxic_blackbox_utility_edit_distance}
    \hspace{-0.5cm}
}
\subfigure[Jaccard Similarity Coefficient]{
    \centering
    \includegraphics[width=0.235\textwidth]{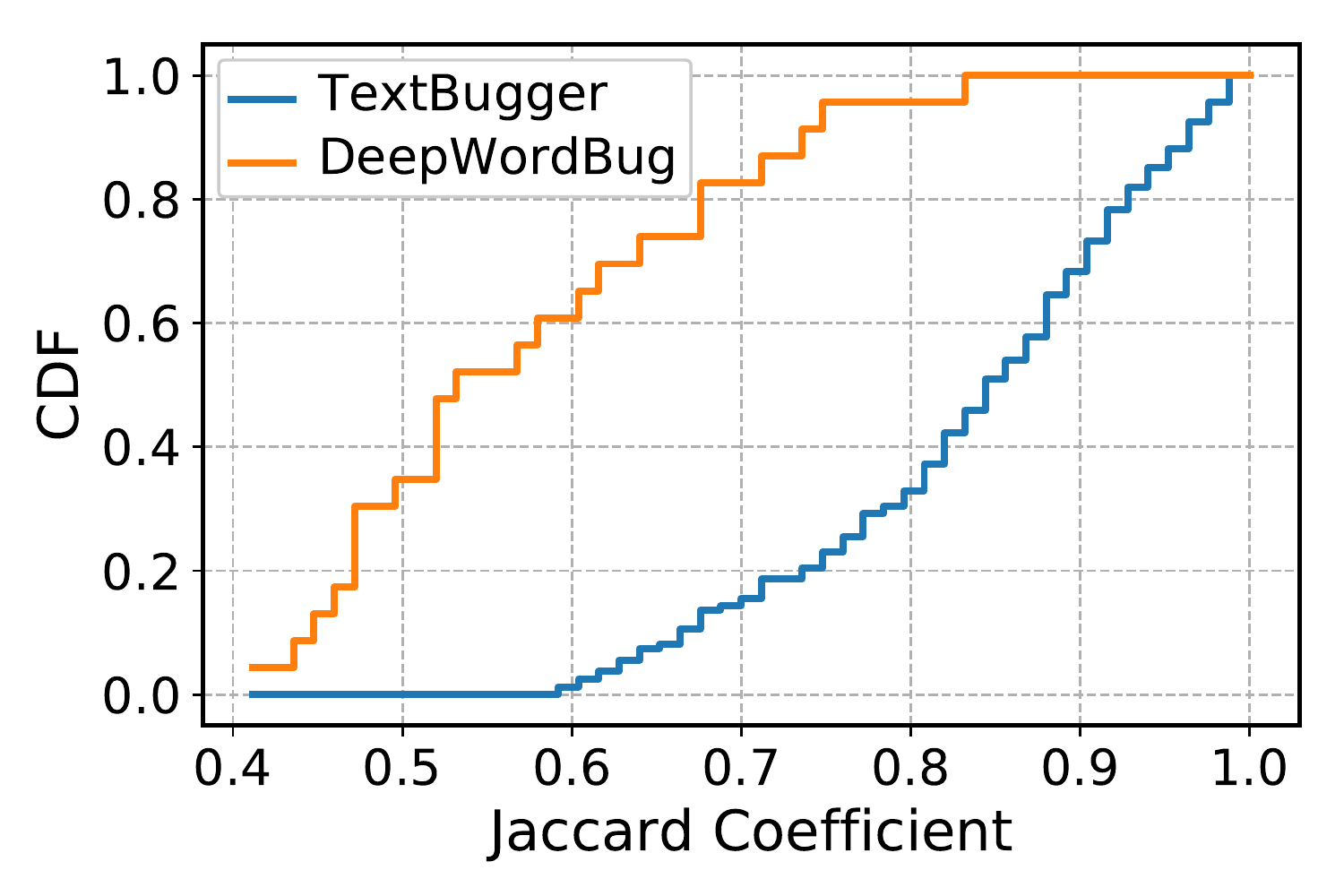}
    \label{fig:toxic_blackbox_utility_Jaccard_coef}
}
\subfigure[Euclidean Distance]{
    \centering
    \includegraphics[width=0.235\textwidth]{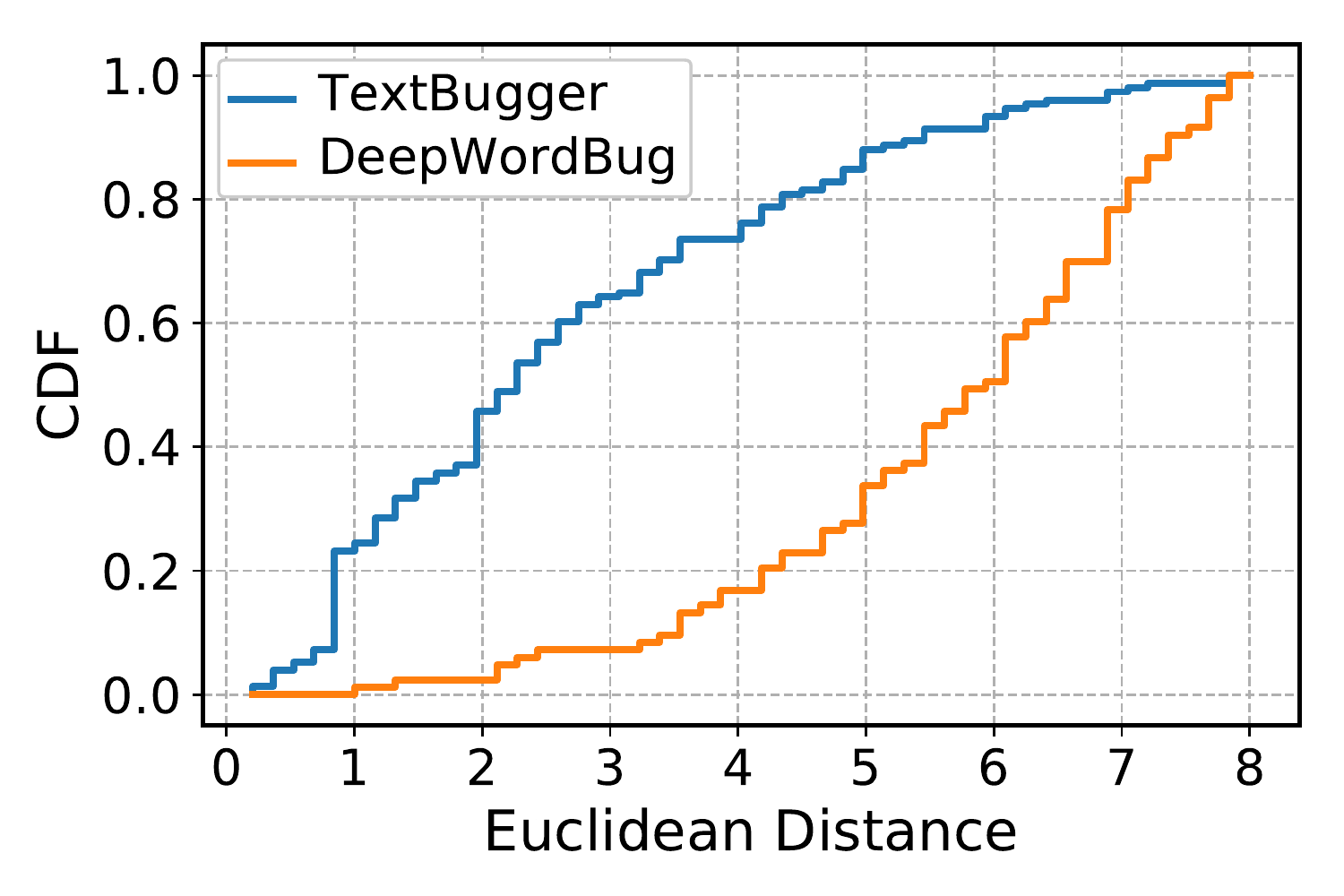}
    \label{fig:toxic_blackbox_utility_Euclidean_distance}
    \hspace{-0.5cm}
}
\subfigure[Semantic Similarity]{
    \centering
    \includegraphics[width=0.235\textwidth]{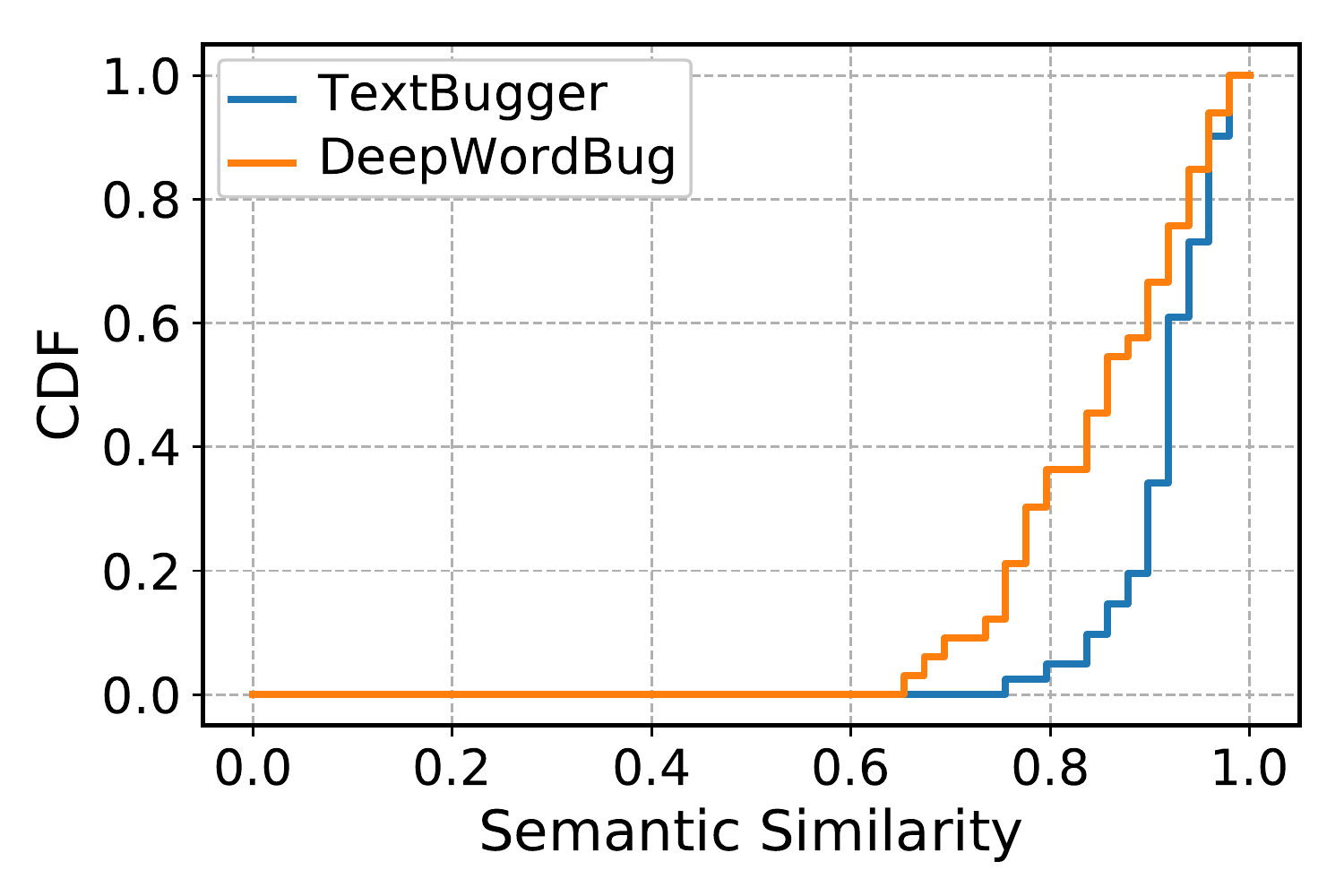}
    \label{fig:toxic_blackbox_utility_semantic_similarity}
}
\caption{The average utility of adversarial texts generated on Kaggle dataset under black-box settings for 5 platforms.}
\label{fig:toxic_blackbox_utility}
\vspace{-0.25cm} 
\end{figure}

\Cref{fig:toxic_whitebox_utility,fig:toxic_blackbox_utility} show the similarity between original texts and adversarial texts under white-box and black-box settings respectively.
First, \Cref{fig:toxic_blackbox_utility} clearly shows that the adversarial texts generated by \textsc{TextBugger} preserve more utility than that generated by DeepWordBug.
Second, from \Cref{fig:toxic_whitebox_utility_edit_distance,fig:toxic_whitebox_utility_Jaccard_coef,fig:toxic_blackbox_utility_edit_distance,fig:toxic_blackbox_utility_Jaccard_coef}, we can observe that the adversarial texts preserve good utility in terms of word-level.
Specifically, \Cref{fig:toxic_whitebox_utility_edit_distance} shows that almost 80\% adversarial texts have no more than 20 edit distance comparing with the original texts for three models.
Meanwhile, \Cref{fig:toxic_whitebox_utility_Euclidean_distance,fig:toxic_whitebox_utility_semantic_similarity,fig:toxic_blackbox_utility_Euclidean_distance,fig:toxic_blackbox_utility_semantic_similarity} show that the generated adversarial texts preserve good utility in terms of vector-level.
Specifically, from \Cref{fig:toxic_whitebox_utility_semantic_similarity}, we can see that almost 90\% adverasrial texts preserve 0.9 semantic similarity of the original texts.
These results imply that \textsc{TextBugger} can fool classifiers with high success rate while preserving good utility of the generated adversarial texts.

\subsection{Discussion}

\begin{figure}[tp]
\centering
\subfigure[Word Cloud]{
    \centering
    \includegraphics[width=0.235\textwidth]{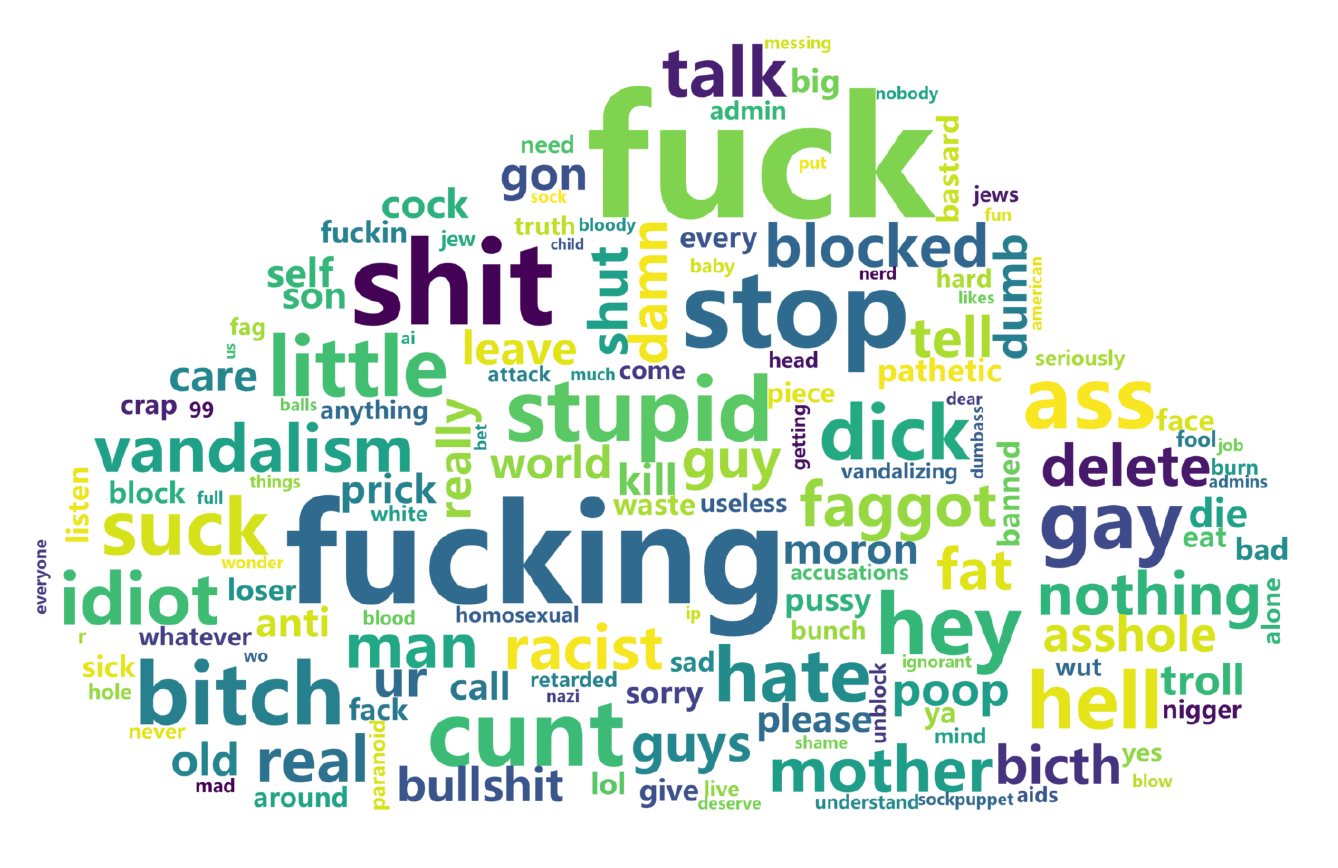}
    \label{fig:toxic_wordcloud}
    \hspace{-0.5cm}
}
\subfigure[Bug Distribution]{
    \centering
    \includegraphics[width=0.235\textwidth]{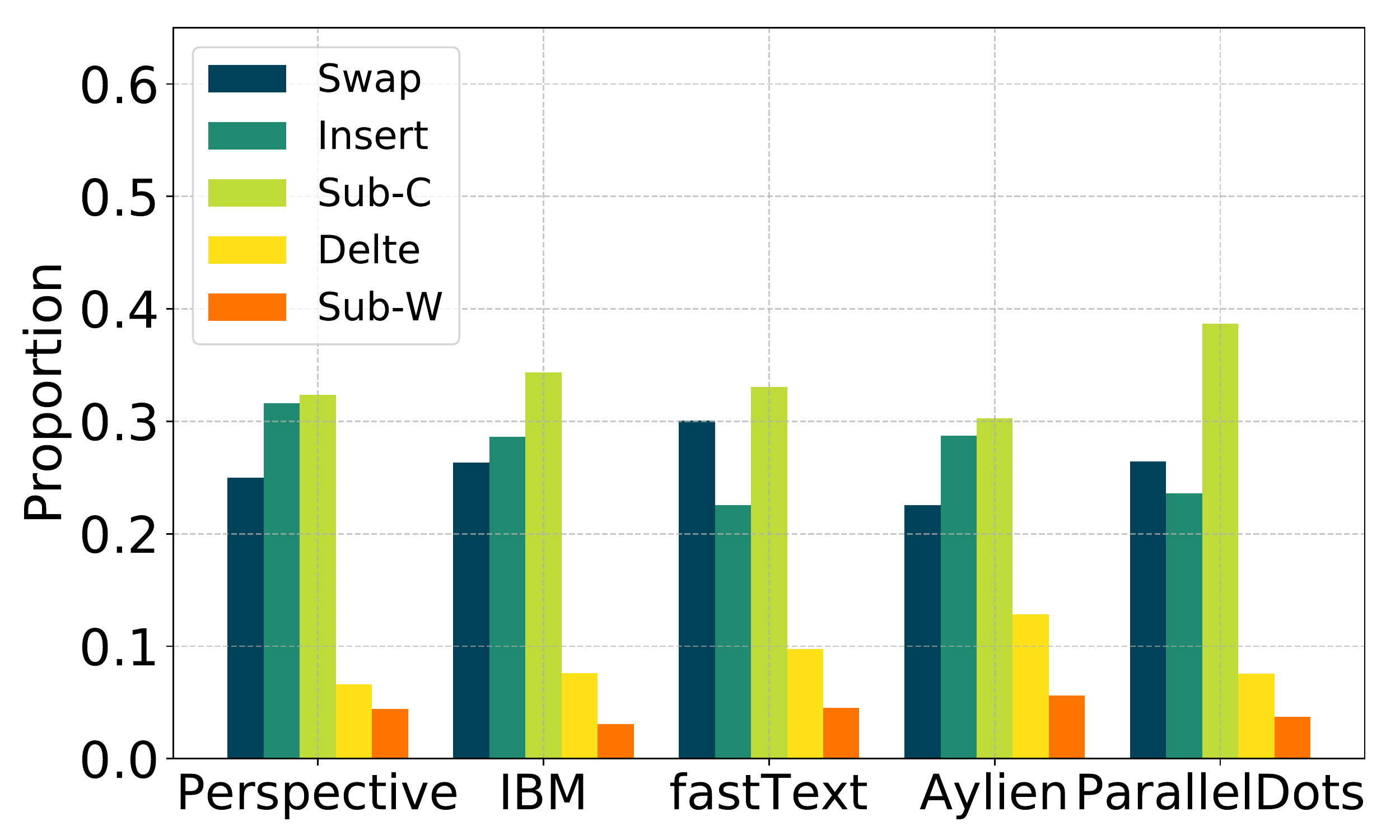}
    \label{fig:kaggle_bug_distribution}
}
\caption{(a) The word cloud is generated from Kaggle dataset against the CNN model. (b) The bug distribution of the adversarial texts is generated from Kaggle dataset against the online platforms.}
\label{fig:kaggle_wordcloud_and_bug_distribution}
\end{figure}

\textbf{Toxic Words Distribution.}
\Cref{fig:toxic_wordcloud} shows the visualization of the found important words according to their frequency, where the higher frequency words have larger font sizes.
Observe that the found important words are indeed toxic words, e.g., ``fuck'', ``dick'', etc.
It is clear that slightly perturbing these toxic words would decrease the toxic score of toxic content.

\textbf{Bug Distribution.}
\Cref{fig:kaggle_bug_distribution} shows the proportion of each operation chosen by the adversary for the black-box attack.
Observe that Sub-C is the dominant operation for all platforms, and Sub-W is still the least used operation.
We do not give detailed analysis since the results are similar to that in \Cref{sec:experiments_white}.

\section{Further Analysis}
\subsection{Transferability}

In the image domain, an important property of adversarial examples is the transferability, i.e., adversarial images generated for one classifier are likely to be misclassified by other classifiers. 
This property can be used to transform black-box attacks to white-box attacks as demonstrated in \cite{papernot2017practical}.
Therefore, we wonder whether adversarial texts also have this property.

In this evaluation, we generated adversarial texts on all three datasets for LR, CNN, and LSTM models.
Then, we evaluated the attack success rate of the generated adversarial texts against other models/platforms.
The experimental results are shown in \Cref{tab:transferability_IMDB_MR,tab:transferability_toxic}.
From \Cref{tab:transferability_IMDB_MR}, we can see that there is a moderate degree of transferability among models.
For instance, the adversarial texts generated on the MR dataset targeting the LR model have 39.5\% success rate when attacking the Azure platform.
This demonstrates that the adversarial texts generated by \textsc{TextBugger} can successfully transfer across multiple models.
From \Cref{tab:transferability_toxic}, we can see that the adversarial texts generated on the Kaggle dataset also has good transferability on Aylien and ParallelDots toxic content detection platforms. 
For instance, the adversarial texts against the LR model has 54.3\% attack success rate on the ParallelDots platform.
This means attackers can use transferability to attack online platforms even they have call limits.

\begin{table}[tp]
    \caption{Transferability on IMDB and MR datasets.}
    \label{tab:transferability_IMDB_MR}
    \centering
    \scalebox{0.92}{
    \begin{tabular}{C{0.55cm}C{0.6cm}C{0.48cm}C{0.48cm}C{0.5cm}C{0.48cm}C{0.48cm}C{0.48cm}C{0.6cm}C{0.5cm}}
    \toprule 
     \multirow{2}{*}{\textbf{Dataset}} & \multirow{2}{*}{\textbf{Model}} & \multicolumn{3}{c}{\textbf{White-box Models}} & \multicolumn{5}{c}{\textbf{Black-box APIs}} \\
     \cmidrule{3-5} \cmidrule(l){6-10}
     & & \textbf{LR} & \textbf{CNN} & \textbf{LSTM} & \textbf{IBM} & \textbf{Azure} & \textbf{Google} & \textbf{fastText} & \textbf{AWS} \\
     \midrule
     \multirow{3}{*}{\textbf{IMDB}}
     & LR & 95.2\% & 20.3\% & 14.5\% & 14.5\% & 24.8\% & 15.1\% & 18.8\% & 19.0\% \\
     & CNN & 28.9\% & 90.5\% & 21.2\% & 21.2\% & 31.4\% & 20.4\% & 25.3\% & 20.0\% \\
     & LSTM & 28.8\% & 23.8\% & 86.6\% & 27.3\% & 26.7\% & 27.4\% & 23.1\% & 25.1\%\\
     \midrule
     \multirow{3}{*}{\textbf{MR}}
     & LR & 92.7\% & 18.3\% & 28.7\% & 22.4\% & 39.5\% & 31.3\% & 19.8\% & 29.8\% \\
     & CNN & 26.5\% & 82.1\% & 31.1\% & 25.3\% & 28.2\% & 21.0\% & 19.1\% & 20.5\% \\
     & LSTM & 21.4\% & 24.6\% & 88.2\% & 21.9\% &  17.7\% & 22.5\% & 16.5\% & 18.7\% \\
    \bottomrule
    \end{tabular}}
\end{table}

\begin{table}[!tp]
    \caption{Transferability on Kaggle dataset.}
    \label{tab:transferability_toxic}
    \centering
    \scalebox{0.94}{
    \begin{tabular}{C{0.5cm}C{0.4cm}C{0.4cm}C{0.5cm}C{1.0cm}cC{0.5cm}C{0.5cm}C{1.0cm}}
    \toprule 
     \multirow{2}{*}{\textbf{Model}} & \multicolumn{3}{c}{\textbf{White-box Models}} & \multicolumn{5}{c}{\textbf{Black-box APIs}} \\
    \cmidrule{2-4} \cmidrule(l){5-9}
    & \textbf{LR} & \textbf{CNN} & \textbf{LSTM} & \textbf{Perspective} & \textbf{IBM} & \textbf{fastText} & \textbf{Aylien} & \textbf{ParallelDots} \\
     \midrule
     LR & 92.3\% & 28.6\% & 32.3\% & 38.1\% & 32.2\% & 29.0\% & 49.7\% & 54.3\% \\
     CNN & 23.7\% & 82.5\% & 35.6\% & 26.4\% & 27.1\% & 25.7\% & 52.6\% & 50.8\% \\
     LSTM & 21.5\% & 26.9\% & 94.8\% & 23.1\% & 26.5\% & 25.9\% & 31.4\% & 28.1\% \\
    \bottomrule
    \end{tabular}}
    \vspace{-0.15cm} 
\end{table}

\subsection{User study}

We perform a user study with human participants on Amazon Mechanical Turk (MTurk) to see whether the applied perturbation will change the human perception of the text's sentiment.
Before the study, we consulted with the IRB office and this study was approved and we did not collect any other information of participants except for necessary result data.

First, we randomly sampled 500 legitimate samples and 500 adversarial samples from IMDB and Kaggle datasets, respectively. 
Among them, half were generated under white-box settings and half were generated under black-box setting. 
All the selected adversarial samples successfully fooled the targeted classifiers.
Then, we presented these samples to the participants and asked them to label the sentiment/toxicity of these samples, i.e., the text is positive/non-toxic or negative/toxic. 
Meanwhile, we also asked them to mark the suspicious words or inappropriate expression in the samples.
To avoid labeling bias, we allow each user to annotate at most 20 reviews and collect 3 annotations from different users for each sample.
Finally, 3,177 valid annotations from 297 AMT workers were obtained in total.

After examining the results, we find that 95.5\% legitimate samples can be correctly classified and 94.9\% adversarial samples can be classified as their original labels. 
Furthermore, we observe that for both legitimate and adversarial samples, almost all the incorrect classifications are made on several specific samples that have some ambiguous expressions.
This indicates that \system did not affect the human judgment on the polarity of the text, i.e., the utility is preserved in the adversarial samples from human perspective, which shows that the generated adversarial texts are of high quality.

\begin{figure}[tp]
\centering
\subfigure[]{
    \centering
    \includegraphics[width=0.23\textwidth]{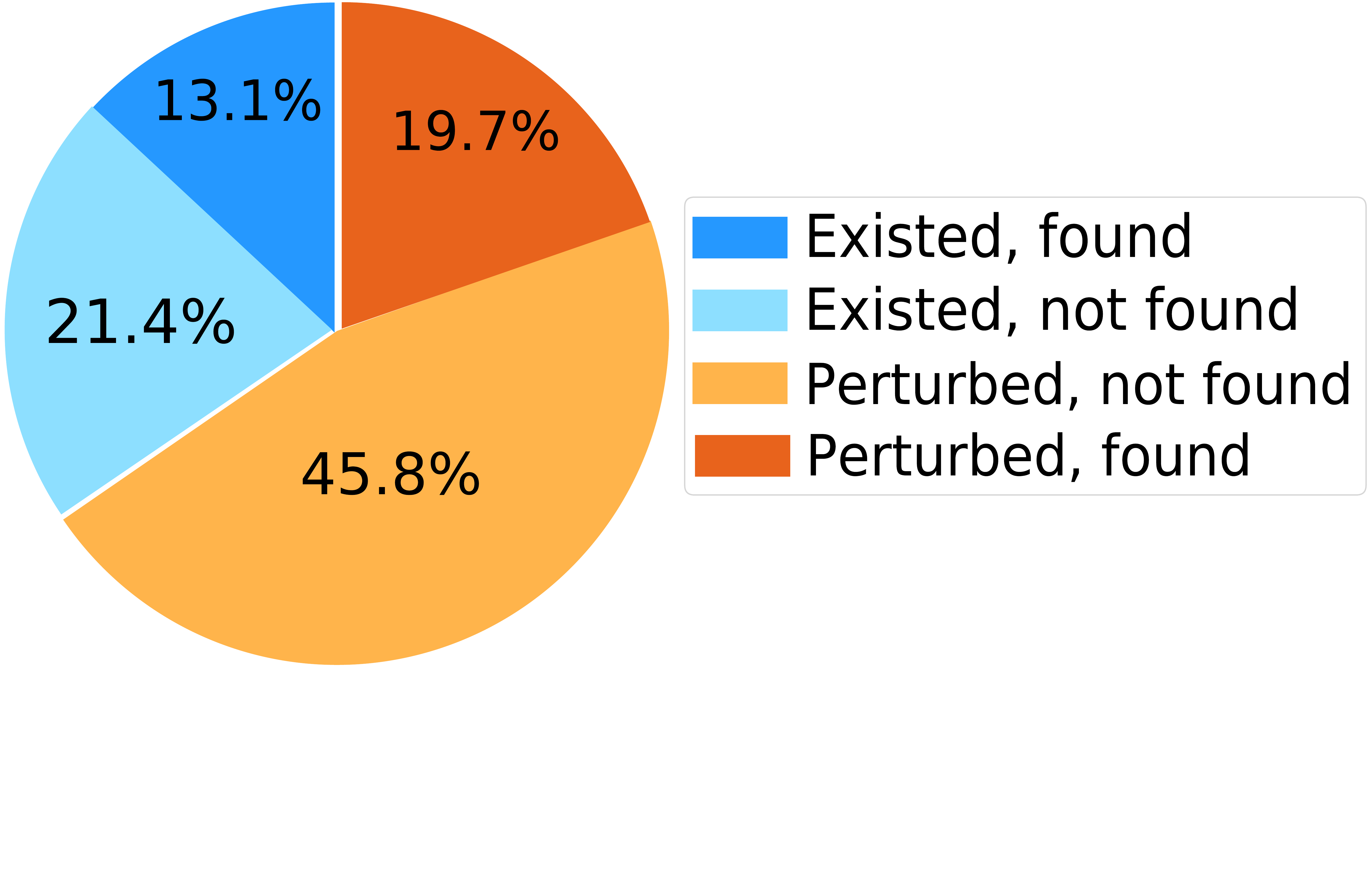}
    \label{fig:user_study_pie}
    \hspace{-0.5cm}
}
\subfigure[]{
    \centering
    \includegraphics[width=0.23\textwidth]{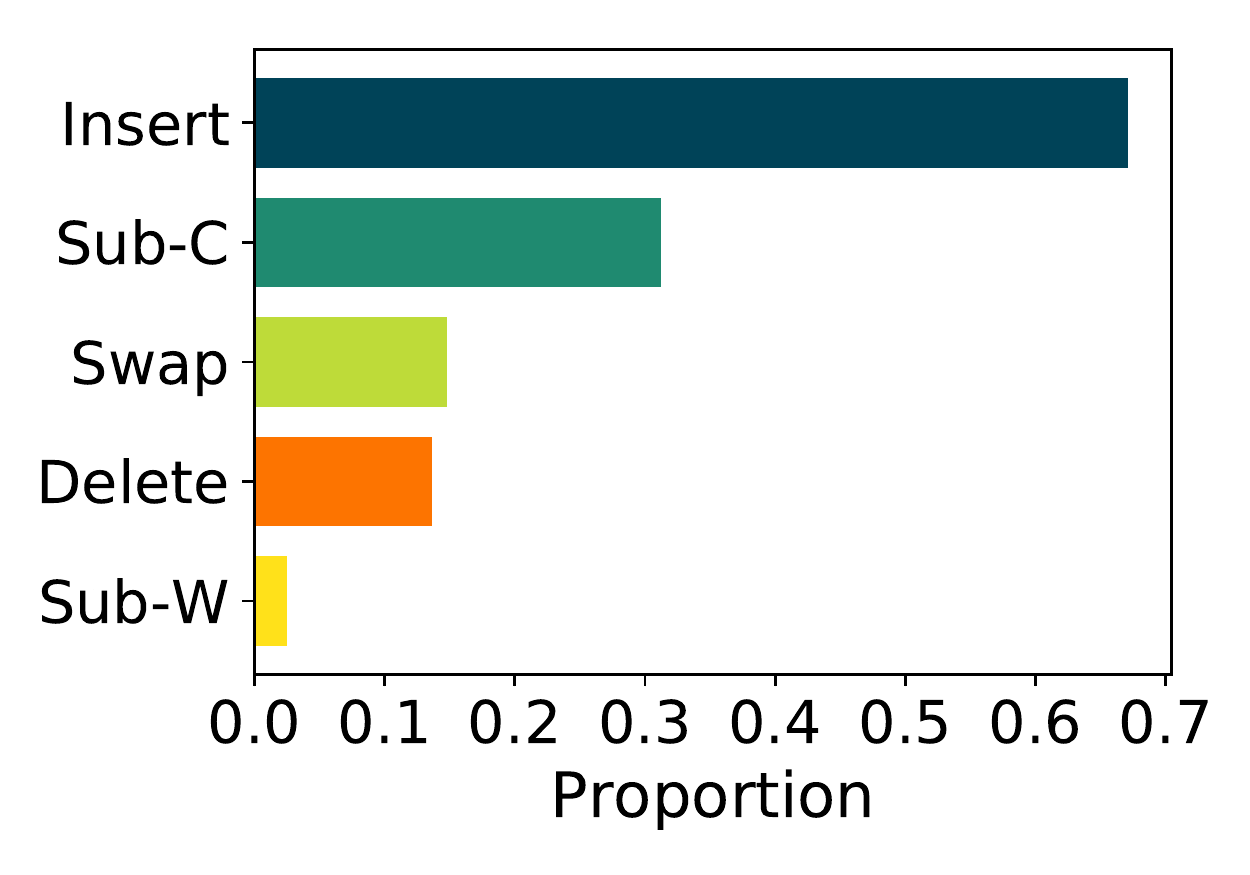}
    \label{fig:user_study_bug_distribution}
}
\caption{The detailed results of user study. (a) The distribution of all mistakes in the samples, including originally existed errors and manully perturbed bugs. (b) The proportion of found bugs accounting for each kind of bug added in the samples. For instance, if there are totally 10 Sub-C perturbations in the samples and we only find 3 of them, the ratio is 3/10=0.3.}
\label{fig:user_study}
\vspace{-0.05cm}
\end{figure}

Some detailed results are shown in \Cref{fig:user_study}.
From \Cref{fig:user_study_pie}, we can see that in our randomly selected samples, the originally existed errors (including spelling mistakes, grammatical errors, etc.) account for 34.5\% of all errors, and the bugs we added account for 65.5\% of all errors. 
Among them, 38.0\% (13.1\%/34.5\%) of existed errors and 30.1\% (19.7\%/65.5\%) of the added bugs are successfully found by participants, which implies that our perturbation is inconspicuous.
From \Cref{fig:user_study_bug_distribution}, we can see that insert is the easiest bug to find, followed by Sub-C.
Specifically, the found Sub-C perturbations are almost the substitution of ``o'' to ``0'', and the substitution of ``l'' to ``1'' is seldom found.
In addition, the Sub-W perturbation is the hardest to find.

\section{Potential Defenses} \label{sec:defense}

To the best of our knowledge, there are few defense methods for the adversarial text attack.
Therefore, we conduct a preliminary exploration of two potential defense schemes, i.e., spelling check and adversarial training.
Specifically, we evaluate the spelling check under the black-box setting and evaluate the adversarial training under the white-box setting.
By default, we use the same implementation settings as that in \Cref{sec:experiments_black}.

\begin{table}[tp]
    \caption{Results of SC on IMDB and MR datasets.}
    \label{tab:defense_spelling_check_sentiment}
    \centering
    \scalebox{0.99}{
    \begin{tabular}{ccccccc}
    \toprule 
    \multirowcell{2}[-.6ex][c]{\centering \textbf{Dataset}} & \multirowcell{2}[-.6ex][c]{\centering \textbf{Method}} & \multicolumn{5}{c}{\textbf{Attack Success Rate}}\\
    \cmidrule{3-7}
     & & \textbf{Google} & \textbf{Watson} & \textbf{Azure} & \textbf{AWS} & \textbf{fastText} \\
     \midrule
     \multirow{2}{*}{\textbf{IMDB}} & \textsc{TextBugger} & 22.2\% & 27.1\% & 32.2\% & 20.8\% & 21.1\% \\
     & DeepWordBug & 15.9\% & 12.2\% & 15.9\% & 9.8\% & 13.6\% \\
     \midrule
     \multirow{2}{*}{\textbf{MR}} & \textsc{TextBugger} & 38.2\% & 36.3\% & 30.8\% & 31.1\% & 28.6\% \\
     & DeepWordBug & 26.9\% & 17.7\% & 13.8\% & 22.1\% & 10.2\% \\
    \bottomrule
    \end{tabular}}
    \vspace{-0.25cm} 
\end{table}

\begin{table}[tp]
    \caption{Results of SC on Kaggle dataset.}
    \label{tab:defense_spelling_check_toxic}
    \centering
    \scalebox{1.01}{
    \begin{tabular}{cccccc}
    \toprule 
     \multirowcell{2}[-.7ex][c]{\centering \textbf{Method}} & \multicolumn{5}{c}{\textbf{Attack Success Rate}}\\
    \cmidrule{2-6}
     & \textbf{Perspective} & \textbf{IBM} & \textbf{fastText} & \textbf{ParallelDots} & \textbf{Aylien} \\
     \midrule
     \textsc{TextBugger} & 35.6\% & 14.8\% & 29.0\% & 40.3\% & 42.7\% \\
     DeepWordBug & 16.5\% & 4.3\% & 13.9\% & 35.1\% & 30.4\% \\
    \bottomrule
    \end{tabular}}
    \vspace{-0.05cm} 
\end{table}

\begin{figure}[!tp]
\centering
\subfigure[IMDB]{
    \centering
    \includegraphics[width=0.235\textwidth]{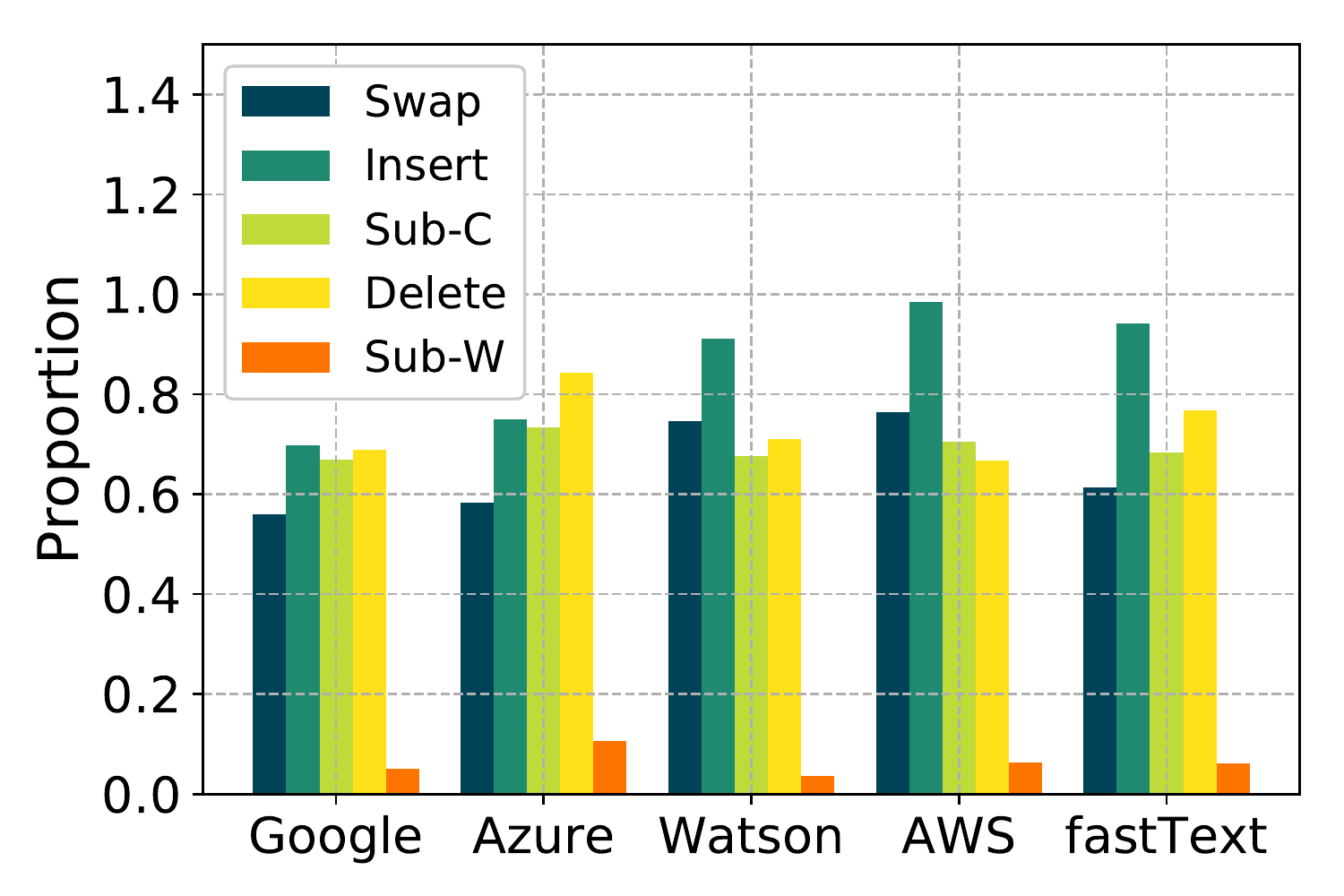}
    \label{fig:imdb_corrected_bug_distribution}
    \hspace{-0.5cm}
}
\subfigure[Kaggle]{
    \centering
    \includegraphics[width=0.235\textwidth]{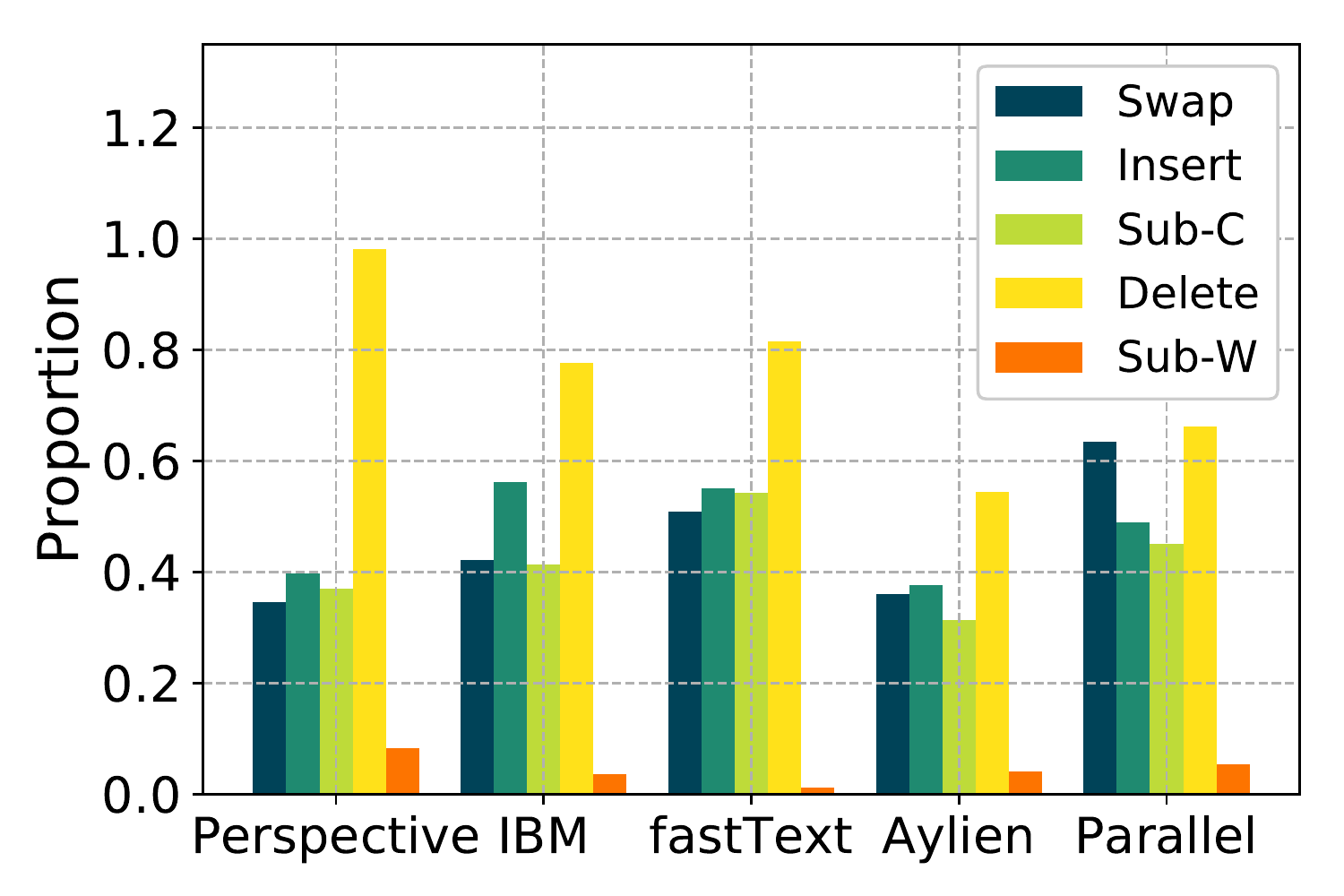}
    \label{fig:kaggle_corrected_bug_distribution}
}
\caption{The ratio of the bugs corrected by spelling check to the total bugs generated on IMDB and Kaggle datasets.}
\label{fig:SC_corrected_bugs_distribution}
\vspace{-0.25cm}
\end{figure}

\textbf{Spelling Check (SC).}
In this experiment, we use a context-aware spelling check service provided by Microsoft Azure\footnote{\url{https://azure.microsoft.com/zh-cn/services/cognitive-services/spell-check/}}.
Experimental results are shown in \Cref{tab:defense_spelling_check_sentiment,tab:defense_spelling_check_toxic}, from which we can see that though many generated adversarial texts can be detected by spell checking, \textsc{TextBugger} still have higher success rate than DeepWordBug on multiple online platforms after correcting the misspelled words.
For instance, when targeting on Perspective API, \textsc{TextBugger} has 35.6\% success rate while DeepWordBug only has 16.5\% after spelling check.
This means \system is still effective and stronger than DeepWordBug.

Further, we analyze the difficulty of correcting each kind of bug.
Specifically, we wonder which kind of bugs is the easiest to correct and which kind of bugs is the hardest to correct.
We count the number of corrected bugs of each kind and show the results in \Cref{fig:SC_corrected_bugs_distribution}.
From \Cref{fig:SC_corrected_bugs_distribution}, we can see that the easiest bug to correct is insert and delete for IMDB and Kaggle respectively.
The hardest bug to correct is Sub-W, which has less than 10\% successfully correction ratio.
This phenomenon partly accounts for why \textsc{TextBugger} is stronger than DeepWordBug.


\textbf{Adversarial Training (AT).}
Adversarial training means training the model with generated adversarial examples.
For instance, in the context of toxic content detection systems, we need to include different modified versions of the toxic documents into the training data.
This method can improve the robustness of machine learning models against adversarial examples \cite{Goodfellow:2014tl}.


In our experiment, we trained the targeted model with the combined dataset for 10 epochs, and the learning rate is set to be 0.0005.
We show the performance of this scheme along with detailed settings in \Cref{tab:defense_adversarial_training}, where accuracy means the prediction accuracy of the new models on the legitimate samples, and success rate with adversarial training (SR with AT) denotes the percentage of the adversarial samples that are misclassified as wrong labels by the new models.
From \Cref{tab:defense_adversarial_training}, we can see that the success rate of adversarial texts decreases while the models' performance on legitimate samples does not change too much with AT.
Therefore, adversarial training might be effective in defending \system.

However, a limitation of adversarial training is that it needs to know the details of the attack strategy and to have sufficient adversarial texts for training. 
In practice, however, attackers usually do not make their approaches or adversarial texts public.
Therefore, adversarial training is limited in defending unknown adversarial attacks.

\begin{table}[tp]
    \caption{Results of AT on three datasets.}
    \label{tab:defense_adversarial_training}
    \centering
    \scalebox{1.05}{
    \begin{tabular}{cccccc}
    \toprule 
     \textbf{Dataset} & \textbf{Model} & \textbf{\# of Leg.} & \textbf{\# of Adv.} & \textbf{Accuracy} & \textbf{SR with AT} \\
     \midrule
     \multirow{3}{*}{IMDB}
     & LR & 25,000 & 2,000 & 83.5\% & 28.0\% \\
     & CNN & 25,000 & 2,000 & 85.3\% & 15.7\% \\
     & LSTM & 25,000 & 2,000 & 88.6\% & 11.6\% \\
     \midrule
     \multirow{3}{*}{MR}
     & LR & 10,662 & 2,000 & 76.3\% & 23.6\% \\
     & CNN & 10,662 & 2,000 & 80.1\% & 16.6\% \\
     & LSTM & 10,662 & 2,000 & 78.5\% & 16.5\% \\
     \midrule
     \multirow{3}{*}{Kaggle}
     & LR & 20,000 & 2,000 & 86.7\% & 27.6\% \\
     & CNN & 20,000 & 2,000 & 91.1\% & 15.4\% \\
     & LSTM & 20,000 & 2,000 & 92.3\% & 11.0\% \\
    \bottomrule
    \end{tabular}}
    \vspace{-0.25cm} 
\end{table}

\textbf{Further Improvement of \textsc{TextBugger}.} 
Though \system can be partly defended by the above methods, attackers can take some strategies to improve the robustness of their attacks.
For instance, attackers can increase the proportion of Sub-W as it is almost cannot be corrected by spelling check.
In addition, attackers can adjust the proportion of different strategies among different platforms.
For instance, attackers can increase the proportion of swap on the Kaggle dataset when targeting the Perspective and Aylien API, since less than 40\% swap modifications have been corrected as shown in \Cref{fig:kaggle_corrected_bug_distribution}.
Attackers can also keep their adversarial attack strategies private and change the parameters of the attack frequently to evade the AT defense.

\section{Discussion} \label{sec:discussion}

\textbf{Extension to Targeted Attack.}
In this paper, we only perform untargeted attacks, i.e., changing the model's output.
However, \system can be easily adapted for targeted attacks (i.e., forcing the model to give a particular output) by modifying Eq.\ref{eq:jacobian_score} from computing the Jacobian matrix with respect to the ground truth label to computing the Jacobian matrix with respect to the targeted label.

\textbf{Limitations and Future Work.}
Though our results demonstrate the existence of natural-language adversarial perturbations, our perturbations could be improved via a more sophisticated algorithm that takes advantage of language processing technologies, such as syntactic parsing, named entity recognition, and paraphrasing.
Furthermore, the existing attack procedure of finding and modifying salient words can be extended to beam search and phrase-level modification, which is an interesting future work.
Developing effective and robust defense schemes is also a promising future work.

\section{Related work} \label{sec:related_work}



\subsection{Adversarial Attacks for Text}


\textbf{Gradient-based Methods.} 
In one of the first attempts at tricking deep neural text classifiers \cite{papernot2016crafting}, Papernot \textit{et al.} proposed a white-box adversarial attack and applied it repetitively to modify an input text until the generated sequence is misclassified.
While their attack was able to fool the classifier, their word-level changes significantly affect the original meaning.
In \cite{ebrahimi2017hotflip}, Ebrahimi \textit{et al.} proposed a gradient-based optimization method that changes one token to another by using the gradients of the model with respect to the one-hot vector input.
In \cite{samanta2017towards}, Samanta \textit{et al.} used the embedding gradient to determine important words. 
Then, heuristic driven rules together with hand-crafted synonyms and typos were designed. 

\textbf{Out-of-Vocabulary Word.}
Some existing works generate adversarial examples for text by replacing a word with one legible but out-of-vocabulary word \cite{belinkov2017synthetic,hosseini2017deceiving,gao2018black}.
In \cite{belinkov2017synthetic}, Belinkov \textit{et al.} showed that character-level machine translation systems are overly sensitive to random character manipulations, such as keyboard typos. 
Similarly, Gao \textit{et al.} proposed DeepWordBug \cite{gao2018black}, which applies character perturbations to generate adversarial texts against deep learning classifiers. 
However, this method is not computationally efficient and cannot be applied in practice.
In \cite{hosseini2017deceiving}, Hosseini \textit{et al.} showed that simple modifications, such as adding spaces or dots between characters, can drastically change the toxicity score from Perspective API.

\textbf{Replace with Semantically/Syntactically Similar Words.} 
In \cite{alzantot2018generating}, Alzantot \textit{et al.} generated adversarial text against sentiment analysis models by leveraging a genetic algorithm and only replacing words with semantically similar ones. 
In \cite{sears:acl18}, Ribeiro \textit{et al.} replaced tokens by random words of the same POS tag with a probability proportional to the embedding similarity. 

\textbf{Other Methods.}
In \cite{jia2017adversarial}, Jia \textit{et al.} generated adversarial examples for evaluating reading comprehension systems by adding distracting sentences to the input document.
However, their method requires manual intervention to polish the added sentences.
In \cite{zhao2017generating}, Zhao \textit{et al.} used Generative Adversarial Networks (GANs) to generate adversarial sequences for textual entailment and machine translation applications.
However, this method requires neural text generation, which is limited to short texts.

\subsection{Defense}
To the best of our knowledge, existing defense methods for adversarial examples mainly focus on the image domain and have not been systematically studied in the text domain.
For instance, the adversarial training, one of the famous defense methods for adversarial images, has been only used as a regularization technique in the DLTU task \cite{li2016learning,miyato2017adversarial}.
These works only focused on improving the accuracy on clean examples, rather than defending textual adversarial examples.

\subsection{Remarks}
In summary, the following aspects distinguish \system from existing adversarial attacks on DLTU systems.
First, we use both character-level and word-level perturbations to generate adversarial texts, in contrast to previous works that use the projected gradient \cite{papernot2016crafting} or linguistic-driven steps \cite{jia2017adversarial}.
Second, we demonstrate that our method has great efficiency while previous works seldom evaluate the efficiency of their methods \cite{gao2018black,ebrahimi2017hotflip}. 
Finally, most if not all previous works only evaluate their method on self-implemented models \cite{gao2018black,gong2018adversarial,samanta2017towards}, or just evaluate them on one or two public offline models \cite{jia2017adversarial,ebrahimi2017hotflip}.
By contrast, we evaluate the generated adversarial examples on 15 popular real-world online DLTU systems, including Google Cloud NLP, IBM Watson, Amazon AWS, Microsoft Azure, Facebook fastText, etc.
The results demonstrate that \system is more general and robust.

\section{Conclusion} \label{conclusion}
Overall, we study adversarial attacks against state-of-the-art sentiment analysis and toxic content detection models/platforms under both white-box and black-box settings.
Extensive experimental results demonstrate that \system is effective and efficient for generating targeted adversarial NLP.
The transferability of such examples hint at potential vulnerabilities in many real applications, including text filtering systems (e.g., racism, pornography, terrorism, and riots), online recommendation systems, etc. 
Our findings also show the possibility of spelling check and adversarial training in defending against such attacks. Ensemble of linguistically-aware or structurally-aware based defense system can be further explored to improve robustness.

\section*{Acknowledgment}
This work was partly supported by NSFC under No. 61772466, the Zhejiang Provincial Natural Science Foundation for Distinguished Young Scholars under No. LR19F020003, the Provincial Key Research and Development Program of Zhejiang, China under No. 2017C01055, and the Alibaba-ZJU Joint Research Institute of Frontier Technologies.
Ting Wang is partially supported by the National Science Foundation under Grant No. 1566526 and 1718787.
Bo Li is partially supported by the Defense Advanced Research Projects Agency (DARPA).

\balance 
\bibliographystyle{IEEEtranS} 
\bibliography{adv_text}

\end{document}